\documentstyle[12pt,graphicx]{article}

\def\mathfrak{\bf}

\def\rI{{{}_{\rm I}}}
\def\rJ{{{}_{\rm J}}}
\def\rK{{{}_{\rm K}}}
\def\rL{{{}_{\rm L}}}
\def\hj{{\hat\jmath}}
\def\hk{{\hat k}}
\def\hi{{\hat\imath}}

\def\vCent#1{\vcenter{\hbox{\hss#1\hss}}}

\def\be{\begin{equation}}
\def\ee{\end{equation}}
\def\bea{\begin{eqnarray}}
\def\eea{\end{eqnarray}}

\def\dt#1{\on{\hbox{\bf .}}{#1}}                
\def\Dot#1{\dt{#1}}
\def\IR{\relax{\rm I\kern-.18em R}}
\def\binomial#1#2{\left(\,{\buildrel
{\raise4pt\hbox{$\displaystyle{#1}$}}\over
{\raise-6pt\hbox{$\displaystyle{#2}$}}}\,\right)}

\def\[{\lfloor{\hskip 0.35pt}\!\!\!\lceil}
\def\]{\rfloor{\hskip 0.35pt}\!\!\!\rceil}

\newcommand{\AmS}{{\protect\the\textfont2
  A\kern-.1667em\lower.5ex\hbox{M}\kern-.125emS}}

\catcode`@=11
\def\un#1{\relax\ifmmode\@@underline#1\else
        $\@@underline{\hbox{#1}}$\relax\fi}
\catcode`@=12

\def\fracm#1#2{\hbox{\large{${\frac{{#1}}{{#2}}}$}}}

\def\ad{{\kern0.5pt
                   \alpha \kern-5.05pt
\raise5.8pt\hbox{$\textstyle.$}\kern
0.5pt}}

\def\Dot#1{{\kern0.5pt
     {#1} \kern-5.05pt \raise5.8pt\hbox{$\textstyle.$}\kern
0.5pt}}

\def\a{\alpha}
\def\b{\beta}

\def\d{\delta}
\def\e{\epsilon}

\def\g{\gamma}

\def\l{\lambda}
\def\m{\mu}
\def\n{\nu}

\def\r{\rho}
\def\s{\sigma}
\def\t{\tau}

\def\D{\Delta}

\def\L{\Lambda}

\def\bo{{\raise.15ex\hbox{\large$\Box$}}}               
\def\pa{\partial}                                       
\def\TH{{\raise.2ex\hbox{$\displaystyle \bigodot$}\mskip-4.7mu \llap H
\;}}
\def\face{{\raise.2ex\hbox{$\displaystyle \bigodot$}\mskip-2.2mu \llap
{$\ddot
        \smile$}}}                                      

\def\bj#1{{}_{\rm #1}}                          
\def\Tilde#1{\widetilde{#1}}                    
\def\Hat#1{\widehat{#1}}                        
\def\Bar#1{\overline{#1}}                       
\def\leftrightarrowfill{$\mathsurround=0pt \mathord\leftarrow \mkern-6mu
        \cleaders\hbox{$\mkern-2mu \mathord- \mkern-2mu$}\hfill
        \mkern-6mu \mathord\rightarrow$}
\def\dvec#1{\vbox{\ialign{##\crcr
        \leftrightarrowfill\crcr\noalign{\kern-1pt\nointerlineskip}
        $\hfil\displaystyle{#1}\hfil$\crcr}}}           
\def\dt#1{{\buildrel {\hbox{\LARGE .}} \over {#1}}}     

\def\fracm#1#2{\hbox{\large{${\frac{{#1}}{{#2}}}$}}}
\def\frac#1#2{{\textstyle{#1\over\vphantom2\smash{\raise.20ex
        \hbox{$\scriptstyle{#2}$}}}}}                   
\def\sfrac#1#2{{\vphantom1\smash{\lower.5ex\hbox{\small$#1$}}\over
        \vphantom1\smash{\raise.4ex\hbox{\small$#2$}}}} 
\def\bfrac#1#2{{\vphantom1\smash{\lower.5ex\hbox{$#1$}}\over
        \vphantom1\smash{\raise.3ex\hbox{$#2$}}}}       
\def\afrac#1#2{{\vphantom1\smash{\lower.5ex\hbox{$#1$}}\over#2}}    
\def\on#1#2{\mathop{\null#2}\limits^{#1}}               

\newskip\humongous \humongous=0pt plus 1000pt minus 1000pt
\def\caja{\mathsurround=0pt}
\def\eqalign#1{\,\vcenter{\openup2\jot \caja
        \ialign{\strut \hfil$\displaystyle{##}$&$
        \displaystyle{{}##}$\hfil\crcr#1\crcr}}\,}
\newif\ifdtup

  \def\pp{{\mathchoice
          {
              \kern 1pt%
              \raise 1pt
              \vbox{\hrule width5pt height0.4pt depth0pt
                    \kern -2pt
                    \hbox{\kern 2.3pt
                          \vrule width0.4pt height6pt depth0pt
                          }
                    \kern -2pt
                    \hrule width5pt height0.4pt depth0pt}%
                    \kern 1pt
           }
            {
              \kern 1pt%
              \raise 1pt
              \vbox{\hrule width4.3pt height0.4pt depth0pt
                    \kern -1.8pt
                    \hbox{\kern 1.95pt
                          \vrule width0.4pt height5.4pt depth0pt
                          }
                    \kern -1.8pt
                    \hrule width4.3pt height0.4pt depth0pt}%
                    \kern 1pt
            }
            {
              \kern 0.5pt%
              \raise 1pt
              \vbox{\hrule width4.0pt height0.3pt depth0pt
                    \kern -1.9pt  
                    \hbox{\kern 1.85pt
                          \vrule width0.3pt height5.7pt depth0pt
                          }
                    \kern -1.9pt
                    \hrule width4.0pt height0.3pt depth0pt}%
                    \kern 0.5pt
            }
            {
              \kern 0.5pt%
              \raise 1pt
              \vbox{\hrule width3.6pt height0.3pt depth0pt
                    \kern -1.5pt
                    \hbox{\kern 1.65pt
                          \vrule width0.3pt height4.5pt depth0pt
                          }
                    \kern -1.5pt
                    \hrule width3.6pt height0.3pt depth0pt}%
                    \kern 0.5pt
            }
        }}

  \def\mm{{\mathchoice
                       {
                             \kern 1pt
               \raise 1pt    \vbox{\hrule width5pt height0.4pt depth0pt
                                  \kern 2pt
                                  \hrule width5pt height0.4pt depth0pt}
                             \kern 1pt}
                       {
                            \kern 1pt
               \raise 1pt \vbox{\hrule width4.3pt height0.4pt depth0pt
                                  \kern 1.8pt
                                  \hrule width4.3pt height0.4pt depth0pt}
                             \kern 1pt}
                       {
                            \kern 0.5pt
               \raise 1pt
                            \vbox{\hrule width4.0pt height0.3pt depth0pt
                                  \kern 1.9pt
                                  \hrule width4.0pt height0.3pt depth0pt}
                            \kern 1pt}
                       {
                           \kern 0.5pt
             \raise 1pt  \vbox{\hrule width3.6pt height0.3pt depth0pt
                                  \kern 1.5pt
                                  \hrule width3.6pt height0.3pt depth0pt}
                           \kern 0.5pt}
                       }}

\def\pd{{\kern0.5pt
                   + \kern-5.05pt \raise5.8pt\hbox{$\textstyle.$}\kern
0.5pt}}

\def\pmd{{\kern0.5pt
                  \pm \kern-5.05pt \raise6.3pt\hbox{$\textstyle.$}\kern1.5pt}}

\def\md{{\mathchoice
   {
      {{\kern 1pt - \kern-6.2pt \raise5pt\hbox{$\textstyle.$}\kern 1pt}}}
    {
      {{\kern 1pt - \kern-6.2pt \raise5pt\hbox{$\textstyle.$}\kern 1pt}}}
    {
      {\kern0.5pt - \kern-5.05pt \raise3.4pt\hbox{$\textstyle.$}\kern0.5pt}}
    {
      {\kern0.5pt - \kern-5.05pt \raise3.4pt\hbox{$\textstyle.$}\kern0.5pt}}}}

\def\ad{{\dot{\alpha}}}

\def\pp{{\mathchoice
          {
              \kern 1pt%
              \raise 1pt
              \vbox{\hrule width5pt height0.4pt depth0pt
                    \kern -2pt
                    \hbox{\kern 2.3pt
                          \vrule width0.4pt height6pt depth0pt
                          }
                    \kern -2pt
                    \hrule width5pt height0.4pt depth0pt}%
                    \kern 1pt
           }
            {
              \kern 1pt%
              \raise 1pt
              \vbox{\hrule width4.3pt height0.4pt depth0pt
                    \kern -1.8pt
                    \hbox{\kern 1.95pt
                          \vrule width0.4pt height5.4pt depth0pt
                          }
                    \kern -1.8pt
                    \hrule width4.3pt height0.4pt depth0pt}%
                    \kern 1pt
            }
            {
              \kern 0.5pt%
              \raise 1pt
              \vbox{\hrule width4.0pt height0.3pt depth0pt
                    \kern -1.9pt  
                    \hbox{\kern 1.85pt
                          \vrule width0.3pt height5.7pt depth0pt
                          }
                    \kern -1.9pt
                    \hrule width4.0pt height0.3pt depth0pt}%
                    \kern 0.5pt
            }
            {
              \kern 0.5pt%
              \raise 1pt
              \vbox{\hrule width3.6pt height0.3pt depth0pt
                    \kern -1.5pt
                    \hbox{\kern 1.65pt
                          \vrule width0.3pt height4.5pt depth0pt
                          }
                    \kern -1.5pt
                    \hrule width3.6pt height0.3pt depth0pt}%
                    \kern 0.5pt
            }
        }}

  \def\mm{{\mathchoice
                       {
                             \kern 1pt
               \raise 1pt    \vbox{\hrule width5pt height0.4pt depth0pt
                                  \kern 2pt
                                  \hrule width5pt height0.4pt depth0pt}
                             \kern 1pt}
                       {
                            \kern 1pt
               \raise 1pt \vbox{\hrule width4.3pt height0.4pt depth0pt
                                  \kern 1.8pt
                                  \hrule width4.3pt height0.4pt depth0pt}
                             \kern 1pt}
                       {
                            \kern 0.5pt
               \raise 1pt
                            \vbox{\hrule width4.0pt height0.3pt depth0pt
                                  \kern 1.9pt
                                  \hrule width4.0pt height0.3pt depth0pt}
                            \kern 1pt}
                       {
                           \kern 0.5pt
             \raise 1pt  \vbox{\hrule width3.6pt height0.3pt depth0pt
                                  \kern 1.5pt
                                  \hrule width3.6pt height0.3pt depth0pt}
                           \kern 0.5pt}
                       }}

\def\pd{{\kern0.5pt
                   + \kern-5.05pt \raise5.8pt\hbox{$\textstyle.$}\kern
0.5pt}}

\def\pmd{{\kern0.5pt
                  \pm \kern-5.05pt \raise6.3pt\hbox{$\textstyle.$}\kern1.5pt}}

\def\md{{\mathchoice
   {
      {{\kern 1pt - \kern-6.2pt \raise5pt\hbox{$\textstyle.$}\kern 1pt}}}
    {
      {{\kern 1pt - \kern-6.2pt \raise5pt\hbox{$\textstyle.$}\kern 1pt}}}
    {
      {\kern0.5pt - \kern-5.05pt \raise3.4pt\hbox{$\textstyle.$}\kern0.5pt}}
    {
      {\kern0.5pt - \kern-5.05pt \raise3.4pt\hbox{$\textstyle.$}\kern0.5pt}}}}

\def\dslash{\not{\hbox{\kern-2pt $\partial$}}}
\def\Dslash{\not{\hbox{\kern-4pt $D$}}}
\def\pslash{\not{\hbox{\kern-2.3pt $p$}}}
 \newtoks\slashfraction
 \slashfraction={.13}
 \def\slash#1{\setbox0\hbox{$ #1 $}
 \setbox0\hbox to \the\slashfraction\wd0{\hss \box0}/\box0 }

\font\ro=cmsy10                          
\def\kcr{{\hbox{\ro \char'170}}}                
\def\ktl{{\hbox{\ro \char'170}}}        
\def\ktr{{\hbox{\ro \char'170}}}        
\def\kbl{{\hbox{\ro \char'170}}}        
\def\kbr{{\hbox{\ro \char'170}}}        

\def\plpl{\raise-2pt\hbox{$\raise3pt\hbox{$_+$}\hskip-6.67pt\raise0.0pt
\hbox{$^+$}\hskip 0.01pt$}}
\def\mimi{\raise-2pt\hbox{$\raise3pt\hbox{$_-$}\hskip-6.67pt\raise0.0pt
\hbox{$^-$}\hskip 0.01pt$}}

\def\bo{{\raise.15ex\hbox{\large$\Box$}}}               
\def\pa{\partial}                                       
\def\TH{{\raise.2ex\hbox{$\displaystyle \bigodot$}\mskip-4.7mu \llap H \;}}
\def\face{{\raise.2ex\hbox{$\displaystyle \bigodot$}\mskip-2.2mu \llap {$\ddot
        \smile$}}}                                      

\def\Tilde#1{\widetilde{#1}}                    
\def\Hat#1{\widehat{#1}}                        
\def\Bar#1{\overline{#1}}                       
\def\leftrightarrowfill{$\mathsurround=0pt \mathord\leftarrow \mkern-6mu
        \cleaders\hbox{$\mkern-2mu \mathord- \mkern-2mu$}\hfill
        \mkern-6mu \mathord\rightarrow$}
\def\dvec#1{\vbox{\ialign{##\crcr
        \leftrightarrowfill\crcr\noalign{\kern-1pt\nointerlineskip}
        $\hfil\displaystyle{#1}\hfil$\crcr}}}           
\def\dt#1{{\buildrel {\hbox{\LARGE .}} \over {#1}}}     

\def\fracm#1#2{\hbox{\large{${\frac{{#1}}{{#2}}}$}}}
\def\frac#1#2{{\textstyle{#1\over\vphantom2\smash{\raise.20ex
        \hbox{$\scriptstyle{#2}$}}}}}                   
\def\sfrac#1#2{{\vphantom1\smash{\lower.5ex\hbox{\small$#1$}}\over
        \vphantom1\smash{\raise.4ex\hbox{\small$#2$}}}} 
\def\bfrac#1#2{{\vphantom1\smash{\lower.5ex\hbox{$#1$}}\over
        \vphantom1\smash{\raise.3ex\hbox{$#2$}}}}       
\def\afrac#1#2{{\vphantom1\smash{\lower.5ex\hbox{$#1$}}\over#2}}    
\def\on#1#2{\mathop{\null#2}\limits^{#1}}               

\parskip=\medskipamount                 
\thicklines                         

\def\oldheadpic{                                
        \setlength{\unitlength}{.4mm}
        \thinlines
        \par
        \begin{picture}(349,16)
        \put(325,16){\line(1,0){4}}
        \put(330,16){\line(1,0){4}}
        \put(340,16){\line(1,0){4}}
        \put(335,0){\line(1,0){4}}
        \put(340,0){\line(1,0){4}}
        \put(345,0){\line(1,0){4}}
        \put(329,0){\line(0,1){16}}
        \put(330,0){\line(0,1){16}}
        \put(339,0){\line(0,1){16}}
        \put(340,0){\line(0,1){16}}
        \put(344,0){\line(0,1){16}}
        \put(345,0){\line(0,1){16}}
        \put(329,16){\oval(8,32)[bl]}
        \put(330,16){\oval(8,32)[br]}
        \put(339,0){\oval(8,32)[tl]}
        \put(345,0){\oval(8,32)[tr]}
        \end{picture}
        \par
        \thicklines
        \vskip.2in}
\def\oldtitle#1#2#3#4{\oldheadpic\begin{center}\vglue.5in{\large\bf #1}\\[.6in]
        {#2}\\[.1in] {\it Department of Physics and Astronomy}\\
        {\it University of Maryland, College Park, MD 20742}\\[.6in]
        Physics Publication \#{#3}\\ {#4}\\[1.5in] {\bf ABSTRACT}\\[.1in]
        \end{center} \begin{quotation}}                 
\def\oldTitle#1#2#3#4#5#6#7{\oldheadpic\begin{center} \vglue .4in
        {\large\bf #1}\\[.4in]
        {#2}\\[.1in] {\it Department of Physics and Astronomy}\\
        {\it University of Maryland, College Park, MD 20742}\\[.1in]
        {#3}\\[.1in] {\it {#4}}\\ {\it {#5}}\\[.4in]
        Physics Publication \#{#6}\\ {#7}\\[.5in] {\bf ABSTRACT}\\[.1in]
        \end{center} \begin{quotation}}                 
\def\border{                                            
        \setlength{\unitlength}{1mm}
        \newcount\xco
        \newcount\yco
        \xco=-21
        \yco=12
        \begin{picture}(140,0)
        \put(\xco,\yco){$\ktl$}
        \advance\yco by-1
        {\loop
        \put(\xco,\yco){$\kcr$}
        \advance\yco by-2
        \ifnum\yco>-240
        \repeat
        \put(\xco,\yco){$\kbl$}}
        \xco=158
        \yco=12
        \put(\xco,\yco){$\ktr$}
        \advance\yco by-1
        {\loop
        \put(\xco,\yco){$\kcr$}
        \advance\yco by-2
        \ifnum\yco>-240
        \repeat
        \put(\xco,\yco){$\kbr$}}
        \put(-20,13){\tiny **University of Maryland * Center for String and
         Particle  Theory* Physics Department***University of Maryland *Center
        for String and Particle  Theory** }
        \put(-20,-241.5){\tiny  The University of Iowa Particle Theory
        Group The University of Iowa Particle Theory Group The
        University of Iowa Particle Theory Group The University}
        \end{picture}
        \par\vskip-8mm}
\def\bordero{                                           
        \setlength{\unitlength}{1mm}
        \newcount\xco
        \newcount\yco
        \xco=-31
        \yco=12
        \begin{picture}(140,0)
        \put(\xco,\yco){$\ktl$}
        \advance\yco by-1
        {\loop
        \put(\xco,\yco){$\kclr}
        \advance\yco by-2
        \ifnum\yco>-240
        \repeat
        \put(\xco,\yco){$\kbl$}}
        \xco=151
        \yco=12
        \put(\xco,\yco){$\ktr$}
        \advance\yco by-1
        {\loop
        \put(\xco,\yco){$\kcr$}
        \advance\yco by-2
        \ifnum\yco>-240
        \repeat
        \put(\xco,\yco){$\kbr$}}
        \put(-20,12){\ooo bacdefghidfghghdhededbihdgdfdfhhdheidhdhebaaahjhhdahba

hgdedge
   hgfdiehhgdigicba}
        \put(-20,-241.5){\ooo ababaighefdbfghgeahgdfgafagihdidihiidhiagfedhadbfd

ecdcdfa
   gdcbhaddhbgfchbgfdacfediacbabab}
        \end{picture}
        \par\vskip-8mm}
\def\headpic{                                           
        \indent
        \setlength{\unitlength}{.4mm}
        \thinlines
        \par
        \begin{picture}(29,16)
        \put(165,16){\line(1,0){4}}
        \put(170,16){\line(1,0){4}}
        \put(180,16){\line(1,0){4}}
        \put(175,0){\line(1,0){4}}
        \put(180,0){\line(1,0){4}}
        \put(185,0){\line(1,0){4}}
        \put(169,0){\line(0,1){16}}
        \put(170,0){\line(0,1){16}}
        \put(179,0){\line(0,1){16}}
        \put(180,0){\line(0,1){16}}
        \put(184,0){\line(0,1){16}}
        \put(185,0){\line(0,1){16}}
        \put(169,16){\oval(8,32)[bl]}
        \put(170,16){\oval(8,32)[br]}
        \put(179,0){\oval(8,32)[tl]}
        \put(185,0){\oval(8,32)[tr]}
        \end{picture}
        \par\vskip-6.5mm
        \thicklines}
\def\title#1#2#3#4{\border\headpic {\hbox to\hsize{#4 \hfill UMDEPP #3}}\par
        \begin{center} \vglue .5in {\large\bf #1}\\[.6in]
        {#2}\\[.1in] {\it Department of Physics and Astronomy}\\
        {\it University of Maryland, College Park, MD 20742}\\[1.5in]
        {\bf ABSTRACT}\\[.1in] \end{center} \begin{quotation}}  
\def\Title#1#2#3#4#5#6#7{\border\headpic
        {\hbox to\hsize{#7 \hfill UMDEPP #6}}\par
        \begin{center} \vglue .4in {\large\bf #1}\\[.4in]
        {#2}\\[.1in] {\it Department of Physics and Astronomy}\\
        {\it University of Maryland, College Park, MD 20742}\\[.1in]
        {#3}\\[.1in] {\it {#4}}\\ {\it {#5}}\\[.5in] {\bf ABSTRACT}\\[.1in]
        \end{center} \begin{quotation}}                 
\def\endtitle{\end{quotation}\newpage}                  

\def\qd{{\kern0.5pt
                   q \kern-5.05pt \raise5.8pt\hbox{$\textstyle.$}\kern
0.5pt}}

\begin{document}

\def\dt#1{\on{\hbox{\bf .}}{#1}}                
\def\Dot#1{\dt{#1}}

\def\gfrac#1#2{\frac {\scriptstyle{#1}}
        {\mbox{\raisebox{-.6ex}{$\scriptstyle{#2}$}}}}
\def\gg{{\hbox{\sc g}}}
\border\headpic {\hbox to\hsize{February 2009 \hfill
{UMDEPP 08-025}}}
\par
{$~$ \hfill
{hep-th/0902.3830}}
\par

\setlength{\oddsidemargin}{0.3in}
\setlength{\evensidemargin}{-0.3in}
\begin{center}
\vglue .10in
{\large\bf 4D, ${\cal N}$ = 1 Supersymmetry Genomics (I)}\\[.5in]

S.\, James Gates, Jr.\footnote{gatess@wam.umd.edu}${}^{\dagger}$,
James Gonzales,
Boanne MacGregor\footnote{boanne-macgregor@uiowa.edu},
James Parker
, Ruben Polo-Sherk\footnote{rpoloshe@umd.edu},
Vincent G.\ J.\ Rodgers\footnote{vrodgers@newton.physics.uiowa.edu}${}^*$
and Luke Wassink\footnote{luke-wassink@uiowa.edu}
\\[0.3in]
${}^\dag${\it Center for String and Particle Theory\\
Department of Physics, University of Maryland\\
College Park, MD 20742-4111 USA}
\\[0.1in]
{\it {and}}
\\[0.1in]
${}^*${\it Department of Physics and Astronomy\\
The University of Iowa\\
Iowa City, IA 52242 USA}
\\[.5in]
{\bf ABSTRACT}\\[.01in]
\end{center}
\begin{quotation}
{Presented in this paper the nature of the supersymmetrical representation theory
behind 4D, $\cal N$ = 1 theories, as described by component fields,  is investigated using the 
tools of Adinkras and Garden Algebras.   A survey of familiar matter multiplets using these
techniques reveals they are described by two fundamental valise Adinkras that are given the 
names of the cis-Valise (c-V) and  the trans-Valise (t-V).  A conjecture is made that all off-shell 4D, $\cal N$ 
= 1 component descriptions of supermultiplets are associated with two integers ($n_c$, $n_t$) 
- the numbers of c-V and t-V Adinkras that occur in the representation.}

\endtitle

\setlength{\oddsidemargin}{0.3in}
\setlength{\evensidemargin}{-0.3in}

\setcounter{equation}{0}
\section{Introduction}

$~~~$ One of the long-standing unsolved problems in discussions of supersymmetrical 
theories is the notorious ``off-shell problem.''  It is meant by this term that for a given set 
of propagating fields, there is currently no generally known prescription for how to augment 
this set with an additional finite number of fields (called ``auxiliary fields'') such that the 
algebra (see the appendix for the conventions used in this work)
\be
\{ \, {\rm Q}_a^{{}^{\rm I}}  \,,\,  {\rm Q}_b^{{}^{\rm J}}  \,  \} \,  
~=~  i\, 2 \, \d{}^{{}^{\rm I}  \, {}^{\rm J} }(\gamma^\mu){}_{a \,b}\,  \partial_\mu  ~~~,
\label{SUSYdef}   
\ee
is satisfied for `supercharges' Q${}_a$ that act non-trivially on both the propagating 
and auxiliary fields.  They should act in such a way so as to {\em {not}} impose any particular 
dynamical equations on the propagating fields nor the auxiliary ones.  Multiplets of both 
propagating and auxiliary fields that satisfy (\ref{SUSYdef}) and the conditions in the 
previous sentence are called ``off-shell representations.''  Though the corresponding problem
without auxiliary fields  has long been resolved (see for example \cite{DZ1}), finding all such sets 
of fields in the off-shell case has been an unsolved problem since the birth of supersymmetry.

There is a general belief that  this is an `impossible' problem to solve.  A widely accepted 
no-go theorem \cite{RS-T} has been derived that would seem to preclude the existence 
of such off-shell representations  for a large class of {\em {interesting}} theories such as 
the 4D, $\cal N$ =  2 Hypermultiplet \cite{FSHypr}, 4D, $\cal N$ = 4 SUSY YM theory 
\cite{N4YM} and all 10D supersymmetrical theories that emerge as the low-energy 
zero-slope limits of superstring theories \cite{LESstrings}.

Two approaches have arisen to surmount the  ``off-shell problem.''  One of these is
known as the `harmonic superspace' approach \cite{GIKOS} and the other is referred
to as `projective superspace' approach \cite{PrjSspace}.  At the time of their creation,
each approach seemed distinct but with the common feature of providing an off-shell
description of the Hypermultiplet at the expense of using an infinite number of auxiliary
fields.   In the language of harmonic superspace the Hypermultiplet is known as the 
`q-hypermultiplet.'  Correspondingly, in the language of projective superspace the 
Hypermultiplet is known as the `polar-hypermultiplet.' 

Differences in the two approaches do exist.  One of the most pointed is that only 
within the projective superspace approach is it possible to {\em {easily}} define a 
4D, $\cal N$ = 1 superfield truncation.  However, there has been presented a proof
\cite{puncT} that any action in the harmonic superspace approach maybe engineered
to yield an equivalent projective superspace formulation.

Though both of these two powerful methods have a long list of accomplishments to
recommend them, it has long been the opinion of one of the authors (SJG) that these
cannot represent the `final' answer to the off-shell problem.  One indication of this is the 
fact that though these infinite auxiliary-field extended technologies work, they only do
so in a limited domain of theories.  To our knowledge, neither of the methods has 
allowed a significant breakthrough for either 4D, $\cal N$ = 4 SUSY YM theory nor
any of the 10D supersymmetrical theories mentioned above.  A final answer must deal
with these cases also.  In our opinion, to reach such a goal requires new tools and 
a new perspective.

For some time now, we have been developing two interlocking approaches in the
effort to make progress on this problem.  The first of these approaches \cite{GRana, ENUF}
involves what we refer to as the ${\cal {GR}}$(d, ${\cal N}$) Algebra (or `Garden Algebra') 
approach. Garden Algebras are real versions of Clifford Algebras that seem to
provide the basic building blocks of a rigorous theory of representations for space-time
SUSY.  Our second approach is based on a set of diagrams we have named `Adinkras'
\cite{FauxG}.  An Adinkra is essentially a weight space diagram (as is known for compact
Lie algebras) but with the added feature of including the orbits of the distinct generators
as they act on the states of any SUSY representation.  Adinkras provide convenient
graphical representations of Garden Algebras.  A growing body of literature on these
 topics is being developed by a collaboration of computer scientists, mathematicians
and theoretical physicists (the DFGHILM collaboration). 

Some of this work has already uncovered unexpected relations between a classification 
of SUSY reps and graph theory \cite{FauxG}, Filtered Clifford Algebras \cite{FCA}, graphical 
topology \cite{DCC}, and self-dual error-correcting codes \cite{SDEC} on the mathematics 
side.  Alternately there has been presented a new off-shell 4D, $\cal N$ = 2 hypermultiplet
(the `hyperplet') \cite{hyprplet},  new models for supersymmetrical quantum systems \cite{
NususyQMdL}, and the first prepotential description of models \cite{AdnkDynam} with an 
arbitrary degree of $\cal N$-extended SUSY on the physics side.

One of the current activities of the DFGHILM collaboration is the construction of a classification 
of supersymmetry representations up to and including $\cal N$ = 32 systems.  In the effort there 
is (what apparently seems to be) an incredible profusion of representations.  So much so that
we have been struck by the analogy with the problem of classifying genomes in biological 
systems.  Building on this analogy, we have chosen to include the word `genomics' in the 
title of this paper.

Although the work of \cite{ENUF} described a method (reduction on a 0-brane) by which the 
Adinkra/Garden Algebra description (the `genetic description) of a supersymmetric 
representation can be uncovered, there has not to this point been a detailed presentation applying 
this technique to well-known 4D, $\cal N$ = 1 systems more generally.  In the following, we will 
obtain the genetic description of; 

(a.) the off-shell \& on-shell chiral multiplet,  \newline \indent
(b.) the off-shell tensor multiplet,  \newline \indent
(c.) the on-shell double-tensor multiplet, and  \newline \indent
(d.) the off-shell \& on-shell vector multiplet.  
\vskip 0.05in
 \noindent
In a separate, but companion work, the complex linear multiplet and some other topics will be treated.

The structure of this work is as follows.  

In chapter 2, reviews are given of the
4D, $\cal N$ = 1 chiral, tensor, double tensor, and vector supermultiplets.  This is mostly done
to establish our notational conventions.  In chapter 3, new results are presented.  We carry
out the reduction on a 0-brane of the 4D, $\cal N$ = 1 supermultiplets discussed in the 
previous chapter.  This truncation leads to 1D, $\cal N$ = 4 supersymmetrical shadows and 
allows us to present an explicit derivation of the Garden Algebra matrices associated with
these distinct multiplets.  Though in the work of \cite{ENUF} it was stated that this procedure
always leads to the discovery of the Garden Algebras matrices associated with each supersymmetric
representation, the current work marks the first time this has been explicitly demonstrated for
these familiar 4D, $\cal N$ = 1 supermultiplets beyond the chiral multiplet.  The work in chapter 
4 is devoted to studying properties of the matrices associated with each of the multiplets.  
It is shown that (as expected) all the off-shell theories belong to a universal class of 
algebras...the Garden Algebras.  On the other hand, the matrices associated with the on-shell
theories do {\em {not}} possess features that lead to a universal identification.   Thus the
mathematical basis for understanding these in the context of matrix representation requires
much more study.  However, it is shown that there is one sharp distinction that can be made
between `generic' on-shell theories and `pathogenic' on-shell theories.
A definition is given for when
two sets of Garden Algebra matrices are members of an equivalence class.  Traces of the
Garden Algebra matrices that respect this definition of equivalence are defined.  Evidence
is shown to support the proposal that the superspin of the 4D multiplets are encoded in the 
Garden Algebra and the initial steps toward defining characters are taken.  A quantity, 
denoted by $\chi_{{}_0}$, is proposed as an actual character for the representations.
The fifth section explores the construction of the Adinkras associated with each multiplet.  By
comparing the case of the chiral multiplet with the vector multiplet, it is shown what property
of the Adinkra can be associated with $\chi_{{}_0}$.  We give our conclusions in chapter six.  
At the end there are three appendices describing conventions, aspects of the structure we 
call ${ \bf{ {\cal GR} ({\rm d}_L, \, {\rm d}_R, \,  {\cal N})}}$, and a primer of Adinkra manipulation.

In closing this section, let us make a clear statement as what is and what is {\em {not}} 
the goal of this work.  It is {\em {not}} a goal here to {\em {solve}} the problem of off-shell 
formulations of supersymmetric field theories in higher dimensions.  The goal of the
present work is much more modest.  By applying simple 4D $\to$ 1D reduction (called
``reduction on the 0-brane'') we want to study the explicit results for a number of familiar
4D $\cal N$ = 1 multiplets when by reduction, they are injected into the `sea' of 1D 
representations that was discovered in the work of \cite{GRana}.  

As was pointed out in \cite{FauxG}, the number of representations in 1D (for a fixed 
number of supercharges) is enormously larger than those that arise as reductions 
of representations from higher dimensions.  This raises the question of what 
distinguishes the generic 1D representations from those that are connected to higher 
D ones?  As there is no over-arching theoretical guide for answering this question, 
it is paramount to know what are the explicit 1D representations that result from 
reduction.  In a sense it is necessary to do a `genomic scan' (i. e. to find the associated 
Adinkras and `root superfield representations' \cite{ENUF}) of the reduced 
representations in order to compare these with generic 1D representations.  We 
have chosen for this arena of study the 4D, $\cal N$ = 1 theories.
 
We should point out that this current paper fills a hole in this line of investigations.
With the exception of the work of \cite{ENUF}, the DFGHILM collaboration has {\em {not}}
produced works looking at the actual injection of higher dimensional multiplets into
1D.  The work of the collaboration has been largely been directed to developing
a firm mathematical background and understanding of the 1D theory.   As a 
consequence, a number of results of this paper have not been seen previously and 
this paper is complementary to the general line of DFGHILM works.

\section{Review of Some 4D, $\cal N$ = 1 Multiplets }\label{Mults}

$~~~$ In each of the following subsections, a supersymmetric multiplet is presented in terms
of its field content and supersymmetry transformation laws.  Presentations are given for three
off-shell multiplets as well as three on-shell multiplets.

\subsection{Review of the 4D, $\cal N$ = 1 Chiral Multiplet }\label{ChiMult}

$~~~$ The 4D, $\cal N$ = 1 chiral multiplet is very well known to consist of a scalar $A$,
a pseudoscalar $B$, a Majorana fermion $\psi_a$, a scalar auxiliary field $F$, and a
pseudoscalar auxiliary field $G$.  A convenient way to express the supersymmetry
variation of these component fields is by first regarding them as the lowest component
of a superfield (denoted by the same symbol) and then expressing the action of the superspace
covariant derivative D${}_a$ acting on each.  As we have included the auxiliary fields
$F$ and $G$, necessarily it is the off-shell theory under consideration.  

The supersymmetry variations can be cast into the form of a set of specifications of the 
superspace `covariant derivative' acting on a set of superfields.  We have in our conventions
\be \eqalign{
{\rm D}_a A ~&=~ \psi_a  ~~~, \cr
{\rm D}_a B ~&=~ i \, (\gamma^5){}_a{}^b \, \psi_b  ~~~, \cr
{\rm D}_a \psi_b ~&=~ i\, (\gamma^\mu){}_{a \,b}\,  \partial_\mu A 
~-~  (\gamma^5\gamma^\mu){}_{a \,b} \, \partial_\mu B ~-~ i \, C_{a\, b} 
\,F  ~+~  (\gamma^5){}_{ a \, b} G  ~~, \cr
{\rm D}_a F ~&=~  (\gamma^\mu){}_a{}^b \, \partial_\mu \, \psi_b   ~~~, \cr
{\rm D}_a G ~&=~ i \,(\gamma^5\gamma^\mu){}_a{}^b \, \partial_\mu \,  
\psi_b  ~~~.
} \label{chi1}
\ee
A direct calculation shows that
\be \eqalign{  {~~~~~}
\{ ~ {\rm D}_a  \,,\,  {\rm D}_b ~\} \, A 
~&=~  i\, 2 \, (\gamma^\mu){}_{a \,b}\,  \partial_\mu \,  A ~~~, ~~~
\{ ~ {\rm D}_a  \,,\,  {\rm D}_b ~\} \, B 
~=~  i\, 2 \, (\gamma^\mu){}_{a \,b}\,  \partial_\mu \, B ~~~, \cr
&~~~~\{ ~ {\rm D}_a  \,,\,  {\rm D}_b ~\} \, \psi{}_{c}  
~=~  i\, 2 \, (\gamma^\mu){}_{a \,b}\,  \partial_\mu \,  \psi{}_{c}    ~~~, 
\cr
\{ ~ {\rm D}_a  \,,\,  {\rm D}_b ~\} \, F
~&=~  i\, 2 \, (\gamma^\mu){}_{a \,b}\,  \partial_\mu \,  F ~~~, ~~~
\{ ~ {\rm D}_a  \,,\,  {\rm D}_b ~\} \, G 
~=~  i\, 2 \, (\gamma^\mu){}_{a \,b}\,  \partial_\mu \, G ~~~.
} \label{chi2}
\ee
As expected, the algebra of (\ref{SUSYdef}) is satisfied independently of
the field upon which it is evaluated.

The simplest version of the on-shell theory occurs by simply setting 
$F$ = $G$ = 0 in (\ref{chi1}) and (\ref{chi2}).  Thus (\ref{chi1}) is replaced
by
\be \eqalign{
{\rm D}_a A ~&=~ \psi_a  ~~~, \cr
{\rm D}_a B ~&=~ i \, (\gamma^5){}_a{}^b \, \psi_b  ~~~, \cr
{\rm D}_a \psi_b ~&=~ i\, (\gamma^\mu){}_{a \,b}\,  \partial_\mu A 
~-~  (\gamma^5\gamma^\mu){}_{a \,b} \, \partial_\mu B   ~~~.
} \label{chi3}
\ee

Using (\ref{chi3}), a direct calculation shows that
\be \eqalign{  {~~~~~}
\{ ~ {\rm D}_a  \,,\,  {\rm D}_b ~\} \, A 
~&=~  i\, 2 \, (\gamma^\mu){}_{a \,b}\,  \partial_\mu \,  A ~~~, ~~~
\{ ~ {\rm D}_a  \,,\,  {\rm D}_b ~\} \, B 
~=~  i\, 2 \, (\gamma^\mu){}_{a \,b}\,  \partial_\mu \, B ~~~, \cr
\{ ~ {\rm D}_a  \,,\,  {\rm D}_b ~\} \, \psi{}_{c}  
~&=~  i\, 2 \, (\gamma^\mu){}_{a \,b}\,  \partial_\mu \,  \psi{}_{c}    
~-~   i \, (\gamma^\mu){}_{a \,b}\, (\gamma_\mu
\gamma^\nu){}_c{}^d  \partial_\nu \,  \psi{}_{d}     ~~~.
} \label{chi4}
\ee
The first two of these equations have the same form as (\ref{SUSYdef}) in the case where
$\cal N$ = 1.  However, the third term immediately above can be expressed as
\be \eqalign{
\{ ~ {\rm D}_a  \,,\,  {\rm D}_b ~\} \, \psi{}_{c}  
~&=~  i\, 2 \, (\gamma^\mu){}_{a \,b}\,  \partial_\mu \,  \psi{}_{c}    ~+~  i\, 2 \, (\gamma^\mu){}_{a \,b}\, 
(\gamma_\mu){}_c{}^d 
{\cal K}{}_{d} (\psi) ~~,~~  \cr
 {\cal K}{}_{c} (\psi) ~&=~ - \, \fracm 12 \,  (
\gamma^\nu){}_c{}^d  \partial_\nu \,  \psi{}_{d}     ~~~,
}  \label{chi5}
\ee
where $ {\cal K}{}_ c$ measures the `non-closure' of the algebra. It is also seen that
the relations
\be \eqalign{
 {\cal K}{}_{c} (\psi)~=~    - \, \fracm 12 \,   \,   {\rm D}_c F  ~~~,~~
  {\cal K}{}_{c} (\psi)~=~    i  \, \fracm 12 \,  (\g^5 ){}_c{}^d   \,   {\rm D}_d G
 ~~~
}  \label{chi6}
\ee
are satisfied.  This is important for the consistency of the truncation in (\ref{chi4}) with
regard to the starting point in (\ref{chi2}).  If we set $F$ = $G$ = 0 in (\ref{chi2}) then it is
consistent to set $ {\cal K}{}_{c}$ = 0 in (\ref{chi4}) - (\ref{chi6}).  

This is the essence of the ``Off-Shell Problem.''  Namely, if we begin only knowing (\ref{chi3}),
how would we systematically go about finding out that it is required to add $F$ and $G$ as
in (\ref{chi1})?  A related question is, ``Is the addition of $F$ and $G$ unique?''  (The answer to
this second question is known to be, ``No.''  This will be discussed in a companion work to
accompany this paper.)

\subsection{Review of the 4D, $\cal N$ = 1 Tensor Multiplet }\label{TensMult}

$~~~$ The 4D, $\cal N$ = 1 tensor multiplet consists of a scalar $\varphi$, a second-rank
skew symmetric tensor, $B{}_{\mu \, \nu }$, and a Majorana fermion $\chi_a$.  Their
supersymmetry variations can be cast in the forms
\be \eqalign{
{\rm D}_a \varphi ~&=~ \chi_a  ~~~, \cr
{\rm D}_a B{}_{\mu \, \nu } ~&=~ -\, \fracm 14 ( [\, \gamma_{\mu}
\, , \,  \gamma_{\nu} \,]){}_a{}^b \, \chi_b  ~~~, \cr
{\rm D}_a \chi_b ~&=~ i\, (\gamma^\mu){}_{a \,b}\,  \partial_\mu \varphi 
~-~  (\gamma^5\gamma^\mu){}_{a \,b} \, \e{}_{\mu}{}^{\r \, \s \, \t}
\partial_\r B {}_{\s \, \t}~~.
} \label{ten1}
\ee
The commutator algebra for the D-operator calculated from (\ref{ten1}) takes the form 
\be \eqalign{
\{ ~ {\rm D}_a  \,,\,  {\rm D}_b ~\} \, {\varphi}  
~&=~  i\, 2 \, (\gamma^\mu){}_{a \,b}\,  \partial_\mu \,  {\varphi}  ~~, \cr
{~~~~~~~}
\{ ~ {\rm D}_a  \,,\,  {\rm D}_b ~\} \, B{}_{\mu \, \nu }  
~&=~  i\, 2 \, (\gamma^\r){}_{a \,b}\,  \partial_\r \, B{}_{\mu \, \nu }  
~+~ \partial_\mu \, q{}_{\nu ~ a \, b} ~-~  \partial_\nu \, q{}_{\mu ~ 
a \, b} ~~, \cr
\{ ~ {\rm D}_a  \,,\,  {\rm D}_b ~\} \, \chi{}_{c}  
~&=~  i\, 2 \, (\gamma^\mu){}_{a \,b}\,  \partial_\mu \,  \chi{}_{c}    ~~,  ~~
q{}_{\mu ~ a \, b} ~\equiv~     i\, 2 \, (\gamma^\nu){}_{a \,b}   \, 
 [ \, B{}_{\mu \, \nu } \,+\, \fracm 12 \eta {}_{\mu \, \nu } \,\varphi \, ] ~~.
} \label{ten2}
\ee

The second line in (\ref{ten2}) is interesting.  On a first glance, it appears that the
two final $q$-dependent parts are `non-closure' terms as seen in the on-shell 
chiral multiplet.  Let us multiply the middle line of (\ref{ten2}) by parameters
$\e^a_1$ and $\e^b_2$ to obtain
\be \eqalign{
\e^a_1 \, \e^b_2 \,
\{ ~ {\rm D}_a  \,,\,  {\rm D}_b ~\} \, B{}_{\mu \, \nu }  
~&=~  i\, 2 \, \e^a_1 \, \e^b_2 \,(\gamma^\r){}_{a \,b}\,  \partial_\r \, B{}_{\mu \, \nu }  
~+~ \partial_\mu \, {\rm v}_{\nu} ~-~  \partial_\nu \, {\rm v}_{\mu} ~   \cr
{\rm {where}} ~\,
{\rm v}_{\m}~&\equiv~     i\, 2 \, \e^a_1 \, \e^b_2 \,  (\gamma^\nu){}_{a \,b}   \, 
 [ \, B{}_{\mu \, \nu } \,+\, \fracm 12 \eta {}_{\mu \, \nu } \,\varphi \, ]  ~~~.
} \label{ten3}
\ee
Since $B{}_{\mu \, \nu }  $ is anti-symmetric, it is possible to define a `gauge' variation 
denoted by $\d_G^{(2)}(\ell)$ that acts upon it according to
\be \eqalign{
\d_G^{(2)}(\ell) \, B{}_{\mu \, \nu } ~=~  \partial_\mu \, {\ell}_{\nu} ~-~  \partial_\nu \, {\ell}_{\mu} 
~~~,} \label{ten4}
\ee
and if we identify ${\ell}_{\mu} $ = v${}_{\mu}$, then (\ref{ten3}) may be expressed as
\be \eqalign{
\e^a_1 \, \e^b_2 \,
\{ ~ {\rm D}_a  \,,\,  {\rm D}_b ~\} \, B{}_{\mu \, \nu }  
~&=~  i\, 2 \, \e^a_1 \, \e^b_2 \,(\gamma^\r){}_{a \,b}\,  \partial_\r \, B{}_{\mu \, \nu }  
~+~ \d_G^{(2)}({\rm v}) \, B{}_{\mu \, \nu }   \cr
~&\equiv~ \xi^\r  \partial_\r \, B{}_{\mu \, \nu }  
~+~ \d_G^{(2)}({\rm v}) \, B{}_{\mu \, \nu } 
~~~.
} \label{ten5}
\ee
These equations inform us that any theory possessing the symmetries described
by (\ref{ten1}) must also possess the symmetries described on the RHS (right hand side)
of (\ref{ten5}).  The first of these is simple translation symmetry.  The second is
identifiable as the gauge symmetry of an anti-symmetric rank two tensor field.  Finally, 
we observe that there are no Lorentz-covariant truncations of the fields in the tensor 
multiplet.  So it is not possible to define an `on-shell' version of this multiplet as it 
was with the chiral multiplet.

\subsection{4D, $\cal N$ = 1 Double Tensor Multiplet}\label{2TenMult}

$~~~$ Though little known, the  4D, $\cal N$ = 1 ``double tensor'' multiplet is quite old 
\cite{DbL10sor}.  To motivate the consideration of this multiplet, it is useful to compare 
(\ref{chi3}) with (\ref{ten1}), looking for differences and similarities.  Immediately, one 
glaring difference is that the pseudoscalar field $B$ in the chiral multiplet is replaced 
by the 2-form $B{}_{\mu \, \nu }  $ in the tensor multiplet.  This obviously motivates the 
query, ``What would occur if {\em {both}} $A$ {\em {and}} $B$ were replaced by 2-forms?''   
The 4D, $\cal N$ = 1 double tensor multiplet consists of  two second-rank skew symmetric 
tensors $X{}_{\mu \, \nu }$ and $Y{}_{\mu \, \nu }$ along with a Majorana fermion $\L_a$.   
Thus we arrive at the double tensor multiplet with supersymmetry variations taking the 
forms
\be \eqalign{
{\rm D}_a X{}_{\mu \, \nu } ~&=~ i \, \fracm 14 (\gamma^5  [\, \gamma_{\mu}
\, , \,  \gamma_{\nu} \,]){}_a{}^b \, \L_b  ~~~, \cr
{\rm D}_a Y{}_{\mu \, \nu } ~&=~ - \, \fracm 14 ( [\, \gamma_{\mu}
\, , \,  \gamma_{\nu} \,]){}_a{}^b \, \L_b  ~~~, \cr
{\rm D}_a \L_b ~&=~ i\, (\gamma^\mu){}_{a \,b}\,  \e{}_{\mu}{}^{\r \, \s \, \t}
\partial_\r X {}_{\s \, \t} ~-~  (\gamma^5\gamma^\mu){}_{a \,b} \, 
\e{}_{\mu}{}^{\r \, \s \, \t} \partial_\r Y {}_{\s \, \t}~~.
} \label{dblten1}
\ee
Upon comparing (\ref{chi4}) with (\ref{dblten1}), it is clear that the first two equations in
the former will become the first two equations of the latter if we perform the replacements
\be \eqalign{
A ~\to~ X{}_{\mu \, \nu } ~~,~~ B ~\to~ Y{}_{\mu \, \nu } ~~,~~ 
\psi_a ~\to~ i \, \fracm 14 (\gamma^5  [\, \gamma_{\mu}
\, , \,  \gamma_{\nu} \,]){}_a{}^b \, \L_b 
~~.
} \label{dblten2}
\ee
Curiously though, if the replacements in (\ref{dblten2}) are inserted into the final line
in (\ref{chi4}), we obtain
\be \eqalign{
{\rm D}_a \L_b ~&=~ i \, \fracm 16 \, 
(\gamma^\mu){}_{a \,b}\, \left[ \, \,  \e{}_{\mu}{}^{\r \, \s \, \t} \, \partial_\r 
X {}_{\s \, \t} ~-~  2 \,  \partial^\n Y {}_{\m \, \n}  \, \, \right]  \cr
&~~~+~ \fracm 16 \,  (\gamma^5\gamma^\mu){}_{a \,b} \, \left[ \, \,  \e{}_{\mu}{}^{\r \, \s 
\, \t} \, \partial_\r Y {}_{\s \, \t}  ~+~  2 \, \partial^\n X {}_{\m \, \n} \,   \, \right]  ~~
~~.
} \label{dblten3}
\ee
which is {\em {not}} the same as the final line in (\ref{dblten1}).  In any event we next 
use (\ref{dblten1}) to calculate the anti-commutator of the D-operator as realized on 
the fields of the double tensor multiplet and find
\be \eqalign{
{~~~~~~~}
\{ ~ {\rm D}_a  \,,\,  {\rm D}_b ~\} \, X{}_{\mu \, \nu }  
~&=~  i\, 2 \, (\gamma^\r){}_{a \,b}\,  \partial_\r \, X{}_{\mu \, \nu }  
~+~ \partial_\mu \, s{}_{\nu ~ a \, b} ~-~  \partial_\nu \, s{}_{\mu ~ 
a \, b}   \cr
&~~~~-~ i \, [~  \eta{}_{\a \, \mu } \, (\gamma_\nu){}_{a \,b} ~-~
\eta{}_{\a \, \nu } \, (\gamma_\mu){}_{a \,b}  ~]  \, \e{}^{\a \, \r \, \s \, \t}
\partial_\r Y {}_{\s \, \t}~~,    \cr
{~~~~~~~}
\{ ~ {\rm D}_a  \,,\,  {\rm D}_b ~\} \, Y{}_{\mu \, \nu }  
~&=~  i\, 2 \, (\gamma^\r){}_{a \,b}\,  \partial_\r \, Y{}_{\mu \, \nu }  
~+~ \partial_\mu \, t{}_{\nu ~ a \, b} ~-~  \partial_\nu \, t{}_{\mu ~
 a \, b}   \cr
&~~~~+~ i \, [~  \eta{}_{\a \, \mu } \, (\gamma_\nu){}_{a \,b} ~-~
\eta{}_{\a \, \nu } \, (\gamma_\mu){}_{a \,b}  ~]  \, \e{}^{\a \, \r \, \s \, \t}
\partial_\r X {}_{\s \, \t}~~,   \cr
 &s{}_{\mu ~ a \, b} ~\equiv~     i\, 2 \, (\gamma^\nu){}_{a \,b}   \, 
 X{}_{\mu \, \nu }   ~~,~~  t{}_{\mu ~ a \, b} ~\equiv~   i\, 2 \, 
 (\gamma^\nu){}_{a \,b} \, Y{}_{\mu \, \nu }  ~~,    \cr
{~~~~~~~}
\{ ~ {\rm D}_a  \,,\,  {\rm D}_b ~\} \, \L_c ~&=~ i\, 2\, (\gamma^\mu){}_{a \,b}\,  
\partial_\mu \,  \L_c ~+~ i \, (\gamma^\mu){}_{a \,b}\, (\gamma_\mu \, 
\gamma^\nu){}_c {}^d\,  \partial_\nu \,  \L_d    ~~~~.
}
\ee \label{dblten5}

We can begin our analysis of (16) by concentrating on the anticommutator
as realized on the fermion $\L_a$.  Similar to (6), we can write
\be \eqalign{
\{ ~ {\rm D}_a  \,,\,  {\rm D}_b ~\} \, \L{}_{c}  ~&=~  i\, 2 \, (\gamma^\mu){}_{a \,b}\, 
\partial_\mu \,  \L{}_{c}    ~+~  i\, 2 \, (\gamma^\mu){}_{a \,b}\, (\gamma_\mu){}_c
{}^d {\Tilde {\cal K}}{}_{d} (\L) ~~,~~  \cr
{\Tilde {\cal K}}{}_{c} (\L) ~&=~  \fracm 12 \,  (\gamma^\nu){}_c{}^d  \partial_\nu 
\,  \L{}_{d}     ~~~.
}  \label{dblten6}
\ee
and we see the emergence of a non-closure function ${\Tilde {\cal K}}{}_{c} (\L)$ as
in the case of the on-shell chiral multiplet.  This equation proves that the double
tensor multiplet is an on-shell construction and no finite set of auxiliary fields
is known to alleviate this. 

The forms of the anticommutator as realized on $X{}_{\mu \, \nu }  $ and $Y{}_{\mu \, \nu }  $
reveal that any theory with the symmetries described by (\ref{dblten1}) must also
possess translation symmetry and the gauge symmetries for both two-form fields.
However, the last term of the first equation  and the last term in the second equation 
in (16) also imply something new.  These theories must also possess a symmetry
under a `Killing vector' of the form
\be \eqalign{
\d{}_Z ~=~ -  i 2\, \xi_{\m} \,   \e_{\n}{}^{\r \, \s \, \t} \left[ \,  (\partial_\r Y {}_{\s \, \t})
{ {\pa {~~~~}} \over {\pa X{}_{\m \n}} } ~-~  (\partial_\r X {}_{\s \, \t})
{ {\pa {~~~~}} \over {\pa Y{}_{\m \n}} } \,  \right] ~~.
}  \label{dblten7}
\ee
Since bosons typically satisfy second order differential equations of motion, this
term (known as a `central charge') {\em {cannot}} be interpreted as a non-closure
term that vanishes upon use of the equations of motion.  In fact, this is a `on-shell
central charge,' meaning that is has a non-trivial effect on the fields even when the
theory obeys its equations of motion.

\subsection{Review of 4D, $\cal N$ = 1 Vector Multiplet }\label{VMult}

$~~~$ The 4D, $\cal N$ = 1 vector multiplet off-shell is described by a vector
 $A{}_{\mu}$, a Majorana fermion $\l_a$, and a pseudoscalar auxiliary field
 d.  Their supersymmetry variations are described by
\be \eqalign{
{\rm D}_a \, A{}_{\mu} ~&=~  (\gamma_\mu){}_a {}^b \,  \l_b  ~~~, \cr
{\rm D}_a \l_b ~&=~   - \,i \, \fracm 14 ( [\, \gamma^{\mu}\, , \,  \gamma^{\nu} 
\,]){}_a{}_b \, (\,  \partial_\mu  \, A{}_{\nu}    ~-~  \partial_\nu \, A{}_{\mu}  \, )
~+~  (\gamma^5){}_{a \,b} \,    {\rm d} ~~,  \cr
{\rm D}_a \, {\rm d} ~&=~  i \, (\gamma^5\gamma^\mu){}_a {}^b \, 
\,  \partial_\mu \l_b  ~~~. \cr
} \label{V1}
\ee
These lead in a straightforward manner to the following anticommutator algebra.
\be \eqalign{
{~~~~~~~}
\{ ~ {\rm D}_a  \,,\,  {\rm D}_b ~\} \, A{}_{\mu}  
~&=~  i\, 2 \, (\gamma^\r){}_{a \,b}\,  \partial_\r \, A{}_{\mu}  
~-~ \partial_\mu \, r{}_{a \, b}   ~~,  ~~
r{}_{a \, b} ~\equiv~     i\, 2 \, (\gamma^\nu){}_{a \,b}   \, A{}_{\nu }
~~,    \cr
\{ ~ {\rm D}_a  \,,\,  {\rm D}_b ~\} \,\l{}_c
~&=~  i\, 2 \, (\gamma^\mu){}_{a \,b}\,  \partial_\mu \,  \l{}_c ~~,    \cr
\{ ~ {\rm D}_a  \,,\,  {\rm D}_b ~\} \, {\rm d}  
~&=~  i\, 2 \, (\gamma^\mu){}_{a \,b}\,  \partial_\mu \,  {\rm d}  ~~.
}   \label{V2}
\ee
The term involving $r{}_{a b}$ implies that any theory involving the vector gauge field 
above must also admit a symmetry of the form
\be \eqalign{
\d_G^{(1)}(\a) \, A{}_{\m} ~=~  \partial_\mu \a 
~~~,}  \label{V3}
\ee
which is easily identifiable as the usual form of a spin-1 gauge transformation.

In the on-shell theory, we set $d$ = 0 but retain all other terms in (\ref{V1})
\be \eqalign{
{\rm D}_a \, A{}_{\mu} ~&=~  (\gamma^\mu){}_a {}^b \,  \l_b  ~~~, \cr
{\rm D}_a \l_b ~&=~   - \,i \, \fracm 14 ( [\, \gamma^{\mu}\, , \,  \gamma^{\nu} 
\,]){}_a{}_b \, (\,  \partial_\mu  \, A{}_{\nu} ~-~  \partial_\nu  \, A{}_{\mu} 
\, )  ~~~, \cr
} \label{V4}
\ee
and once again we calculate the anticommutator as realized on the remaining fields
to find
\be \eqalign{
{~~~~~~~}
\{ ~ {\rm D}_a  \,,\,  {\rm D}_b ~\} \, A{}_{\mu}  
~&=~  i\, 2 \, (\gamma^\r){}_{a \,b}\,  \partial_\r \, A{}_{\mu}  
~-~ \partial_\mu \, r{}_{a \, b}   ~~,  ~~
r{}_{a \, b} ~\equiv~     i\, 2 \, (\gamma^\nu){}_{a \,b}   \, A{}_{\nu }
~~,    \cr
\{ ~ {\rm D}_a  \,,\,  {\rm D}_b ~\} \,\l{}_c
~&=~  i\, 2 \, (\gamma^\mu){}_{a \,b}\,  \partial_\mu \,  \l{}_c   ~-~ i \,
\fracm 12 \,  (\gamma^\mu){}_{a \,b}\,  (\gamma_\mu  
\gamma^\nu){}_c{}^d \,   \partial_\nu \,  \l{}_d   \cr
&~~~~+~ i \, \fracm 1{16} \,  ([ \, \gamma^\a \, , \, \gamma^\b \,]){}_{a \,b}\,  
( [ \, \gamma_\a \, , \, \gamma_\b \,] \gamma^\nu){}_c{}^d \,   \partial_\nu \,  \l{}_d   
~~.
} \label{V5}
\ee
The final equation of (\ref{V5}) shows the presence of {\em {two}} non-closure terms.  We may
rewrite the final line as
\be \eqalign{
\{ ~ {\rm D}_a  \,,\,  {\rm D}_b ~\} \,\l{}_c
~&=~  i\, 2 \, (\gamma^\mu){}_{a \,b}\,  \partial_\mu \,  \l{}_c   ~+~ i 2 \,
 (\gamma^\mu){}_{a \,b}\,  (\gamma_\mu  ){}_c{}^d \,  {\Hat K}{}_d(\l)   \cr
&~~~~-~ i \, \fracm 1{4} \,  ([ \, \gamma^\a \, , \, \gamma^\b \,]){}_{a \,b}\,  
( [ \, \gamma_\a \, , \, \gamma_\b \,] ){}_c{}^d \,  {\Hat K}{}_d(\l)  ~~, \cr
  {\Hat K}{}_c(\l) ~&\equiv ~  -\, \fracm 14 (\gamma^\nu){}_c{}^d \,   \partial_\nu \,  \l{}_d
  ~=~  i \, \fracm 14 \, (\g^5){}_c{}^d {\rm D}_d {\rm d}
~~,
} \label{V6}
\ee
where the non-closure term $  {\Hat K}{}_c(\l)$ is introduced.  Once more, it is seen
to be consistent to set d = 0 if the non-closure term vanishes, i.e. the fermion obeys
an equation of motion.

\newpage
\section{Garden Algebra Matrices From 0-Brane \\ Reduction }

$~~~$ The four previous chapters have presented a review of well known results.  In 
this chapter, we will undertake  to uncover the form of the Garden Algebra matrices 
associated with each supermultiplet previously discussed.  According to the technique 
proposed in \cite{ENUF} this goal can be achieved by first performing a toroidal 
compactification of any  higher D supersymmetircal multiplet on a 0-brane and thus 
retain only the temporal dependence of all fields in the supermultiplet.

\subsection{4D, $\cal N$ = 1 Chiral Multiplet On The 0-Brane}

$~~~$ The supersymmetry transformation laws in (\ref{chi1}) are generally valid independent
of the coordinate dependence of the various functions that appear in the equations.
These equations remain valid if we restrict the functions so that they remain dependent
only on the $\tau$-coordinate.  Under this restriction these equations can be recast in
the form
\be
\begin{array}{ccccccccc}
{\rm D}_1 A =& \psi_1 ~~~~~~~& {\rm D}_2 A =& \psi_2 ~~~~~&  ~~&{\rm D}_3A=&\psi_3  ~~~~~~&  
{\rm D}_4A=&\psi_4 ~~~~~~~\\
{\rm D}_1 B =& - \, \psi_4  ~~~~~~~& {\rm D}_2B=&\psi_3 ~~~~&  &{\rm D}_3B=&-\psi_2 ~~~~~~~&  
{\rm D}_4B=&\psi_1~~~~~~~ \\
{\rm D}_1 F =& \pa_0 \psi_2  ~~~~~& {\rm D}_2F=&-\pa_0\psi_1 &  &{\rm D}_3F=&-\pa_0\psi_4 
 ~~~~ &  {\rm D}_4F=&\pa_0\psi_3 ~~~~\\
{\rm D}_1 G =& - \,  \pa_0 \psi_3 ~~~& {\rm D}_2G=&-\pa_0\psi_4 &  &{\rm D}_3G=&\pa_0\psi_1  
 ~~~~&  {\rm D}_4G=&\pa_0\psi_2  ~~.~~
\end{array}
 \label{chiD0A}
\ee
Next a set of re-definitions can be carried out on the fermions  according to
\be
\psi_1 ~\to~ i \, \Psi_1 ~~~,~~~ \psi_2 ~\to~ i \, \Psi_2 ~~~,~~~ 
\psi_3 ~\to~   i \, \Psi_3 ~~~,~~~ \psi_4 ~\to~ i \, \Psi_4 ~~~,
 \label{chiD0B}
\ee
so the previous equations take the forms of \newline 
\be
\begin{array}{ccccccccc}
{\rm D}_1 A =& i \, \Psi_1 ~~~~~~~& {\rm D}_2 A =& i \, \Psi_2 ~~~~~&  ~~&{\rm D}_3A=&i \, \Psi_3  ~~~~~~&  
{\rm D}_4 A=&i \, \Psi_4 ~~~~~~~\\
{\rm D}_1 B =& - \, i \, \Psi_4  ~~~~~~~& {\rm D}_2B=&i \, \Psi_3 ~~~~&  &{\rm D}_3B=&-i \, \Psi_2 ~~~~~~~&  
{\rm D}_4B=&i \, \Psi_1~~~~~~~ \\
{\rm D}_1 F =&  i \, \pa_0\Psi_2  ~~~~~& {\rm D}_2F=&-i \pa_0 \, \Psi_1 &  &{\rm D}_3F=&-i \pa_0 \, \Psi_4 
 ~~~~ &  {\rm D}_4F=&i \pa_0 \, \Psi_3 ~~~~\\
{\rm D}_1 G =& - \, i \pa_0  \Psi_3 ~~~& {\rm D}_2G=&-i \pa_0 \, \Psi_4 &  &{\rm D}_3G=&i \pa_0 \, \Psi_1  
 ~~~~&  {\rm D}_4G=&i \pa_0 \, \Psi_2  ~~.~~
\end{array}
   \label{chiD0C}
\ee
Now we define
\be
\Phi_1 ~=~ A ~~~,~~~ \Phi_2 ~=~ B ~~~,~~~ \pa_0 \Phi_3 ~=~  F ~~~,~~~
 \pa_0 \Phi_4 ~=~  G ~~~,
  \label{chiD0D}
\ee
and note the above system of equations can be written in the form
\be
{\rm D}{}_{{}_{\rm I}} \Phi_i ~=~ i \, \left( {\rm L}{}_{{}_{\rm I}}\right) {}_{i \, {\hat k}}  \,  \Psi_{\hat k}
~~.
 \label{chiD0E}
\ee
The explicit form of the L-matrices that appear here are given by
$$
\left( {\rm L}{}_{1}\right) {}_{i \, {\hat k}}   ~=~
\left[\begin{array}{cccc}
~1 & ~~0 &  ~~0  &  ~~0 \\
~0 & ~~0 &  ~~0  &  ~-\, 1 \\
~0 & ~~1 &  ~~0  &  ~~0 \\
~0 & ~~0 &  ~-\, 1  &  ~~0 \\
\end{array}\right] ~~~,~~~
\left( {\rm L}{}_{2}\right) {}_{i \, {\hat k}}   ~=~
\left[\begin{array}{cccc}
~0 & ~~1 &  ~~0  &  ~ \, \, 0 \\
~0 & ~~ 0 &  ~~1  &  ~~0 \\
-\, 1 & ~~ 0 &  ~~0  &  ~~0 \\
~ 0 & ~~~0 &  ~~0  &   -\, 1 \\
\end{array}\right]  ~~~,
$$
\be
\left( {\rm L}{}_{3}\right) {}_{i \, {\hat k}}   ~=~
\left[\begin{array}{cccc}
~0 & ~~0 &  ~~1  &  ~~0 \\
~0 & ~- \, 1 &  ~~0  &  ~~0 \\
~0 & ~~0 &  ~~0  &  -\, 1 \\
~1 & ~~0 &  ~~0  &  ~~0 \\
\end{array}\right] ~~~,~~~
\left( {\rm L}{}_{4}\right) {}_{i \, {\hat k}}   ~=~
\left[\begin{array}{cccc}
~0 & ~~0 &  ~~0  &  ~ \, \, 1 \\
~1 & ~~ 0 &  ~~0  &  ~~0 \\
~0 & ~~ 0 &  ~~1  &  ~~0 \\
~ 0 & ~~~1 &  ~~0  &   ~~0  \\
\end{array}\right]  ~~.
 \label{chiD0F}
\ee

After writing the results for the fermions we find
\be
\begin{array}{ccccccccc}
{\rm D}_1 \Psi_1 =&  \pa_0 A ~~~~~~~& {\rm D}_2 \Psi_1 =& - F ~~&  ~~&{\rm D}_3\Psi_1=& G  ~~~~~~&  
{\rm D}_4\Psi_1=&  \pa_0 B ~~~~~~~\\
{\rm D}_1 \Psi_2 =&  F  ~~~~~~~& {\rm D}_2\Psi_2=& \pa_0 A ~~&  &{\rm D}_3\Psi_2=&-  \pa_0 
B ~~~~~~~&  {\rm D}_4\Psi_2=& G~~~~~~~~~ \\
{\rm D}_1 \Psi_3 =& -  G  ~~~~~~~& {\rm D}_2\Psi_3=& \pa_0 B &  &{\rm D}_3\Psi_3=& \pa_0 A
 ~~~~ &  {\rm D}_4\Psi_3=&  F ~~~~~~~~~\\
{\rm D}_1 \Psi_4 =& -  \pa_0 B ~~~~~& {\rm D}_2\Psi_4=&-  G &  &{\rm D}_3\Psi_4=& -  F  
 ~~~~&  {\rm D}_4\Psi_4=& \pa_0 A  ~~.~~~~
\end{array}
  \label{chiD0H}
\ee
Once more we use the definitions in (\ref{chiD0D}) and note that the above system of equations can be
written in the form
\be
{\rm D}{}_{{}_{\rm I}} \Psi_{\hat k} ~=~  \left( {\rm R}{}_{{}_{\rm I}}\right) {}_{{\hat k} \, i}  \,
{{d ~} \over { d t}} \, \Phi_{i}  ~~.
 \label{chiD0J}
\ee
The explicit form of the matrices that appear here are given by
$$
\left( {\rm R}{}_{1}\right) {}_{i \, {\hat k}}   ~=~
\left[\begin{array}{cccc}
~1 & ~~0 &  ~~0  &  ~~0 \\
~0 & ~~0 &  ~~1  &  ~~ 0 \\
~0 & ~~0 &  ~ 0  &  ~- 1 \\
~0 & -1 &  ~ 0  &  ~~0 \\
\end{array}\right] ~~~,~~~
\left( {\rm R}{}_{2}\right) {}_{i \, {\hat k}}   ~=~
\left[\begin{array}{cccc}
0 & ~~0 &  ~-\, 1  &  ~ \, \, 0 \\
~1 & ~~ 0 &  ~~0  &  ~~0 \\
~0 & ~~ 1 &  ~~0  &  ~~0 \\
~ 0 & ~~~0 &  ~~0  &   -\, 1 \\
\end{array}\right]  ~~~,
$$
\be
\left( {\rm R}{}_{3}\right) {}_{i \, {\hat k}}   ~=~
\left[\begin{array}{cccc}
~0 & ~~0 &  ~~0  &  ~1 \\
~0 & - \, 1 &  ~~0  &  ~0 \\
~1 & ~~0 &  ~~0  &  ~0 \\
~0 & ~~0 &   - 1  &  ~0 \\
\end{array}\right] ~~~,~~~
\left( {\rm R}{}_{4}\right) {}_{i \, {\hat k}}   ~=~
\left[\begin{array}{cccc}
~0 & ~~1 &  ~~0  &  ~ \, \, 0 \\
~0 & ~~ 0 &  ~~0  &  ~~1 \\
~0 & ~~ 0 &  ~~1  &  ~~0 \\
~ 1 & ~~~0 &  ~~0  &   ~~0  \\
\end{array}\right]  ~~~.
 \label{chiD0K}
\ee
It is now seen that the  set of L-matrices (\ref{chiD0F}) and R-matrices (\ref{chiD0K}) satisfy 
the equation
\be
\left( {\rm R}{}_{{}_{\rm I}}\right) ~\equiv~  \left[ \, \left( {\rm L}{}_{{}_{\rm I}}\right) \, \right]{}^{t}
 \label{chiD0L}
\ee
where the $t$ superscript stands for transposition.

We now turn our attention to the on-shell case.  This begins by setting $F$ = $G$ = 0.
The consistency of these conditions implies $\pa_0 \psi_{\hk}$ = $\pa_0^2 A$ = $\pa_0^2 B$ = 0.
Further consistency conditions imply that $\Phi_i$ be defined by
\be
\Phi_1 ~=~ A ~~~,~~~ \Phi_2 ~=~ B  ~~~,
  \label{chiD0M}
\ee
while $\Psi_{\hk}$ is still defined by (\ref{chiD0B}).  In these considerations of the on-shell
theory,  (\ref{chiD0E}) and  (\ref{chiD0H}) are still valid.  However, the definition of the 
L-matrices and R-matrices must now be changed to
$$
\left( {\rm L}{}_{1}\right) {}_{i \, {\hat k}}   ~=~
\left[\begin{array}{cccc}
~1 & ~~0 &  ~~0  &  ~~0 \\
~0 & ~~0 &  ~~0  &  ~-\, 1 \\
\end{array}\right] ~~~,~~~
\left( {\rm L}{}_{2}\right) {}_{i \, {\hat k}}   ~=~
\left[\begin{array}{cccc}
~0 & ~~1 &  ~~0  &  ~ \, \, 0 \\
~0 & ~~ 0 &  ~~1  &  ~~0 \\
\end{array}\right]  ~~~,
$$
\be
\left( {\rm L}{}_{3}\right) {}_{i \, {\hat k}}   ~=~
\left[\begin{array}{cccc}
~0 & ~~0 &  ~~1  &  ~~0 \\
~0 & ~- \, 1 &  ~~0  &  ~~0 \\
\end{array}\right] ~~~,~~~
\left( {\rm L}{}_{4}\right) {}_{i \, {\hat k}}   ~=~
\left[\begin{array}{cccc}
~0 & ~~0 &  ~~0  &  ~ \, \, 1 \\
~1 & ~~ 0 &  ~~0  &  ~~0 \\
\end{array}\right]  ~~~,
 \label{chiD0N}
\ee
$$
\left( {\rm R}{}_{1}\right) {}_{i \, {\hat k}}   ~=~
\left[\begin{array}{cc}
~1 & ~~0  \\
~0 & ~~0  \\
~0 & ~~0  \\
~0 & -1  \\
\end{array}\right] ~~~,~~~
\left( {\rm R}{}_{2}\right) {}_{i \, {\hat k}}   ~=~
\left[\begin{array}{cc}
0 & ~~0  \\
~1 & ~~ 0  \\
~0 & ~~ 1  \\
~ 0 & ~~~0  \\
\end{array}\right]  ~~~,
$$
\be
\left( {\rm R}{}_{3}\right) {}_{i \, {\hat k}}   ~=~
\left[\begin{array}{cc}
~0 & ~~0  \\
~0 & - \, 1  \\
~1 & ~~0  \\
~0 & ~~0  \\
\end{array}\right] ~~~,~~~
\left( {\rm R}{}_{4}\right) {}_{i \, {\hat k}}   ~=~
\left[\begin{array}{cc}
~0 & ~~1  \\
~0 & ~~ 0  \\
~0 & ~~ 0  \\
~ 1 & ~~~0   \\
\end{array}\right]  ~~.
 \label{chiD0O}
\ee

\subsection{4D, $\cal N$ = 1 Tensor Multiplet On The 0-Brane}

$~~~$ We now repeat the process of the last subsection.  However, we now take as our
starting point the results in (\ref{ten1}).  Carrying out the reduction  yields the following
for the bosons 
\be
\begin{array}{cccccccccccc}
{\rm D}_1 \phi ~~~~=& \chi_1& ~& {\rm D}_2 \phi ~~~~=& \chi_2& ~&  ~&{\rm D}_3\phi~~~~=
&\chi_3&  ~&  {\rm D}_4\phi ~~~~=&\chi_4 ~~~~~~~\\
2 \, {\rm D}_1 B_{12} =& - \, \chi_3&  ~& 2 \, {\rm D}_2B_{12}=&\chi_4& ~&  &2 \, {\rm D}_3B_{12}=
&\chi_1& ~&  2 \, {\rm D}_4B_{12}=&- \chi_2~~~~~~~ \\
2 \, {\rm D}_1 B_{23} =& -  \chi_4&  ~& 2 \, {\rm D}_2B_{23}=&- \chi_3& ~&  &2 \, {\rm D}_3B_{23}=
& \chi_2&  ~ &  2 \, {\rm D}_4B_{23}=& \chi_1 ~~~~~~\\
2 \, {\rm D}_1 B_{31} =& - \,    \chi_2& ~& 2 \, {\rm D}_2B_{31}=& \chi_1& ~&  &2 \, {\rm D}_3B_{31}=
&- \chi_4&   ~&  2 \, {\rm D}_4B_{31}=& \chi_3  ~~,~~~~
\end{array}
\label{tenD0A}
\ee
and for the fermions the analogous results,
\be
\begin{array}{cccccccccc}
{\rm D}_1 \chi_1 =& i \pa_0 \phi&   {\rm D}_2 \chi_1 =& i 2 \pa_0  B_{31}&  &  {\rm D}_3\chi_1=&  
i 2 \pa_0  B_{12}&    {\rm D}_4\chi_1=& i  2 \pa_0  B_{23} &\\
{\rm D}_1 \chi_2 =&-i  2  \pa_0  B_{31}&   {\rm D}_2\chi_2=& i \pa_0 \phi&  &  {\rm D}_3\chi_2=& 
i 2 \pa_0  
B_{23}&    {\rm D}_4\chi_2=& - i 2 \pa_0  B_{12}& \\
{\rm D}_1 \chi_3 =& -  2  \pa_0  B_{12}&   {\rm D}_2\chi_3=& -  2 \pa_0  B_{23}& &  {\rm D}_3\chi_3=&  
i \pa_0 \phi &    {\rm D}_4\chi_3=&  i 2  \pa_0  B_{31}& \\
{\rm D}_1 \chi_4 =& - i 2\pa_0  B_{23}&   {\rm D}_2\chi_4=& i 2 \pa_0  B_{12}& &  {\rm D}_3\chi_4=& 
- i 2 \pa_0  B_{31}&    {\rm D}_4\chi_4=& i \pa_0 \phi&
\end{array} ~~.
\label{tenD0B}
\ee

Next the fermions are re-defined according to
\be
 \chi_1 ~\to~ i \, \Psi_1 ~~~,~~~  \chi_2 ~\to~ i \, \Psi_2 ~~~,~~~ 
 \chi_3 ~\to~   i \, \Psi_3 ~~~,~~~  \chi_4 ~\to~ i \, \Psi_4
\label{tenD0C}
\ee
and the bosons are re-defined according to
\be
\Phi_1 ~=~ \phi ~~~,~~~ \Phi_2 ~=~ 2 \, B_{12} ~~~,~~~  \Phi_3 ~=~  2 \, B_{23} ~~~,~~~
 \Phi_4 ~=~  2 \, B_{31} ~~,
\label{tenD0D}
\ee
so the above system of equations ((\ref{tenD0A}) and (\ref{tenD0B})) respectively can be written 
in the forms
\be
{\rm D}{}_{{}_{\rm I}} \Phi_i ~=~ i \, \left( {\rm L}{}_{{}_{\rm I}}\right) {}_{i \, {\hat k}}  \,  \Psi_{\hat k}
~~,~~ {\rm D}{}_{{}_{\rm I}} \Psi_{\hat k} ~=~  \left( {\rm R}{}_{{}_{\rm I}}\right) {}_{{\hat k} \, i}  \,
{{d ~} \over { d t}} \, \Phi_{i}
\label{tenD0E}
\ee
where the explicit form of the L-matrices that appear here are given by
$$
\left( {\rm L}{}_{1}\right) {}_{i \, {\hat k}}   ~=~
\left[\begin{array}{cccc}
~1 & ~0 &  ~0  &  ~0 \\
~0 & ~0 &  -\, 1  &  ~ 0 \\
~0 & ~0 &  ~0  &  -\,1 \\
~0 & -\,1 &  ~ 0  &  ~0 \\
\end{array}\right] ~~~,~~~
\left( {\rm L}{}_{2}\right) {}_{i \, {\hat k}}   ~=~
\left[\begin{array}{cccc}
~0 & ~1 &  ~0  &  ~  0 \\
~0 & ~ 0 &  ~0  &  ~ 1 \\
~0 & ~ 0 &  -\,1  &  ~ 0 \\
 ~ 1 & ~0 &  ~0  &   ~ 0 \\
\end{array}\right]  ~~~,
$$
\be {~~~~}
\left( {\rm L}{}_{3}\right) {}_{i \, {\hat k}}   ~=~
\left[\begin{array}{cccc}
~0 & ~0 &  ~1  &  ~0 \\
~1 & ~0 &  ~ 0  &  ~ 0 \\
~0 & ~1 &  ~0  &   ~0 \\
~0 & ~0 &  ~ 0  &  -\, 1 \\
\end{array}\right] ~~~~~~,~~~
\left( {\rm L}{}_{4}\right) {}_{i \, {\hat k}}   ~=~
\left[\begin{array}{cccc}
~0 & ~0 &  ~0  &  ~  1 \\
~0 & -\, 1 &  ~0  &  ~ 0 \\
~1 & ~ 0 &  ~0  &  ~ 0 \\
 ~0 & ~0 &  ~1  &   ~ 0 \\
\end{array}\right]  ~~~,
\label{tenD0F}
\ee
and
$$
\left( {\rm R}{}_{1}\right) {}_{i \, {\hat k}}   ~=~
\left[\begin{array}{cccc}
~1 & ~~0 &  ~~0  &  ~0 \\
~0 & ~~0 &  ~~0  &  -\, 1 \\
~0 & ~-1 &  ~~ 0  &  ~ 0 \\
~0 & ~~0 &   -\,1  &  ~0 \\
\end{array}\right] ~~~,~~~
\left( {\rm R}{}_{2}\right) {}_{i \, {\hat k}}   ~=~
\left[\begin{array}{cccc}
~0 & ~~0 &  ~~ 0  &  ~ \, \, 1 \\
~1 & ~~ 0 &  ~~ 0  &  ~~ 0 \\
~0 & ~~ 0 &  ~- 1  &  ~~0 \\
~ 0 & ~~1 &  ~~0  &   ~~0  \\
\end{array}\right]  ~~~,
$$
\be  {~}
\left( {\rm R}{}_{3}\right) {}_{i \, {\hat k}}   ~=~
\left[\begin{array}{cccc}
~0 & ~1 &  ~~0  &  ~\,0 \\
~0 & ~0 &  ~~1  &  ~\,0 \\
~1 & ~~0 &  ~~0  &  ~\,0 \\
~0 & ~0 &   ~~0  &  -\, 1 \\
\end{array}\right] ~~~,~~~
\left( {\rm R}{}_{4}\right) {}_{i \, {\hat k}}   ~=~
\left[\begin{array}{cccc}
~0 & ~~0 &  ~~1  &  ~ \, \, 0 \\
~0 & ~-\, 1 &  ~~0  &  ~~0 \\
~0 & ~~ 0 &  ~~0  &  ~~1 \\
~ 1 & ~~~0 &  ~~0  &   ~~0  \\
\end{array}\right]  ~~.
\label{tenD0G}
\ee

There is also another feature that is noticable from (\ref{ten1}).  It is clear that we also
obtain
\be \eqalign{
{\rm D}_1 B_{0 \, 1} ~&=~ \fracm 12 \chi_1     ~~~~~~,~~   {\rm D}_1 B_{0 \, 2} ~=~  
\fracm 12 \chi_3  ~~,~~   {\rm D}_1 B_{0 \, 3} ~=~  \fracm 12 \chi_2   \cr
{\rm D}_2 B_{0 \, 1} ~&=~  \fracm 12 \chi_2   ~~~~~~,~~   {\rm D}_2 B_{0 \, 2}  ~=~ 
 \fracm 12 \chi_4   ~~,~~   {\rm D}_2 B_{0 \, 3} ~=~ \fracm 12 \chi_2 \cr
{\rm D}_2 B_{0 \, 1} ~&=~ -\,  \fracm 12 \chi_3   ~~,~~  {\rm D}_3 B_{0 \, 2}  ~=~ 
 \fracm 12 \chi_1   ~~,~~   {\rm D}_3 B_{0 \, 3} ~=~ - \fracm 12 \chi_4 \cr
{\rm D}_4 B_{0 \, 1} ~&=~ -\,  \fracm 12 \chi_4  ~~,~~   {\rm D}_4 B_{0 \, 2} ~=~ 
 \fracm 12 \chi_2  ~~,~~   {\rm D}_4 B_{0 \, 3} ~=~ - \fracm 12 \chi_3 ~~,
} \label{tenD0H}
\ee
in addition to the results in (\ref{tenD0A}).  However, on the right hand 
side of the equations in (\ref{tenD0B}) there are {\em {no}} appearances
of terms that depend on $B_{0 \, 1}$, $B_{0 \, 2}$, or $B_{0 \, 3}$.  From
(\ref{ten4}) it can be seen that
\be \eqalign{
\d_{G}^{(2)} B_{0 \, 1} ~&=~  \pa_0  \,  \ell_1  ~~,~~ 
\d_{G}^{(2)} B_{0 \, 2} ~=~  \pa_0  \,  \ell_2  ~~,~~ 
\d_{G}^{(2)} B_{0 \, 3} ~=~  \pa_0  \,  \ell_3 
~~
} \label{tenD0I}
\ee
expresses the form of the gauge transformation on the 0-brane.  The two equations (\ref
{tenD0H}) and (\ref{tenD0I}) together imply that:
\newline $~~~~~$ (a.) the gauge transformations make it possible to choose a gauge
where 
\newline $~~~~~$ $~~~~~$ $B_{0 \, 1}$ =  $B_{0 \, 2}$ =  $B_{0 \, 3}$ = 0 (ignoring issues
related to zero-modes),
\newline $~~~~~$ (b.) the supersymmetry variations described by (\ref{tenD0H}) take
one out of this 
\newline $~~~~~$ $~~~~~$ 
gauge, and
\newline $~~~~~$ (c.) there exist further gauge transformations that can be used to 
restore
\newline $~~~~~$ $~~~~~$ 
the $B_{0 \, 1}$ =  $B_{0 \, 2}$ =  $B_{0 \, 3}$ = 0 gauge condition.  \newline \noindent
So on the 0-brane
it is consistent to simply ignore the field components ($B_{0 \, 1}$, $B_{0 \, 2}$,  $B_{0 \, 3}$)
and work in a `Coulomb gauge.'  

\subsection{4D, $\cal N$ = 1 Double Tensor Multiplet On The 0-Brane}

$~~~$ Starting from (\ref{dblten1}) we find carrying out the reduction  for the bosons leads to
\be
\begin{array}{cc}
{\rm D}_1 X_{12} = -(\frac{1}{2} \L_2) \hspace{15pt} & {\rm D}_2 X_{12} = -(\frac{1}{2} \L_1) \\
{\rm D}_1 X_{23} = +(\frac{1}{2} \L_1) \hspace{15pt}& {\rm D}_2 X_{23} = -(\frac{1}{2} \L_2)  \\
{\rm D}_1 X_{31} = +(\frac{1}{2} \L_3) \hspace{15pt}& {\rm D}_2 X_{31} = +(\frac{1}{2} \L_4)  \\
{\rm D}_1 Y_{12} = -(\frac{1}{2} \L_3) \hspace{15pt}& {\rm D}_2 Y_{12} = +(\frac{1}{2} \L_4)  \\
{\rm D}_1 Y_{23} = -(\frac{1}{2} \L_4) \hspace{15pt}& {\rm D}_2 Y_{23} = -(\frac{1}{2} \L_3)  \\
{\rm D}_1 Y_{31} = -(\frac{1}{2} \L_2) \hspace{15pt}& {\rm D}_2 Y_{31} = +(\frac{1}{2} \L_1)  \\
\end{array}
\label{DTD0A}
\ee
\be
\begin{array}{cc}
{\rm D}_3 X_{12} = +(\frac{1}{2} \L_4) \hspace{15pt} & {\rm D}_4 X_{12} = +(\frac{1}{2} \L_3) \\
{\rm D}_3 X_{23} = -(\frac{1}{2} \L_3) \hspace{15pt}& {\rm D}_4 X_{23} = +(\frac{1}{2} \L_2)  \\
{\rm D}_3 X_{31} = +(\frac{1}{2} \L_1) \hspace{15pt}& {\rm D}_4 X_{31} = +(\frac{1}{2} \L_2)  \\
{\rm D}_3 Y_{12} = +(\frac{1}{2} \L_1) \hspace{15pt}& {\rm D}_4 Y_{12} = -(\frac{1}{2} \L_2)  \\
{\rm D}_3 Y_{23} = +(\frac{1}{2} \L_2) \hspace{15pt}& {\rm D}_4 Y_{23} = +(\frac{1}{2} \L_1)  \\
{\rm D}_3 Y_{31} = -(\frac{1}{2} \L_4) \hspace{15pt}& {\rm D}_4 Y_{31} = +(\frac{1}{2} \L_3)  \\
\end{array}
\label{DTD0B}
\ee
and for the fermions
\be
\begin{array}{ll}
{\rm D}_1 \,(\frac{1}{2}\,\L_{1}) = +i\hspace{2pt} \partial_0 X_{23} \hspace{15pt}
& {\rm D}_2 \,(\frac{1}{2}\,\L_{1}) = -i \hspace{2pt}\partial_0 X_{12}+i\hspace{2pt} 
\partial_0 Y_{31} \hspace{15pt}\\
{\rm D}_1 \,(\frac{1}{2}\,\L_{2}) = -i \hspace{2pt}\partial_0 X_{12}-i\hspace{2pt} 
\partial_0 Y_{31} \hspace{15pt} & {\rm D}_2\, (\frac{1}{2}\,\L_{2}) = -i \hspace{2pt}
\partial_0 X_{23}\hspace{15pt} \\
{\rm D}_1 \,(\frac{1}{2}\,\L_{3}) = +i \hspace{2pt}\partial_0 X_{31}-i\hspace{2pt} 
\partial_0 Y_{12}\hspace{15pt} & {\rm D}_2 \,(\frac{1}{2}\,\L_{3}) = -i \hspace{2pt}
\partial_0 Y_{23} \hspace{15pt}\\
{\rm D}_1 \,(\frac{1}{2}\,\L_{4}) = -i\hspace{2pt} \partial_0 Y_{23} \hspace{15pt}
& {\rm D}_2 \,(\frac{1}{2}\,\L_{4}) = +i\hspace{2pt} \partial_0 X_{31}+i\hspace{2pt} 
\partial_0 Y_{12}\hspace{15pt}
\end{array}
\label{DTD0C}
\ee
\be
\begin{array}{ll}
{\rm D}_3 \,(\frac{1}{2}\,\L_{1}) = +i\hspace{2pt} \partial_0 X_{31}+i \hspace{2pt}
\partial_0 Y_{12} \hspace{15pt} & {\rm D}_4 \,(\frac{1}{2}\,\L_{1}) = +i\hspace{2pt} 
\partial_0 Y_{23} \hspace{15pt}\\
{\rm D}_3 \,(\frac{1}{2}\,\L_{2}) = +i\hspace{2pt} \partial_0 Y_{23} \hspace{15pt}
& {\rm D}_4 \,(\frac{1}{2}\,\L_{2}) = +i\hspace{2pt} \partial_0 X_{31}-i\hspace{2pt} 
\partial_0 Y_{12}\hspace{15pt} \\ 
{\rm D}_3 \,(\frac{1}{2}\,\L_{3}) = -i\hspace{2pt} \partial_0 X_{23}\hspace{15pt}
& {\rm D}_4 \,(\frac{1}{2}\,\L_{3}) = +i \hspace{2pt}\partial_0 X_{12}+i\hspace{2pt} 
\partial_0 Y_{31} \hspace{15pt}\\ 
{\rm D}_3 \,(\frac{1}{2}\,\L_{4}) = +i \hspace{2pt}\partial_0 X_{12}-i\hspace{2pt} 
\partial_0 Y_{31} \hspace{15pt} & {\rm D}_4 \,(\frac{1}{2}\,\L_{4}) = +i \hspace{2pt}
\partial_0 X_{23}\hspace{15pt}
\end{array}  ~~.
\label{DTD0D}
\ee

Using the notation:
\be
 \Phi_i ~=~\left( X_{12},~~~ X_{23},~~~  X_{31},~~~
 Y_{12}, ~~~ Y_{23},~~~ Y_{31}\right) ~~,
\label{DTD0E}
\ee
and for the fermions 
\be
 \frac{1}{2}\,\L_1 ~\to~ i \, \Psi_1 ~~~,~~~  \frac{1}{2}\,\L_2 ~\to~ i \, \Psi_2 ~~~,~~~ 
 \frac{1}{2}\,\L_3 ~\to~   i \, \Psi_3 ~~~,~~~  \frac{1}{2}\,\L_4 ~\to~ i \, \Psi_4 ~~,
\label{DTD0F}
\ee
the above systems of equations can be written in the form of (\ref{chiD0E}) and  (\ref{chiD0J}).
The explicit form of the matrices that appear here are given by

$$
\left( {\rm L}{}_{1}\right) {}_{i \, {\hat k}}   ~=~
\left[\begin{array}{cccc}
~0 & -1 &  ~~0  &  ~~0 \\
~1 & ~~0 &  ~~0  &  ~~ 0 \\
~0 & ~~0 &  ~~1  &  ~~0 \\
~0 & ~~0 &  ~-1  &  ~~0 \\
~0 & ~~0 &  ~~0  &  ~-1 \\
~0 & ~-1 &  ~~ 0  &  ~~0
\end{array}\right],\hspace{15pt}
\left( {\rm L}{}_{2}\right) {}_{i \, {\hat k}}   ~=~
\left[\begin{array}{cccc}
-1 & ~~0 &  ~~0  &  ~~0 \\
~0 & ~-1 &  ~~0  &  ~~ 0 \\
~0 & ~~0 &  ~~0  &  ~~1 \\
~0 & ~~0 &  ~~0  &  ~~1 \\
~0 & ~~0 &  ~-1  &  ~~0 \\
~1 & ~~0 &  ~~ 0  &  ~~0
\end{array}\right],
$$

\be
\left( {\rm L}{}_{3}\right) {}_{i \, {\hat k}}   ~=~
\left[\begin{array}{cccc}
~0 & ~~0 &  ~~0  &  ~~1 \\
~0 & ~~0 &  ~-1  &  ~~ 0 \\
~1 & ~~0 &  ~~0  &  ~~0 \\
~1 & ~~0 &  ~~ 0  &  ~~0 \\
~0 & ~~1 &  ~~0  &  ~~0 \\
~0 & ~~0 &  ~~ 0  &  ~-1\\
\end{array}\right],\hspace{15pt}
\left( {\rm L}{}_{4}\right) {}_{i \, {\hat k}}   ~=~
\left[\begin{array}{cccc}
~0 & ~~0 &  ~~1  &  ~~0 \\
~0 & ~~0 &  ~~0  &  ~~1 \\
~0 & ~~1 &  ~~0  &  ~~0 \\
~0 & ~-1 &  ~~ 0  &  ~~0 \\
~1 & ~~0 &  ~~0  &  ~~0 \\
~0 & ~~0 &  ~~ 1  &  ~~0\\
\end{array}\right].
\label{DTD0G}
\ee

$$
\left( {\rm R}{}_{1}\right) {}_{i \, {\hat k}}   ~=~
\left[\begin{array}{cccccc}
~0 & ~~1 &~0 & ~~0 &  ~~0  &  ~0 \\
-1 & ~~0 &~0 & ~~0 &  ~~0  &  -1 \\
~0 & ~~0 &~1 & ~-1 &  ~~0  &  ~0 \\
~0 & ~~0 &~0 & ~~0 &  ~-1  &  ~0 \\
\end{array}\right],
$$

$$
\left( {\rm R}{}_{12}\right) {}_{i \, {\hat k}}   ~=~
\left[\begin{array}{cccccc}
-1 & ~~0 & ~0 & ~~0 &  ~~0  &  ~1\\ 
~0 & ~-1 & ~0 & ~~0 &  ~~0  &  ~0 \\
~0 & ~~0 & ~0 & ~~0 &  ~-1  &  ~0 \\
~0 & ~~0 & ~1 & ~~1 &  ~~0  &  ~0 \\
\end{array}\right],
$$
$$
\left( {\rm R}{}_{3}\right) {}_{i \, {\hat k}}   ~=~
\left[\begin{array}{cccccc}
~0 & ~~0 & ~1 & ~~1 &  ~~0  &  ~0\\ 
~0 & ~~0 & ~0 & ~~0 &  ~~1  &  ~0 \\
~0 & ~-1 & ~0 & ~~0 &  ~~0  &  ~0 \\
~1 & ~~0 &~ 0 & ~~0 &  ~~0  &  -1 \\
\end{array}\right]  ~~~,
$$
\be
\left( {\rm R}{}_{4}\right) {}_{i \, {\hat k}}   ~=~
\left[\begin{array}{cccccc}
~0 & ~~0 & ~0 & ~~0 &  ~~1  &  ~0\\ 
~0 & ~~0 & ~1 & ~-1 &  ~~0  &  ~0 \\
~1 & ~~0 & ~0 & ~~0 &  ~~0  &  ~1 \\
~0 & ~~1 & ~0 & ~~0 &  ~~0  &  ~0 \\
\end{array}\right]  ~~.
\label{DTD0H}
\ee

\subsection{4D, $\cal N$ = 1 Vector Multiplet On The 0-Brane}

$~~~$ Starting from (\ref{V1}) we find that carrying out the reduction  for the bosons leads to
\be
\begin{array}{cccccccccccc}
 {\rm D}_1 A_{1} =&  \l_2&  ~&  {\rm D}_2A_{1}=&\l_1& ~&  & {\rm D}_3A_{1}=&\l_4& ~&   
{\rm D}_4A_{1}=& \l_3~~~~~~~ \\
 {\rm D}_1 A_{2} =& -  \l_4&  ~&  {\rm D}_2A_{2}=&\l_3& ~&  & {\rm D}_3A_{2}=& \l_2& 
 ~ &   {\rm D}_4A_{2}=& - \l_1 ~~~~~~\\
 {\rm D}_1 A_{3} =&     \l_1& ~&  {\rm D}_2A_{3}=&-  \l_2& ~&  & {\rm D}_3A_{3}=& \l_3&  
 ~&   {\rm D}_4A_{3}=&- \l_4  ~~~~~~ \\
 {\rm D}_1 {\rm d} \,~=& -  \pa_0 \l_3& ~& {\rm D}_2 {\rm d} ~~=&  -  \pa_0 \l_4 & ~&  ~
 &{\rm D}_3{\rm d}~~=&  \pa_0 \l_1&  ~&  
{\rm D}_4{\rm d} ~~=&  \pa_0 \l_2 ~~~~~~~
\end{array}
\label{V1D0A}
\ee
and for the fermions
\be
\begin{array}{cccccccccc}
{\rm D}_1 \l_1 =& i \pa_0 A_{3} &   {\rm D}_2 \l_1 =& i \pa_0  A_{1}&  &  {\rm D}_3\l_1=&  i \, {\rm d}  
&    {\rm D}_4\l_1=& - i   \pa_0  A_{2} &\\
{\rm D}_1 \l_2 =& i \pa_0  A_{1}&   {\rm D}_2\l_2=&- i  \pa_0 A_{3}&  &  {\rm D}_3\l_2=& i \pa_0  
A_{2}&    {\rm D}_4\l_2=& i \, {\rm d}& \\
{\rm D}_1 \l_3 =& - i {\rm d}&   {\rm D}_2\l_3=& i  \pa_0  A_{2}& &  {\rm D}_3\l_3=&i  \pa_0 
A_{3} &    {\rm D}_4\l_3=&  i  \pa_0  A_{1}& \\
{\rm D}_1 \l_4 =& - i \pa_0  A_{2}&   {\rm D}_2\l_4=&- i {\rm d}& &  {\rm D}_3\l_4=& i  \pa_0 
 A_{1}&    {\rm D}_4\l_4=& - i \pa_0 A_{3}&
\end{array}
\label{V1D0B}
\ee
so these suggest the following identifications for the $\Phi$'s and $\Psi$'s
\be {~~}
 \l_1 ~\to~ i \, \Psi_1 ~~~,~~~  \l_2 ~\to~ i \, \Psi_2 ~~~,~~~ 
 \l_3 ~\to~   i \, \Psi_3 ~~~,~~~  \l_4 ~\to~ i \, \Psi_4 ~~~,
\label{V1D0C}
\ee
\be \eqalign{ {~~~~~~~~~~~~~}
 \Phi_1 ~=~ A_{1} ~~~~~,~~~
\Phi_2 ~=~  A_{2} ~~~~~,~~~  
\Phi_3 ~=~  A_{3} ~~~~~,~~~
\pa_0 \Phi_4 ~=~  {\rm d} ~~~~~.  }
\label{V1D0D}
\ee

We continue as in the previous discussion to define the L-matrices and R-matrices.  Given the 
equations (\ref{V1D0A}) - (\ref{V1D0D}) we find the results below for the L-matrices
$$
\left( {\rm L}{}_{1}\right) {}_{i \, {\hat k}}   ~=~
\left[\begin{array}{cccc}
~0 & ~1 &  ~ 0  &  ~ 0 \\
~0 & ~0 &  ~0  &  -\,1 \\
~1 & ~0 &  ~ 0  &  ~0 \\
~0 & ~0 &  -\, 1  &  ~0 \\
\end{array}\right] ~~~,~~~
\left( {\rm L}{}_{2}\right) {}_{i \, {\hat k}}   ~=~
\left[\begin{array}{cccc}
~1 & ~ 0 &  ~0  &  ~ 0 \\
~0 & ~ 0 &  ~1  &  ~ 0 \\
 ~0 & - \, 1 &  ~0  &   ~ 0 \\
~0 & ~0 &  ~0  &  -\, 1 \\
\end{array}\right]  ~~~,
$$
\be {~~~~}
\left( {\rm L}{}_{3}\right) {}_{i \, {\hat k}}   ~=~
\left[\begin{array}{cccc}
~0 & ~0 &  ~ 0  &  ~ 1 \\
~0 & ~1 &  ~0  &   ~0 \\
~0 & ~0 &  ~ 1  &  ~0 \\
~1 & ~0 &  ~0  &  ~0 \\
\end{array}\right] ~~~~~~,~~~
\left( {\rm L}{}_{4}\right) {}_{i \, {\hat k}}   ~=~
\left[\begin{array}{cccc}
~0 & ~0 &  ~1  &  ~ 0 \\
-\,1 & ~ 0 &  ~0  &  ~ 0 \\
 ~0 & ~0 &  ~0  &   - \, 1 \\
~0 & ~1 &  ~0  &  ~  0 \\
\end{array}\right]  ~~~,
\label{V1D0E}
\ee
and the associated R-matrices are found by the relation in (\ref{chiD0H}).

Similarity to the case of the tensor multiplet can also be seen.  In addition to the results
in (\ref{V1D0A}) we also have
\be
{\rm D}_1 A_0 ~=~ -\, \l_2  ~~,~~ {\rm D}_2 A_0 ~=~ \l_1  ~~,~~ {\rm D}_3 A_0 ~=~ \l_4  ~~,~~
{\rm D}_4 A_0 ~=~  -\, \l_3 ~~.
 ~~~,
\label{V1D0F}
\ee
Furthermore, there is no appearance of $A_0$ in the equations of (\ref{V1D0B}) and there
is the gauge transformation as stated in (\ref{V3}). Thus it is consistent to work in the
Coulomb gauge where we set $A_0$ = 0 throughout our considerations of the vector multiplet.

As with the chiral multiplet, it is possible in the case of the vector multiplet to consider
the on-shell theory.   This begins by setting $d$ = 0.  The consistency of these conditions 
imply $\pa_0 \l_{\hk}$ = $\pa_0^2 A_1$  = $\pa_0^2 A_2$  = $\pa_0^2 A_3$  = 0.  Further
consistency conditions implies that $\Phi_i$ be defined by 
\be \eqalign{ {~~~~~~~~~~~~~}
 \Phi_1 ~=~ A_{1} ~~~~~,~~~
\Phi_2 ~=~  A_{2} ~~~~~,~~~  
\Phi_3 ~=~  A_{3}  ~~~~~, }
\label{V1D0G}
\ee
while the $\Psi$-fermions are still defined by (\ref{V1D0C}).  Using these definitions, the
on-shell vector multiplet satisfies equations as in (\ref{tenD0E}), but with the L-matrices 
define by
$$
\left( {\rm L}{}_{1}\right) {}_{i \, {\hat k}}   ~=~
\left[\begin{array}{cccc}
~0 & ~1 &  ~ 0  &  ~ 0 \\
~0 & ~0 &  ~0  &  -\,1 \\
~1 & ~0 &  ~ 0  &  ~0 \\
\end{array}\right] ~~~,~~~
\left( {\rm L}{}_{2}\right) {}_{i \, {\hat k}}   ~=~
\left[\begin{array}{cccc}
~1 & ~ 0 &  ~0  &  ~ 0 \\
~0 & ~ 0 &  ~1  &  ~ 0 \\
 ~0 & - \, 1 &  ~0  &   ~ 0 \\
\end{array}\right]  ~~~,
$$
\be {~~~~}
\left( {\rm L}{}_{3}\right) {}_{i \, {\hat k}}   ~=~
\left[\begin{array}{cccc}
~0 & ~0 &  ~ 0  &  ~ 1 \\
~0 & ~1 &  ~0  &   ~0 \\
~0 & ~0 &  ~ 1  &  ~0 \\
\end{array}\right] ~~~~~~,~~~
\left( {\rm L}{}_{4}\right) {}_{i \, {\hat k}}   ~=~
\left[\begin{array}{cccc}
~0 & ~0 &  ~1  &  ~ 0 \\
-\,1 & ~ 0 &  ~0  &  ~ 0 \\
 ~0 & ~0 &  ~0  &   - \, 1 \\
\end{array}\right]  ~~~,
\label{V1D0H}
\ee
and the R-matrices are found to satisfy (\ref{chiD0L}).  

\subsection{Summary Of Multiplet Reduction On The 0-Brane}

$~~~$ Earlier in this chapter, the Garden Algebra matrices associated with four 4D,
$\cal N$ = 1 supermultiplets were derived for: \newline \indent
(a.) the off-shell chiral multiplet where the associated L-matrices 
 \newline \indent $~~~~~$
and R-matrices appear in (\ref{chiD0F}) and (\ref{chiD0K}) (case $I$), 
\newline \indent
(b.) the on-shell chiral multiplet where the associated L-matrices 
 \newline \indent $~~~~~$
and R-matrices appear in (\ref{chiD0N}) and (\ref{chiD0O}) (case $II$), 
\newline \indent
(c.) the off-shell tensor multiplet where the associated L-matrices 
 \newline \indent $~~~~~$
and R-matrices appear in (\ref{tenD0F}) and (\ref{tenD0G}) (case $III$), 
\newline \indent
(d.) the double tensor multiplet where the associated L-matrices 
 \newline \indent $~~~~~$
and R-matrices appear in (\ref{DTD0G}) and (\ref{DTD0H}) (case $IV$), 
\newline \indent
(e.) the off-shell vector multiplet where the associated L-matrices 
 \newline \indent $~~~~~$
and R-matrices appear in (\ref{V1D0E}) and (\ref{chiD0J}) (case $V$), 
\newline \indent
(f.) and the on-shell vector multiplet where the associated L-matrices 
 \newline \indent $~~~~~$
and R-matrices appear in (\ref{V1D0H}) and (\ref{chiD0J})  (case $VI$). 
\newline
For later convenience we will refer to these as case $I$ through case $VI$. 

Before the reduction procedure that reveals the matrices, the multiplets describe four 
1D, $\cal N$ = 4 theories.  The matrices associated with each multiplet in the cases
of $I$, $III$ and $V$ (the off-shell representations) share some common features. They 
all satisfy the equations
\be \eqalign{
 (\,{\rm L}_\rI\,)_i{}^\hj\>(\,{\rm R}_\rJ\,)_\hj{}^k + (\,{\rm L}_\rJ\,)_i{}^\hj\>(\,{\rm R}_\rI\,)_\hj{}^k
  &= 2\,\d_{\rI\rJ}\,\d_i{}^k~~,\cr
 (\,{\rm R}_\rJ\,)_\hi{}^j\>(\, {\rm L}_\rI\,)_j{}^\hk + (\,{\rm R}_\rI\,)_\hi{}^j\>(\,{\rm L}_\rJ\,)_j{}^\hk
  &= 2\,\d_{\rI\rJ}\,\d_\hi{}^\hk~~.
}  \label{GarDNAlg1}
 \ee
\be
~~~~
 (\,{\rm R}_\rI\,)_\hj{}^k\,\d_{ik} = (\,{\rm L}_\rI\,)_i{}^\hk\,\d_{\hj\hk}~~,
\label{GarDNAlg2}
\end{equation}
which we have named as the ``$\cal {GR}$(d, $\cal N$) Algebras'' or ``Garden Algebras.''
Here the indices have ranges that correspond to $\rm I$, $\rm J$, etc.  = 1, $\dots$, $\cal 
N$, i, j, etc. = 1, $\dots$, d${}_L$, and $\hi$, $\hj$, etc. =  1, $\dots$, d${}_R$ for some 
integers $\cal N$, d${}_L$, and d${}_R$.  

Throughout most previous discussions, there has only been consideration of the case 
where d${}_L$ = d${}_R$ = d.  In this case, the L-matrices and R-matrices may be assembled 
according to
\be
\g_\rI  ~=~  \left[\begin{array}{cc}
~0 & ~~  \,{\rm L}_\rI\,  \\
~\,{\rm R}_\rI\, & ~0 \\
\end{array}\right]
\label{GarDNAlg3}
 \ee
 and we may introduce one additional 2d $\times$ 2d matrix $(-1)^{\cal F}$ where
 \be
(-1)^{\cal F}~=~  \left[\begin{array}{cc}
~{\rm I} & ~~  0  \\
~0 & ~ - \, {\rm I} \\
\end{array}\right]  ~~.
\label{GarDNAlg4}
 \ee
Thus, due to (\ref{GarDNAlg1}), the $\g_\rI$'s together with $(-1)^{\cal F}$ satisfy the Clifford 
Algebra $Cl({\cal N} + 1)$ over the reals.

However, the case where d${}_L$ $\ne$ d${}_R$ (i.e. cases $II$, $IV$ and $VI$),
can also be considered.  In this more 
general case, the matrices may be described as belonging to a mathematical structure 
denoated by the symbol $\cal {GR}$(d${}_L$, d${}_R$, $\cal N$).  In the case of on-shell 
theories, it is the case that d${}_L$ $\ne$ d${}_R$ so in order to extend the discussion of 
the previous works to the on-shell cases, we will have to consider $\cal {GR}$(d${}_L$, 
d${}_R$, $\cal N$) matrices for the on-shell theories as well as the Double Tensor Multiplet.
We should add that since we have not studied $\cal {GR}$(d${}_L$, d${}_R$, $\cal N$), 
its precise nature is not understood.  However, calculations involving this structure will
be presented in an appendix.

\newpage
\section{Considering Some Traces}

$~~~$ As we have seen from the discussions of the previous chapters, each
supersymmetrical multiplet has an associated set of L-matrices and R-matrices that
are revealed upon reduction on a 0-brane.  In general, however, these matrices
do not have to be square.  What we have shown is that that when the supermultiplet
is off-shell, the matrices will be square.  A question that might be interesting to consider
is, ``For a given multiplet, how unique are such matrices?''

Clearly, to obtain the matrices, we have made many arbitrary choices along the way.  
So the uniqueness question can also be cast in as the following form.  Let us begin 
with the assumption that there exists two sets (linearly independent of one another) 
of real matrices such that ${\rm L}{}_{\bj I}$ and ${\Hat {\rm L}}{}_{\bj I}$ that satisfy\footnote{No 
summations over indices are implied for the equations in (\ref{eq:E01}).}
\be \eqalign{
{\rm L}_{\bj I} ({\rm L}_{\bj I})^t  ~&=~  ({\rm L}_{\bj I})^t {\rm L}_{\bj I} ~=~   {\bf I} ~~~,~~~
{\Hat {\rm L}}{}_{\bj I} ({\Hat {\rm L}}{}_{\bj I})^t  ~=~  ({\Hat {\rm L}}{}_{\bj I})^t  
{\Hat {\rm L}}{}_{\bj I} ~=~   {\bf I} 
~~~.
}
\label{eq:E01}
\ee
\be \eqalign{
&{\rm L}_{\bj I} ({\rm L}_{\bj J})^t ~+~ {\rm L}_{\bj J} ({\rm L}_{\bj I})^t  ~=~ 0 ~~~,~~~
({\Hat {\rm L}}{}_{\bj I})^t  {\Hat {\rm L}}{}_{\bj J}~+~  ({\Hat {\rm L}}{}_{\bj J})^t 
{\Hat {\rm L}}{}_{\bj I}  ~=~ 0  ~~~. 
}
\label{eq:E02}
\ee
We say that ${\rm L}{}_{\bj I}$ and ${\Hat {\rm L}}{}_{\bj I}$ are members of the same 
equivalence class
if there exists real square matrices $\cal X$ and $\cal Y$ such that
\be \eqalign{
{\Hat {\rm L}}{}_{\bj I} ~=~ {\cal  X} \, {\rm L}{}_{\bj I} {\cal Y}
}
\label{eq:E03}
\ee
and where
\be \eqalign{
{\cal  X} \, ({\cal X})^t ~=~   ({\cal X})^t  \, {\cal  X} ~=~
{\cal  Y} \, ({\cal Y})^t ~=~   ({\cal Y})^t  \, {\cal  Y} 
~=~   {\bf I}   ~~~.
}
\label{eq:E04}
\ee
These last equations imply that ${\cal  X}$ is an element of the O(d${}_L$) group
while ${\cal  Y}$ is an element of the O(d${}_R$) group\footnote{The curious reader
may well ask, ``Why are these groups relevant?''  Some insight into \newline $~~~~~~$ 
this comes from the work of \cite{AdnkDynam}. There it was shown that for valise
Adinkras, it is al- \newline $~~~~~~$ ways possible to construct a supersymmetrical 
invariant that is quadratic in the fields \newline $~~~~~~$ of the Adinkra.  There are 
a large set of linear field redefinitions that do {\em {not}} mix bo-  \newline $~~~~~~$
sons and fermions, under which this supersymmetrical invariant remains unchanged.   
\newline $~~~~~~$ The  O(d${}_L$) and O(d${}_R$) groups are related to these
symmetries.}.  Using (\ref{eq:E03}), we next observe that
\be \eqalign{
{\Hat {\rm L}}{}_{\bj I} ({\Hat {\rm L}}{}_{\bj J})^t   ~&=~   {\cal X} \, [ \, {\rm L}{}_{\bj I} ({\rm L}{}_{\bj J})^t  \, ] \,  
 ({\cal X})^t~~~~~, \cr
({\Hat {\rm L}}{}_{\bj I})^t  {\Hat {\rm L}}{}_{\bj J} ~&=~  ({\cal Y})^t \, [\, ({\rm L}{}_{\bj I})^t  {\rm L}{}_{\bj J}\,]
 \,   {\cal Y}
~~~~~~,
}
\label{eq:E20}
\ee
or on taking traces we see
\be \eqalign{
{\rm {Tr}} \left[ \, {\Hat {\rm L}}{}_{{\bj I}_{{}_1} } ({\Hat {\rm L}}{}_{{\bj J}_{{}_1} })^t  \, \right]
 ~&=~   {\rm {Tr}} \left[ \,  {\rm L}{}_{{\bj I}_{{}_1} }({\rm L}{}_{{\bj J}_{{}_1}})^t    \, \right] ~~~~~, \cr
{\rm {Tr}} \left[ \,  ({\Hat {\rm L}}{}_{{\bj I}_{{}_1} })^t  {\Hat {\rm L}}{}_{{\bj J}_{{}_1} } \, \right] 
 ~&=~  {\rm {Tr}} \left[ \,  ({\rm L}{}_{{\bj I}_{{}_1}})^t    {\rm L}{}_{{\bj J}_{{}_1} }
  \, \right] 
~~~~~.
}
\label{eq:E21}
\ee
This property is shared by more general expressions of the form
\be \eqalign{
\varphi^{(p)} {}_{{\bj I}_{{}_1} } {}_{{\bj J}_{{}_1} }{}_{...}  {}_{{\bj I}_{{}_p} } {}_{{\bj J}_{{}_p} }
~&=~  {\rm {Tr}} \left[ \,  {\rm L}{}_{{\bj I}_{{}_1} }({\rm L}{}_{{\bj J}_{{}_1}})^t   
\cdots \,  {\rm L}{}_{{\bj I}_{{}_p} }({\rm L}{}_{{\bj J}_{{}_p}})^t   
 \, \right] 
~~~~~,  \cr
{\Tilde {\varphi}}{}^{(p)} {}_{{\bj I}_{{}_1} } {}_{{\bj J}_{{}_1} }{}_{...}  {}_{{\bj I}_{{}_p} } {}_{{\bj J}_{{}_p} }
~&=~   {\rm {Tr}} \left[ \, ({\rm L}{}_{{\bj I}_{{}_1}})^t   {\rm L}{}_{{\bj J}_{{}_1} }   \cdots \, 
({\rm L}{}_{{\bj I}_{{}_p}})^t   
 {\rm L}{}_{{\bj J}_{{}_p} }
 \, \right] 
~~~~~.
}
\label{eq:E22}
\ee
We note that for the present case under consideration, we will not consider $p$ $>$
2.  Furthermore, using the cyclicity of the trace operation we have
\be \eqalign{
{\Tilde {\varphi}}{}^{(p)} {}_{{\bj I}_{{}_1} } {}_{{\bj J}_{{}_1} }{}_{...}  {}_{{\bj I}_{{}_p} } {}_{{\bj J}_{{}_p} }
~=~ \varphi^{(p)} {}_{{\bj J}_{{}_1} } {}_{{\bj I}_{{}_2} }{}_{...}  {}_{{\bj J}_{{}_p} } {}_{{\bj I}_{{}_1} }
~~~~~.
}
\label{eq:E23}
\ee
The collection of all such objects shares some of the properties of characters as for
groups.  Due to the identities in (\ref{eq:E20}) the value of these objects is independent
of the linear field redefinitions that leave a quadratic super-invariant (see \cite{AdnkDynam})
unchanged.  We will call these ``chromocharacters'' because their values still depend
on the choices made to describe the supersymmetry generators.  So these objects
still depend on how the colors in an Adinkra are picked.

Since we have derived the L-matrices and R-matrices for six distinct cases,
$I$, $II$, $III$, $IV$, $V$, and $VI$ (as delineated in above equation (\ref{GarDNAlg1})),
we will denote the distinct cases by including a roman numeral after the symbol
for the chromocharacter.  Our calculations reveal
\be \eqalign{
{ {\varphi}}^{(1)}  {}_{{\bj I}} {}_{{\bj J} }(I)
~&=~ 4 \, \delta_{ {}_{\bj I }  {}_{\bj J }}
 ~~~, \cr
{ {\varphi}}^{(1)}  {}_{{\bj I}} {}_{{\bj J} } (II)
~&=~  2 \,  \, \delta_{ {}_{\bj I }  {}_{\bj J }} ~~~, \cr
{ {\varphi}}^{(1)}  {}_{{\bj I}} {}_{{\bj J} } (III)
~&=~   4 \, \delta_{ {}_{\bj I }  {}_{\bj J }}  ~~~, \cr
{ {\varphi}}^{(1)}  {}_{{\bj I}} {}_{{\bj J} }(IV)
~&=~   6 \,  \, \delta_{ {}_{\bj I }  {}_{\bj J }}  ~~~, \cr
{ {\varphi}}^{(1)}  {}_{{\bj I}} {}_{{\bj J} }(V)
~&=~   4 \, \delta_{ {}_{\bj I }  {}_{\bj J }}~~~, \cr
{ {\varphi}}^{(1)} {}_{{\bj I}} {}_{{\bj J} } (VI)
~&=~    3 \,  \, \delta_{ {}_{\bj I }  {}_{\bj J }}
~~~~~.
}
\label{Chr0M1}
\ee
The behavior of the $p$ = 2 chromocharacters is very different for the off-shell cases
($I$, $II$, $V$) versus on-shell cases ($II$, $IV$, $VI$).
We present the off-shell cases first:
\be \eqalign{
\varphi^{(2)} {}_{\bj I }  {}_{\bj J }  {}_{\bj K }  {}_{\bj L } (I)
~&=~  4 \, \left[~ \delta_{ {}_{\bj I }  {}_{\bj J }}\delta_{ 
{}_{\bj K }  {}_{\bj L }} ~-~  \delta_{ {}_{\bj I }  {}_{\bj K }}
\delta_{ {}_{\bj J }  {}_{\bj L }} ~+~ \delta_{ {}_{\bj I }  {}_{
\bj L }}\delta_{ {}_{\bj J }  {}_{\bj K }} ~+~ 
\epsilon_{ {}_{\bj I }  {}_{\bj J }  {}_{\bj K }  {}_{\bj L }}  
~ \right] ~~~, \cr
\varphi^{(2)}  {}_{\bj I }  {}_{\bj J }  {}_{\bj K }  {}_{\bj L }(III)
~&=~    4 \, \left[~ \delta_{ {}_{\bj I }  {}_{\bj J }}\delta_{ 
{}_{\bj K }  {}_{\bj L }} ~-~  \delta_{ {}_{\bj I }  {}_{\bj K }}
\delta_{ {}_{\bj J }  {}_{\bj L }} ~+~ \delta_{ {}_{\bj I }  {}_{
\bj L }}\delta_{ {}_{\bj J }  {}_{\bj K }} ~-~  
\epsilon_{ {}_{\bj I }  {}_{\bj J }  {}_{\bj K }  {}_{\bj L }}  
~ \right]   ~~~, \cr
\varphi^{(2)}  {}_{\bj I }  {}_{\bj J }  {}_{\bj K }  {}_{\bj L }(V)
~&=~    4 \, \left[~ \delta_{ {}_{\bj I }  {}_{\bj J }}\delta_{ 
{}_{\bj K }  {}_{\bj L }} ~-~  \delta_{ {}_{\bj I }  {}_{\bj K }}
\delta_{ {}_{\bj J }  {}_{\bj L }} ~+~ \delta_{ {}_{\bj I }  {}_{
\bj L }}\delta_{ {}_{\bj J }  {}_{\bj K }} ~-~
\epsilon_{ {}_{\bj I }  {}_{\bj J }  {}_{\bj K }  {}_{\bj L }}  
~ \right]  
~~~~~.
}
\label{Chr0M2}
\ee
One of the striking features of these results is their correlation with an issue
about the construction of 4D, $\cal N$ = 2 supermultiplets from 4D, $\cal N$ = 1 
supermultiplets.
In particular, the pattern of the signs of the coefficients multiplying the $\e$-tensors 
is quite revealing.  The off-shell chiral multiplet sign ($ \chi{}_{{}_0} (I)$ = +1) is 
opposite to that of the off-shell tensor multiplet ($ \chi{}_{{}_0} (III)$ = -1) and off-shell vector 
multiplet ($ \chi{}_{{}_0} (V)$ = -1) signs.

An off-shell 4D, $\cal N$ = 1 chiral multiplet may be combined with an off-shell  4D, 
$N$ = 1 tensor multiplet to form an off-shell  4D, $\cal N$ = 2 tensor multiplet.
An off-shell 4D, $\cal N$ = 1 chiral multiplet may be combined with an off-shell  4D, 
$N$ = 1 vector multiplet to form an off-shell  4D, $\cal N$ = 2 vector multiplet.
However, an off-shell  4D, $\cal N$ = 1 tensor multiplet when combined with a
4D, $\cal N$ = 1 vector multiplet  forms the so-called `Vector-Tensor' Multiplet
\cite{VT}. The Vector-Tensor Multiplet is {\em {not}} an off-shell 4D, $\cal N$ = 2 
representation.   The statement above may be confusing to some of our readers.  So 
let us make clear what we are saying.

The work of \cite{GRana} implies something that seems to have escaped the
general notice of the community familiar with this class of problems.  These 
works in the middle nineties showed that for all values of $\cal N$, but only in
1D, it is possible to find supermultiplets that have the properties of:
\newline \indent
$~~$ (a.)  no off-shell central charges, 
\newline \indent
$~~$ (b.) no use of equations of motion, and 
\newline \indent
$~~$ (c.) no infinite sets of auxiliary fields.  
\newline 
In other words, the off-shell problem is solved in 1D.  Since all the work of the
related to the Adinkra/Garden Algebra investigations rests on these fundamental
observations, all these studies are within the assumptions (a.), (b.) and (c.)
immediately above.  Within these restrictions the statement above about the
Vector-Tensor Multiplet is correct.

We suspect that this failure on the part of the Vector-Tensor Multiplet
to form an off-shell 4D, $\cal N$ = 2 representation is related to the values of
$ \chi{}_{{}_0}$ for the two 4D, $\cal N$ =  1 supermultiplets.

We are led to make some conjectures:

 $~~~$ {\it {For all off-shell 4D, $\cal N$ = 1 multiplets, the}} $p$ = 1 {\it {chromocharacters
 take}}
\newline \indent
$~~~$ {\it {the form}} 
\be  \eqalign{
\varphi^{(1)} {}_{\bj I }  {}_{\bj J } 
~&=~    {\rm d}\,  \delta_{ {}_{\bj I }  {}_{\bj J }} ~~~, \cr
}
\label{Chr0M3a}
\ee
where 2d is the number of bosonic plus fermionic degrees of freedom minus gauge degrees
of freedom.
 
 $~~~$ {\it {For all off-shell 4D, $\cal N$ = 1 multiplets, the}} $p$ = 2 {\it {chromocharacters
 take}}
\newline \indent
$~~~$ {\it {the form}} 
\be  \eqalign{
\varphi^{(2)} {}_{\bj I }  {}_{\bj J }  {}_{\bj K }  {}_{\bj L } 
~&=~    {\rm d}\, \left[~ \delta_{ {}_{\bj I }  {}_{\bj J }}\delta_{ 
{}_{\bj K }  {}_{\bj L }} ~-~  \delta_{ {}_{\bj I }  {}_{\bj K }}
\delta_{ {}_{\bj J }  {}_{\bj L }} ~+~ \delta_{ {}_{\bj I }  {}_{
\bj L }}\delta_{ {}_{\bj J }  {}_{\bj K }} ~ \right] ~+~  \chi{}_{{}_0} \, 
\epsilon_{ {}_{\bj I }  {}_{\bj J }  {}_{\bj K }  {}_{\bj L }}  
 ~~~, \cr
}
\label{Chr0M3}
\ee
where $ \chi{}_{{}_0}$ is a {\em {true}} character for classifying the representations of 
4D, $\cal N$ = 1 supersymmetry.  It is interesting to also note that this character distinguishes
between the 2D, $\cal N$ = 2 chiral multiplet versus the twisted chiral multiplet.

Since in the cases of the on-shell chiral multiplet and the on-shell vector multiplet
the second line of (\ref{GarDNAlg1}) is not satisfied, and also since for the case of the double
tensor multiplet neither equation of (\ref{GarDNAlg1}) is satisfied, we relegate the
calculations of the replacements of these equations to Appendix B.  Using the results
from this appendix we find
\be \eqalign{ {~~~}
\varphi^{(2)}  {}_{\bj I }  {}_{\bj J }  {}_{\bj K }  {}_{\bj L }(II)
~&=~  2\, \d_{\rI\rJ} \,   \d_{\rK\rL}    
+~ 2 \, [\, \s^1 \otimes \s^2 \, ]_{\rI\rJ}\,  [\, \s^1 \otimes \s^2 \, ]_{\rK \rL}   ~~~, \cr
\varphi^{(2)}  {}_{\bj I }  {}_{\bj J }  {}_{\bj K }  {}_{\bj L }(IV)
~&=~  6 \, \d_{\rI\rJ} \, \d_{\rK\rL}  ~+~ 4 \, [\, \s^3 \otimes \s^3 \, ]_{\rI\rJ}\, 
[\, \s^3 \otimes \s^3 \, ]_{\rK \rL} ~+~  \cr
&  ~~~~+~  6 \, [\, \s^1 \otimes \s^2 \, ]_{\rI\rJ}\,  [\, \s^1 \otimes \s^2 \, ]_{\rK \rL} 
 ~+~ 4 \, [\, \s^2 \otimes \s^1 \, ]_{\rI\rJ}\,  [\, \s^2 \otimes \s^1 \, ]_{\rK \rL}
\cr
&  ~~~~+~  4 \, [\, \s^3 \otimes \s^1 \, ]_{\rI\rJ}\,  [\, \s^3 \otimes \s^1 \, ]_{\rK \rL} 
 ~+~ 4 \, [\, {\bf I} \otimes \s^2 \, ]_{\rI\rJ}\,  [\, {\bf I} \otimes \s^2 \, ]_{\rK \rL}
\cr
&  ~~~~+~  4 \, [\, \s^1 \otimes {\bf I} \, ]_{\rI\rJ}\,  [\, \s^1 \otimes {\bf I} \, ]_{\rK \rL} 
 ~+~ 4 \, [\, \s^2 \otimes \s^3 \, ]_{\rI\rJ}\,  [\, \s^2 \otimes \s^3 \, ]_{\rK \rL}
\cr
\varphi^{(2)}  {}_{\bj I }  {}_{\bj J }  {}_{\bj K }  {}_{\bj L }(VI)
~&=~    2\, \d_{\rI\rJ} \,   \d_{\rK\rL} ~+~ 2 \, [\, {\bf I} \otimes \s^2 \, ]_{\rI\rJ}\, 
[\, {\bf I} \otimes \s^2 \, ]_{\rK \rL} ~+~ 2 \, [\, \s^2 \otimes {\bf I} \, ]_{\rI\rJ}\,  
[\, \s^2 \otimes {\bf I} \, ]_{\rK \rL}   \cr
&  ~~~~+~ 2 \, [\, \s^2 \otimes \s^1 \, ]_{\rI\rJ}\,  [\, \s^2 \otimes \s^1 \, ]_{\rK \rL} 
~~~~~.
}
\label{Chr0M4}
\ee
The forms of the $p$ = 2 chromocharacters in the even cases may seem very different
from those in the odd cases.  But in fact there are similarities.  

These similarities become obvious with the use of the generators of the SO(4) 
rotation group.  The six generators of SO(4) can be denoted by $i \,[\a{}^1]_{\rI\rJ}$,  
$i\, [\a{}^2]_{\rI\rJ}$, $i\, [\a{}^3]_{\rI\rJ}$, $i \, [\b{}^1]_{\rI\rJ}$, $i\, [\b{}^2]_{\rI\rJ}$, and 
$i\, [\b{}^3]_{\rI\rJ}$ where
 \be
\begin{array}{cccc}
&[{\a}^1]_{\rI \rJ } =~ [\s^2 \otimes \s^1]_{\rI \rJ } ~~, & ~~ [{\a}^2]_{\rI \rJ } = [
{\bf I}  \otimes \s^2 ]_{\rI \rJ } ~~, &  
~~[{\a}^3]_{\rI \rJ } = [\s^2 \otimes \s^3]_{\rI \rJ } ~~, \\
&[{\b}^1]_{\rI \rJ }  =~ [\s^1 \otimes \s^2]_{\rI \rJ } ~~, & ~~ [{\b}^2]_{\rI \rJ } = [
\s^2 \otimes {\bf I}]_{\rI \rJ }  ~~, &  
~~[{\b}^3]_{\rI \rJ } =  [\s^3 \otimes \s^2]_{\rI \rJ } ~~, \\
\end{array}
\label{Chr0M$a}
\ee
and these correspond to the fact that locally SO(4) = SU(2) $\times$ SU(2).

In terms of these, the results in (\ref{Chr0M2}) take the forms
\be \eqalign{
\varphi^{(2)} {}_{\bj I }  {}_{\bj J }  {}_{\bj K }  {}_{\bj L } (I) ~&=~  4 \, 
\left[~ \delta_{ {}_{\bj I }  {}_{\bj J }}\delta_{{}_{\bj K }  {}_{\bj L }} ~+~  
[\vec {\b}]_{\rI \rJ } \cdot \,   [\vec {\b}]_{\rK \rL } ~ \right] ~~~, \cr
\varphi^{(2)}  {}_{\bj I }  {}_{\bj J }  {}_{\bj K }  {}_{\bj L }(III) ~&=~ 4 \, 
\left[~ \delta_{ {}_{\bj I }  {}_{\bj J }}\delta_{ {}_{\bj K }{}_{\bj L }} ~+~  
[\vec {\a}]_{\rI \rJ } \cdot \,  [\vec {\a}]_{\rK \rL } ~ \right]   ~~~, \cr
\varphi^{(2)}  {}_{\bj I }  {}_{\bj J }  {}_{\bj K }  {}_{\bj L }(V) ~&=~  4 \, 
\left[~ \delta_{ {}_{\bj I }  {}_{\bj J }}\delta_{{}_{\bj K }{}_{\bj L }}
~+~  [\vec {\a}]_{\rI \rJ } \cdot \,  [\vec {\a}]_{\rK \rL } ~ \right]  
~~~,
}
\label{Chr0M5}
\ee
and (\ref{Chr0M4}) becomes
\be \eqalign{ {~~~}
\varphi^{(2)}  {}_{\bj I }  {}_{\bj J }  {}_{\bj K }  {}_{\bj L }(II)
~&=~  2\, \d_{\rI\rJ} \,   \d_{\rK\rL}    
+~ 2 \, [\, \b^1  \, ]_{\rI\rJ}\,  [\, \b^1 \, ]_{\rK \rL}   ~~~, \cr
\varphi^{(2)}  {}_{\bj I }  {}_{\bj J }  {}_{\bj K }  {}_{\bj L }(IV)
~&=~  6 \, \d_{\rI\rJ} \, \d_{\rK\rL}  ~+~ 6 \, [\,  \b^1 \, ]_{\rI\rJ}\,  [\,  \b^1 
\, ]_{\rK \rL}     \cr
&  ~~~~+~ 4 \, [\, {\vec {\a}} \, \b^1 \, ]_{\rI\rJ}\,  \cdot \,  [\, {\vec {\a}} \, \b^1 \, ]_{\rK \rL}
~+~ 4 \, [\, {\vec {\a}}  \, ]_{\rI\rJ}\,  \cdot \,  [\, {\vec {\a}} \, ]_{\rK \rL}   \cr
\varphi^{(2)}  {}_{\bj I }  {}_{\bj J }  {}_{\bj K }  {}_{\bj L }(VI)
~&=~    3 \, \d_{\rI\rJ} \,   \d_{\rK\rL} ~+~ 2 \, [\,  \a^2 \, ]_{\rI\rJ}\, 
[\, \a^2 \, ]_{\rK \rL} ~+~ 2 \, [\, \b^2  \, ]_{\rI\rJ}\,  
[\, \b^2 \, ]_{\rK \rL}   \cr
&  ~~~
~+~ 2 \, [\,  \a^1 \, ]_{\rI\rJ}\,  [\,  \a^1 \, ]_{\rK \rL} 
~~~~~.
}
\label{Chr0M6}
\ee

Thus written, the $p$ = 2 chromocharacters for the off-shell theories are seen to have the
form of terms dependent on tensor products of the 4 $\times$ 4 identity matrix plus terms that 
are tensor products in other  4 $\times$ 4 matrices.  We see a nice correlation between the 
spin of the 4D fields and the $p$ = 2 chromocharacters.  The chiral supermultiplet contained 
only Lorentz scalars, and the corresponding chromocharacter depends on the $\b$-generators.  The 
vector and tensor supermultiplets contained fields that carried one or more Lorentz vector 
indices and their chromocharacters depend on the $\a$-generators.

This is the strongest evidence to date that the fourth conjecture made in \cite{ENUF} (though
modified now for our change in conventions) is correct and higher dimensional off-shell
supersymmetric models can be {\em {faithfully}} represented as 1D SUSY models.  The spin
information of the higher dimensional theory is apparently carried in the chromocharacters
associated with the 1D models.
Only the off-shell models realize SU(2) symmetries by rotating either the $\a$'s among themselves
or the $\b$'s among themselves.  
We thus make another conjecture\footnote{See appendix D for an expanded discussion.}:

 $~~~$ {\it {For all off-shell 4D, $\cal N$ = 1 multiplets,}} {\underline {\it {all}}} {\it {chromocharacters
 must}}
\newline \indent
$~~~$ {\it {possess an SU(2)}} $\times $ {\it {SU(2) symmetry.}} 
 \vskip .1pt  \noindent

There are strong {\em {purely}} algebraic distinctions that must be made between the on-shell 
and off-shell cases.

In all off-shell representations, the L-matrices and R-matrices are square.  This is a consequence
of having equal numbers of bosonic and fermionic fields in off-shell supersymmetry representations.
The L-matrices and R-matrices satisfy both conditions in (\ref{GarDNAlg1}) and that in 
(\ref{GarDNAlg2}).  Consequently in off-shell representations, the L-matrices and R-matrices 
for 1D, $\cal N$-extended SUSY models are
obtained by a projection of $Cl({\cal N} + 1)$.  In all off-shell representations each row or column 
of the L-matrices and R-matrices, when regarded as vectors, form an orthonormal basis set of
vectors.  

Among the on-shell cases, there is also a strong distinction
to be made between the $II$ and $VI$ cases (generic on-shell) and the  $IV$ case 
(``pathogenic'' on-shell).

In `generic' on-shell representations, the L-matrices and R-matrices are {\em {not}} square.  This 
is a consequence of having unequal numbers of bosonic and fermionic fields in on-shell supersymmetry 
representations.  The L-matrices and R-matrices satisfy only the first conditions in (\ref{GarDNAlg1}) 
and that in (\ref{GarDNAlg2}).  In all generic on-shell representations each row or column 
of the L-matrices and R-matrices, when regarded as vectors, have unit length.

In `pathogenic' on-shell representations, the  L-matrices and R-matrices satisfy only the conditions in 
(\ref{GarDNAlg2}) but not those in (\ref{GarDNAlg1}).  The L-matrices and R-matrices are generally
{\em {not}} square.  In some `pathogenic'  on-shell representations, each row or column 
of the L-matrices and R-matrices, when regarded as vectors, do {\em {not}} have unit length.

Though we have not discussed them here, there are special pathogenic on-shell representations.
Two of the most familiar of these are the 4D, $\cal N$ = 2 Fayet Hypermultiplet \cite{FSHypr} and
4D, $\cal N$ = 2 Vector-Tensor Multiplet.   Their L-matrices and R-matrices {\em {are}} square.
However, in these cases, the terms that are the analogs of that given in (\ref{dblten7}) have the
property of being dependent on the equations of motion of the bosonic fields in the multiplet.
In this case, these terms are called ``off-shell central charges.''  The initial paper on the 
Fayet Hypermultiplet introduced such models into the physics literature.

One of our main motivations for including the little known case of the double tensor multiplet
was to show that while in 4D theories may superficially appear very similar, after reduction
on a 0-brane sharp differences can be seen.  It is only in the pathogenic case that the 
chromocharacters depend on the {\em {products}} of $\a$-matrices times $\b$-matrices.
This is also a distinction to keep in mind when applying Poincar\' e duality arguments to 
supersymmetrical theories.  Case $VI$ only differs from case $III$ by the application of a 
Poincar\' e duality of one of the spin-0 fields.  

In the next chapter, we are going to discuss a graphical representation of the results
of the current chapter.  This discussion will include all the multiplets seen so far.  It
should be kept in mind that the double-tensor  multiplet has many peculiarities and as no 
off-shell  formulation is known these may not follow the same relations as appear for the 
other on-shell representations.  So many of the comments made about the on-shell multiplets
do no apply to the double-tensor multiplet.   This should be recalled as the reader goes
through the subsequent discussion. 

Let us close this section by noting that for the off-shell multiplets, which possess gauge
symmetries, the method of 0-brane reduction used has a preferred basis of working in the
Coulomb gauge $A_0$ = $B_{0 \, 1}$ = $B_{0 \, 2}$ = $B_{0 \, 3}$ = 0 and this is likely a 
general feature of this technique.

\newpage
\section{Adinkras From Garden Algebra Matrices}

$~~~$ So we have seen from the brief survey of some well--known (and one not well-
known) multiplets how the reduction of a supermultiplet on a 0-brane leads to an
algebraic association between a given supermultiplet and a set of L-matrices and
R-matrices.  This was one of the basic observations of \cite{ENUF}.  However, {\em
{Adinkras}} \cite{FauxG} provide a graphical (and vivid) tool that is often convenient
as a replacement for the Garden Algebra matrices.  We refer the reader to these 
previous works for detailed explanation of how Adinkras are obtained from reduction
on a 0-brane.

We now present the Adinkras for each of the cases $I$ - $VI$.

$$
\vCent
 {\setlength{\unitlength}{1mm}
  \begin{picture}(-20,-140)
   \put(-73,-45){\includegraphics[width=5.6in]{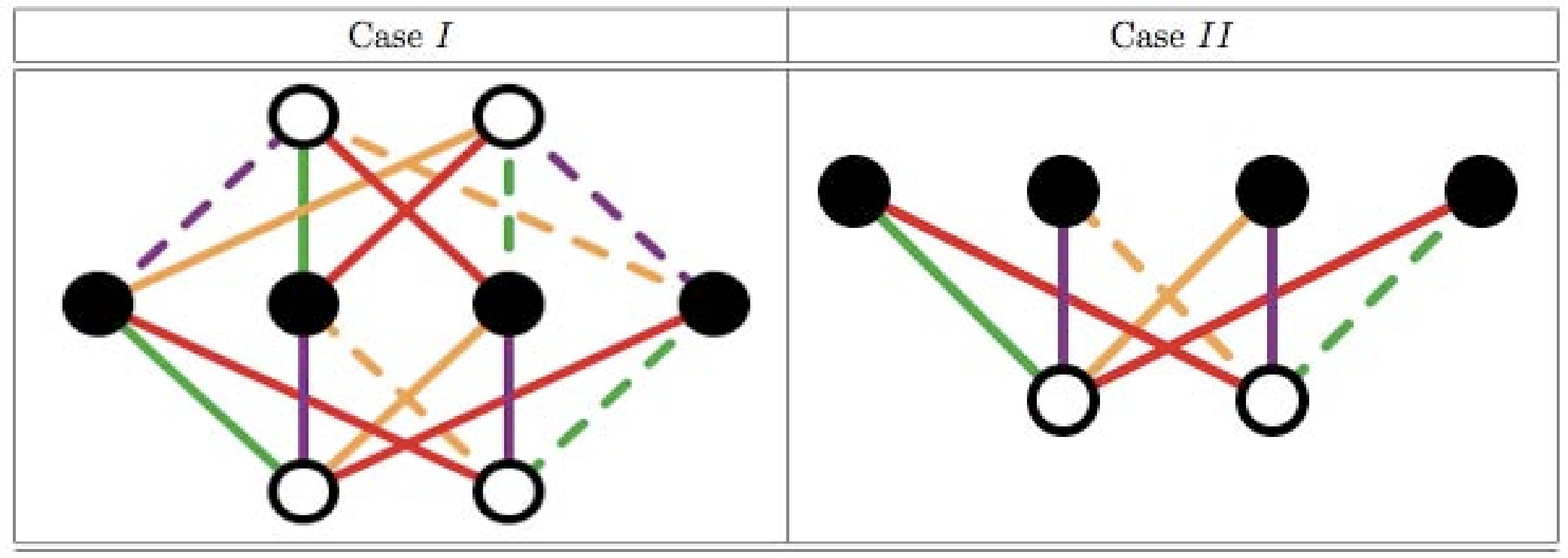}}
  \end{picture}}.
  $$ \newline
  $$
\vCent
 {\setlength{\unitlength}{1mm}
  \begin{picture}(-20,-140)
   \put(-73,-73){\includegraphics[width=5.6in]{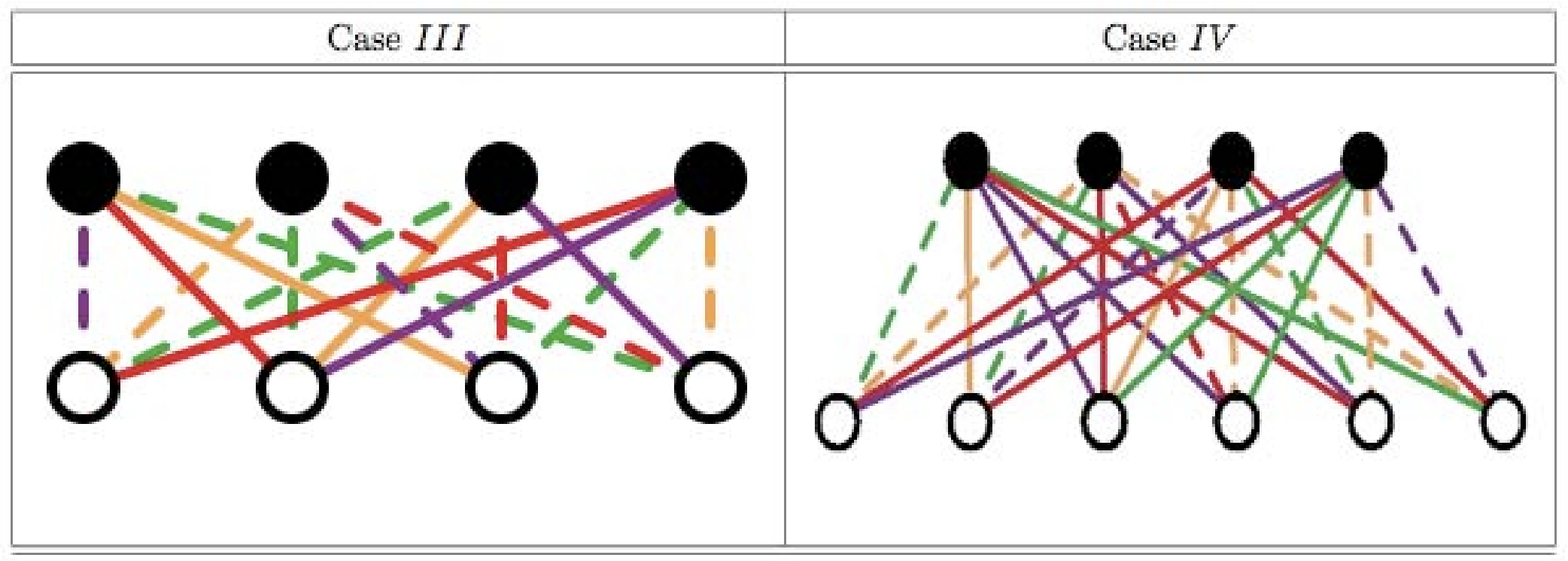}}
  \end{picture}}.
  $$ \newline
    $$
\vCent
 {\setlength{\unitlength}{1mm}
  \begin{picture}(-20,-140)
   \put(-73.2,-101){\includegraphics[width=5.6in]{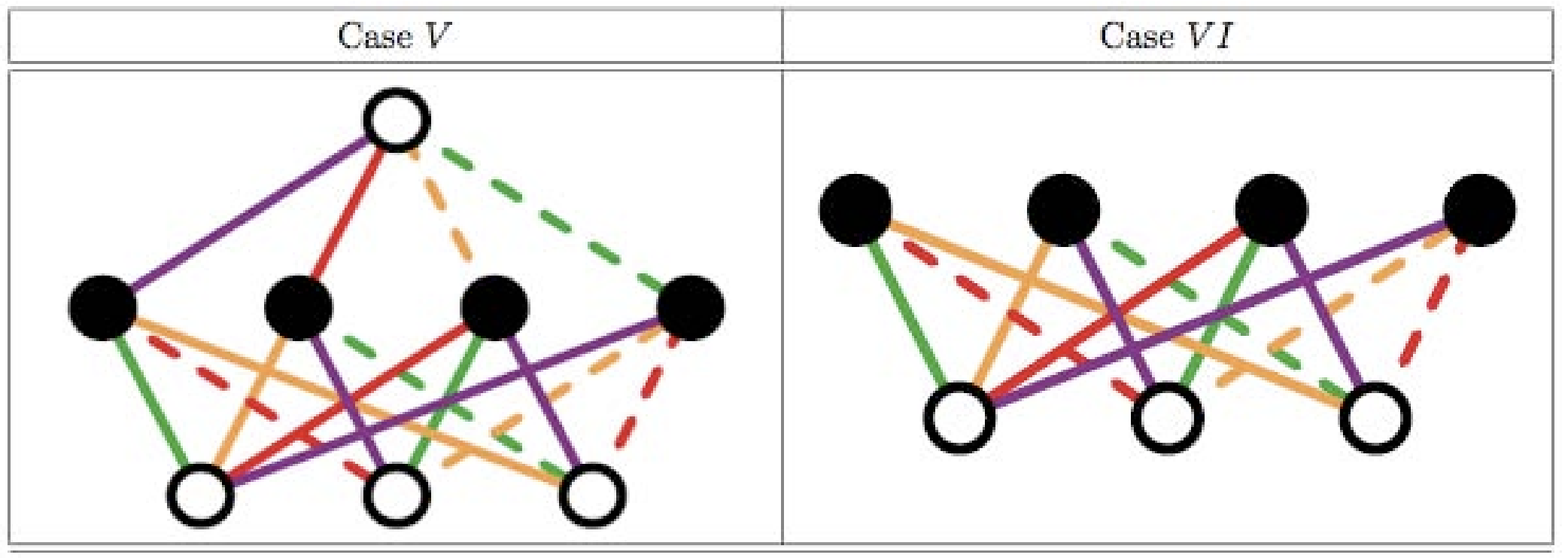}}
  \end{picture}}.
  $$ \newline
  \vskip3in
\begin{equation}
\begin{array}{c}
{}~\\
\end{array}
\label{Adnk1}
\end{equation}
It is easily seen that  the right hand column entries all contain height-two Adinkras. 
\newpage \noindent This is the 
case for {\em {all}} on-shell theories, as all such theories are valises.  The condition of going on-shell
corresponds to the erasure of all nodes and links above the second level.  The problem of
classifying all valise representations is solved and leads to the spectrum of on-shell
supersymmetrical theories, a well developed topic in the physics literature.

Going beyond on-shell theories and valises requires Adinkras of greater height, as these
describe off-shell representations.  For a fixed value of  $\cal N$,  the maximum height of an 
Adinkra that realizes $\cal N$-extended supersymmetry is given by max. height = $\cal N$ + 1.  
In the discussion of this paper\footnote{It must be understood that $\cal N$ refers to the 
world-line supersymmetries.  Thus for four \newline $~~~~~~$ dimensional theories with 
${\Tilde {\cal N}}$-extended supersymmetry,  $\cal N $ = 4 ${\Tilde {\cal N}}$.  A theory with sim- 
\newline $~~~~~~$ ple supersymmetry in four dimensions requires $\cal N$ = 4 on the world-line.}, 
max. height = 5.  Using a slight modification of the argument given in \cite{AdnkDynam}, it can 
be proven that no height-5 Adinkra can possess dynamics defined by an action quadratic in the 
fields of the Adinkra.  At height-4, there are known to be two dynamical theories.  The most familiar 
is the complex linear multiplet \cite{CLM}, which will be discussed in a work \cite{Gnom2} that is 
the companion to this paper.  Also at height-4, there is the matter gravitino multiplet \cite{Spn3/2} 
and some forms of supergravity.  The height-3 Adinkras correspond to the familiar off-shell chiral 
multiplet, the off-shell vector multiplet, and the minimal off-shell supergravity multiplet as the most 
familiar representatives.   The only known off-shell height-2 Adinkra corresponds to the tensor 
multiplet we have seen in our earlier discussion.

The discussion above also points toward future studies that need to be undertaken to find the
Adinkraic representations for all 4D, $\cal N$ = 1 off-shell multiplets.  For example,
there are many `variant' representations \cite{VaR} that are known to exist.  The case of supergravity
and matter gravitino multiplets will have more information on how higher spin manifests itself at the
layers of Adinkras.

\subsection{The Adinkra Transformation Group}

$~~~$ With Adinkras in hand, there is the possibility to give simplified discussions of some 
aspects  of $\cal {GR}$(d, $\cal N$) formulations.  One such issue that is much simplified 
is that of  changing the basis of the representation.
Adinkras may be regarded as playing a role similar to Feynman graphs and providing a tool
to replace matrix manipulations.  To illustrate this, we return to the chiral multiplet and the 
vector multiplet.  From the work of the third chapter, we have for the chiral multiplet
and for the vector multiplet Adinkras given by the following respective images.   \newpage
\be
\vCent
 {\setlength{\unitlength}{1mm}
  \begin{picture}(-20,-140)
   \put(-68,-58){\includegraphics[width=4.8in]{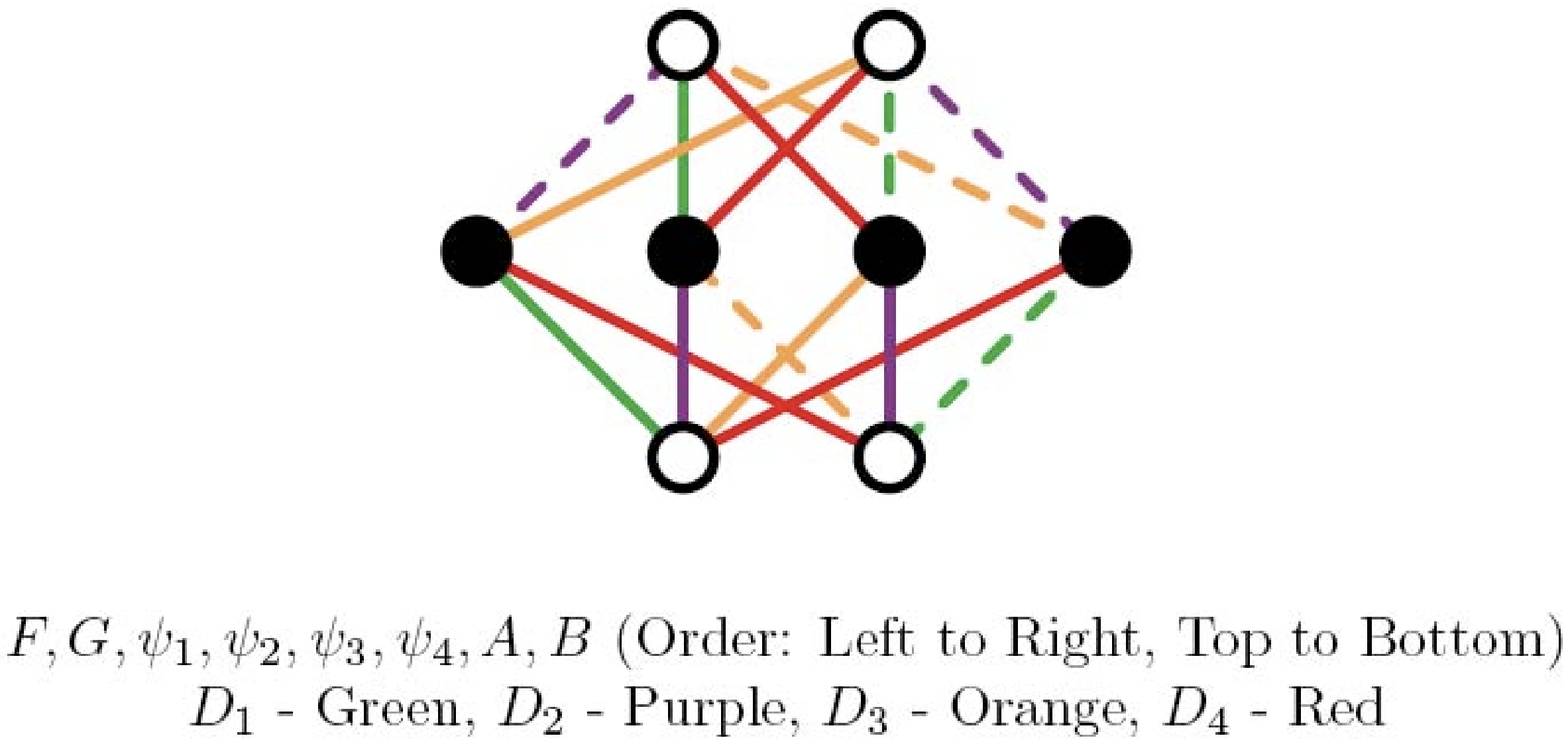}}
  \end{picture}}.
  \label{Adnkchipx}
\ee 
$~$ \vskip 1.8in \noindent

\be
\vCent
 {\setlength{\unitlength}{1mm}
  \begin{picture}(-20,-140)
   \put(-68,-51){\includegraphics[width=4.8in]{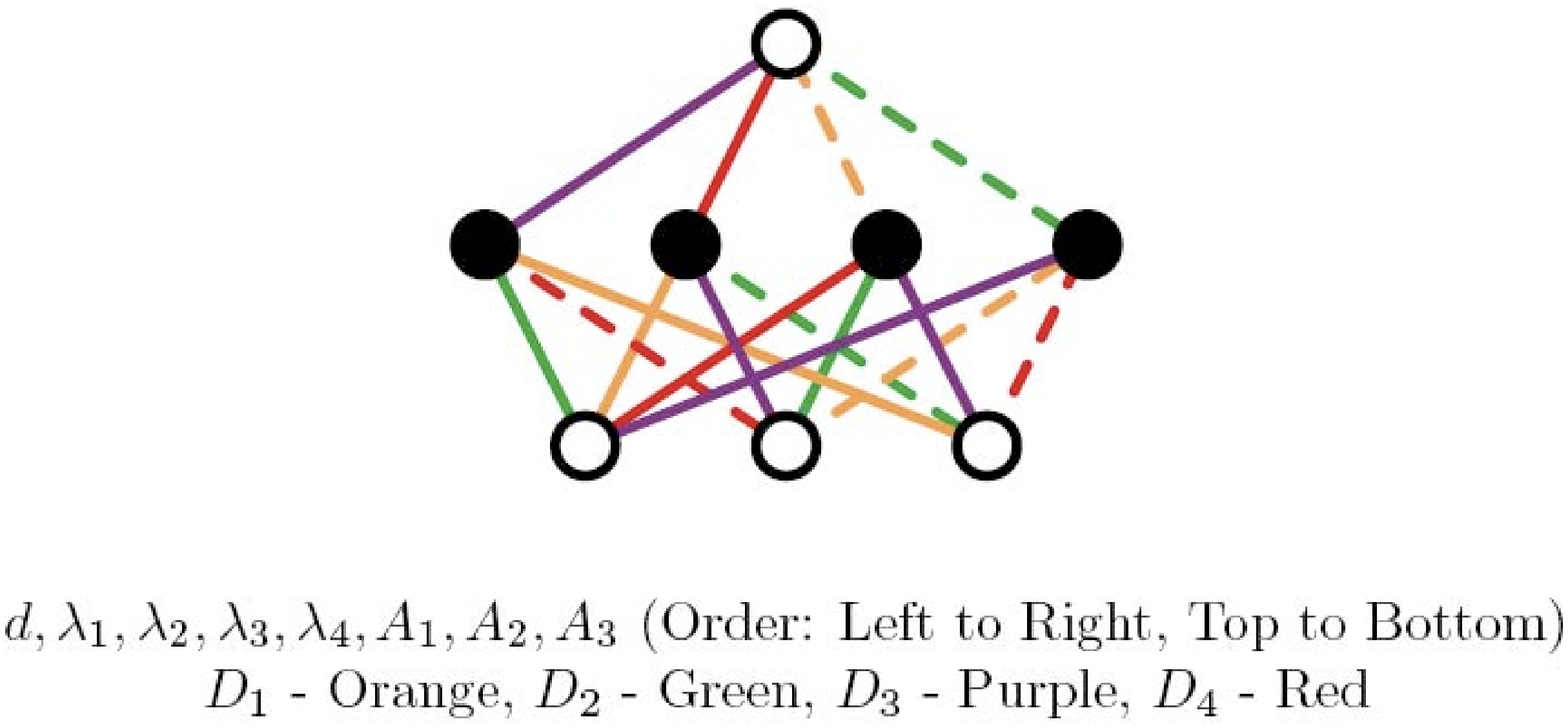}}
  \end{picture}}.
  \label{Adnkvecpx}
\ee \vskip 1.8in \noindent
Using `root superfields' \cite{ENUF}, it is possible to write algebraic expressions
for each of these.  However, we will eschew such a path and pursue a graphical route to
understand the reasons for the different values of $\chi{}_{{}_0}$ for the two multiplets.  This
will provide a graph-theoretical basis for this distinction.

For the work of the DFGHILM collaboration, a graphical piece of software (the Adinkramat - 
see acknowledgments) was developed for the investigation and  manipulation of Adinkras.  
Using this, one can `evolve' a given Adinkra into another.  The second Adinkra is related
to the first by a change of basis and other operations such as `node raising' and `node
lowering.'  Below we will use the Adinkramat to 
cast the chiral Adinkra into a valise using a maximally symmetric basis.  This
is shown in the following sequence of operations\footnote{For the benefit of the reader
following closely, there is an appendix in which the Adinkra \newline 
$~~~~~~$ manipulation and standard
column and row operations are compared side by side.}. \newpage
\be
\vCent
 {\setlength{\unitlength}{1mm}
  \begin{picture}(-20,-140)
   \put(-68,-45){\includegraphics[width=4.8in]{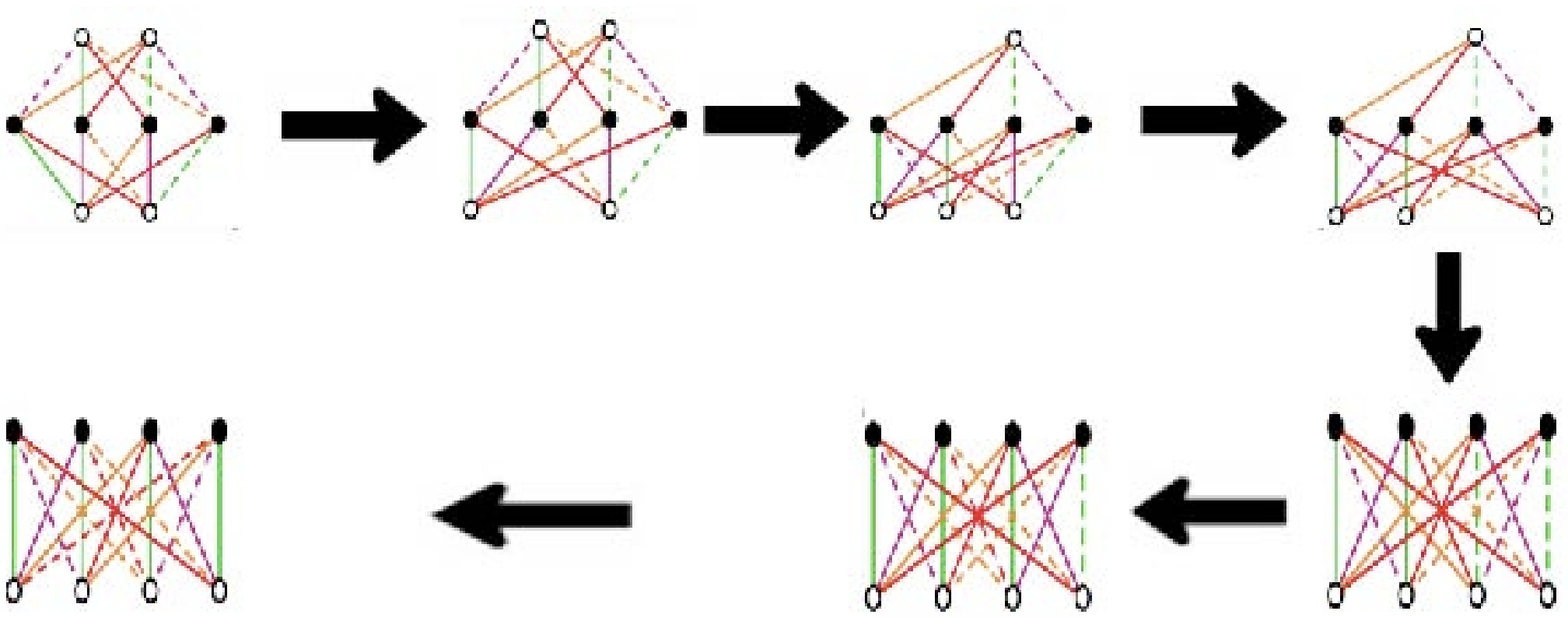}}
  \end{picture}}.
  \label{ChitoVal}
\ee \vskip 1.5in \noindent
Let us describe the sequence of operations:  \newline $~~$
(a.) In the first of these, the identity map is applied.  Horizontal transla-
\newline $~~~~~~~~$  tions of Adinkra nodes only describe the
identity map,  {\em {unless}} the
\newline $~~~~~~~~$  horizontal ordering of nodes is changed.
\newline $~~$ (b.) The second operation is a `node-lowering' one.  The exponent of
the 
\newline $~~~~~~~~$ 
$F$-node in the corresponding root superfield is increased by one 
\newline $~~~~~~~~$ 
unit.  Also an
element of O${}_B$(4) that exchanges the second and 
\newline $~~~~~~~~$ third
bosonic nodes was used.   Here, the $B$ subscript denotes the  
\newline $~~~~~~~~$ O(4) group that acts
on bosonic nodes.
\newline $~~$ (c.) The third operation is an identity map. 
\newline $~~$ (d.) The fourth operation is a `node-lowering' one.  The exponent of
\newline $~~~~~~~~$ 
the $G$-node in the corresponding root superfield is increased by 
\newline $~~~~~~~~$ 
one unit.  Also an
element of O${}_B$(4) that exchanges the third
\newline $~~~~~~~~$ and fourth
bosonic nodes was used.
\newline $~~$ (e.) The fifth operation is an element of O${}_B$(4) that changes the
sign
\newline $~~~~~~~~$
of the third and fourth bosonic nodes.
 
A similar sequence of operations may be carried out on the vector multiplet 
Adinkra using the following sequence of operations.
\be
\vCent
 {\setlength{\unitlength}{1mm}
  \begin{picture}(-20,-140)
   \put(-68,-45){\includegraphics[width=4.8in]{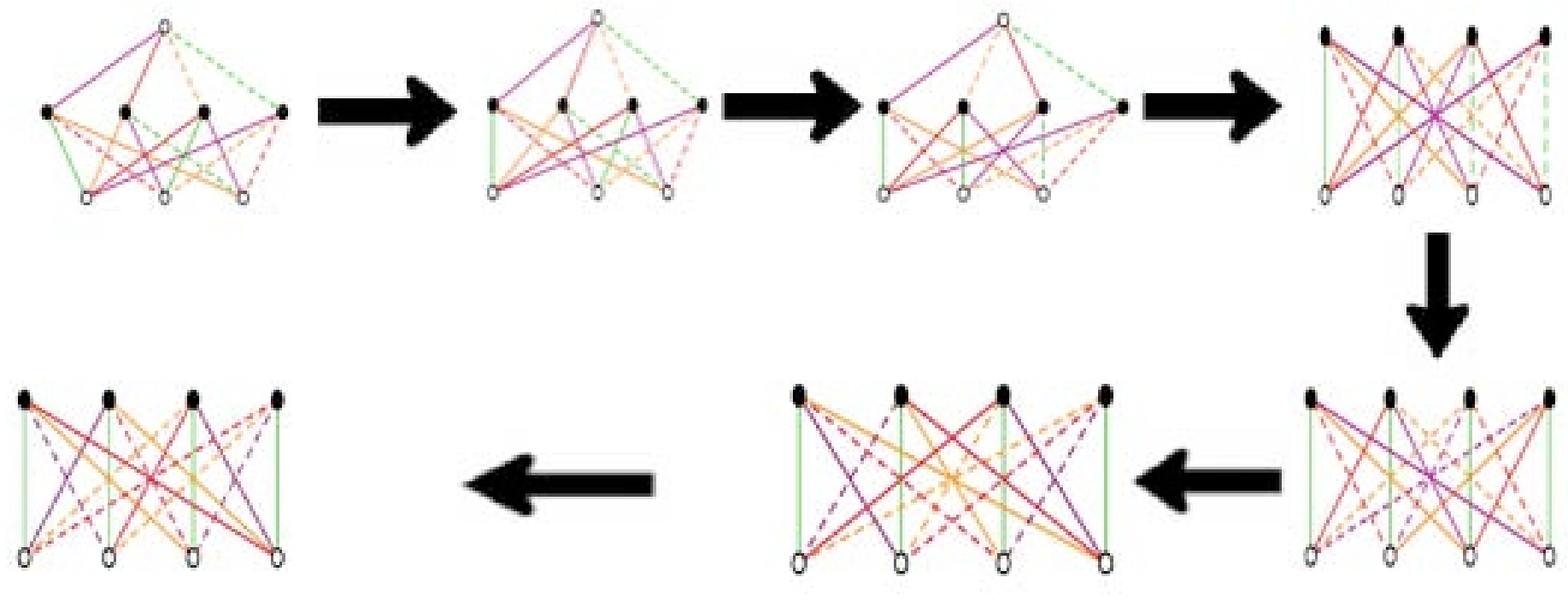}}
  \end{picture}}.
  \label{VectoVal}
\ee \vskip 1.5in \noindent
We again describe the sequence of operations:  \newline $~~$
(a.) In the first of these, the identity map is applied.  
\newline $~~$ (b.) The second operation is an element of O${}_F$(4) that exchanges the
\newline $~~~~~~~~$ the second and third fermionic nodes.   Here, the $F$ subscript 
\newline $~~~~~~~~$
denotes the O(4) that acts on fermionic nodes.
\newline $~~$ (c.) The third operation is a `node-lowering' one.  The exponent of
the 
\newline $~~~~~~~~$ 
$\rm d$-node in the corresponding root superfield is increased by one unit.
\newline $~~$ (d.) The fourth operation is an element of O${}_F$(4) that changes the
signs 
\newline $~~~~~~~~$ of the third and fourth fermionic nodes.
\newline $~~$ (e.) In the fifth operation, elements of O${}_B$(4)  and O${}_F$(4) are used to
ex-
\newline $~~~~~~~~$
change the location of the first and third bosonic nodes as well 
\newline $~~~~~~~~$
as the location of the first and third fermionic nodes.  Moreover,
\newline $~~~~~~~~$
 some signs were
changed.
\newline $~~$ (f.) The sixth operation is an element of O${}_B$(4) and  O${}_F$(4) 
 that ex-
\newline $~~~~~~~~$ changes the first and second bosonic nodes along with the 
\newline $~~~~~~~~$
first and second fermionic nodes.

Thus, under the action of the O${}_B$(4) and  O${}_F$(4) groups described in (\ref{eq:E03})
along with the node-raising and node-lowering group noted for root superfields, we find it
possible to implement the following transformations on the Chiral Multiplet and Vector Multiplet
Adinkras.
\be
\vCent
 {\setlength{\unitlength}{1mm}
  \begin{picture}(-20,-140)
   \put(-68,-90){\includegraphics[width=4.8in]{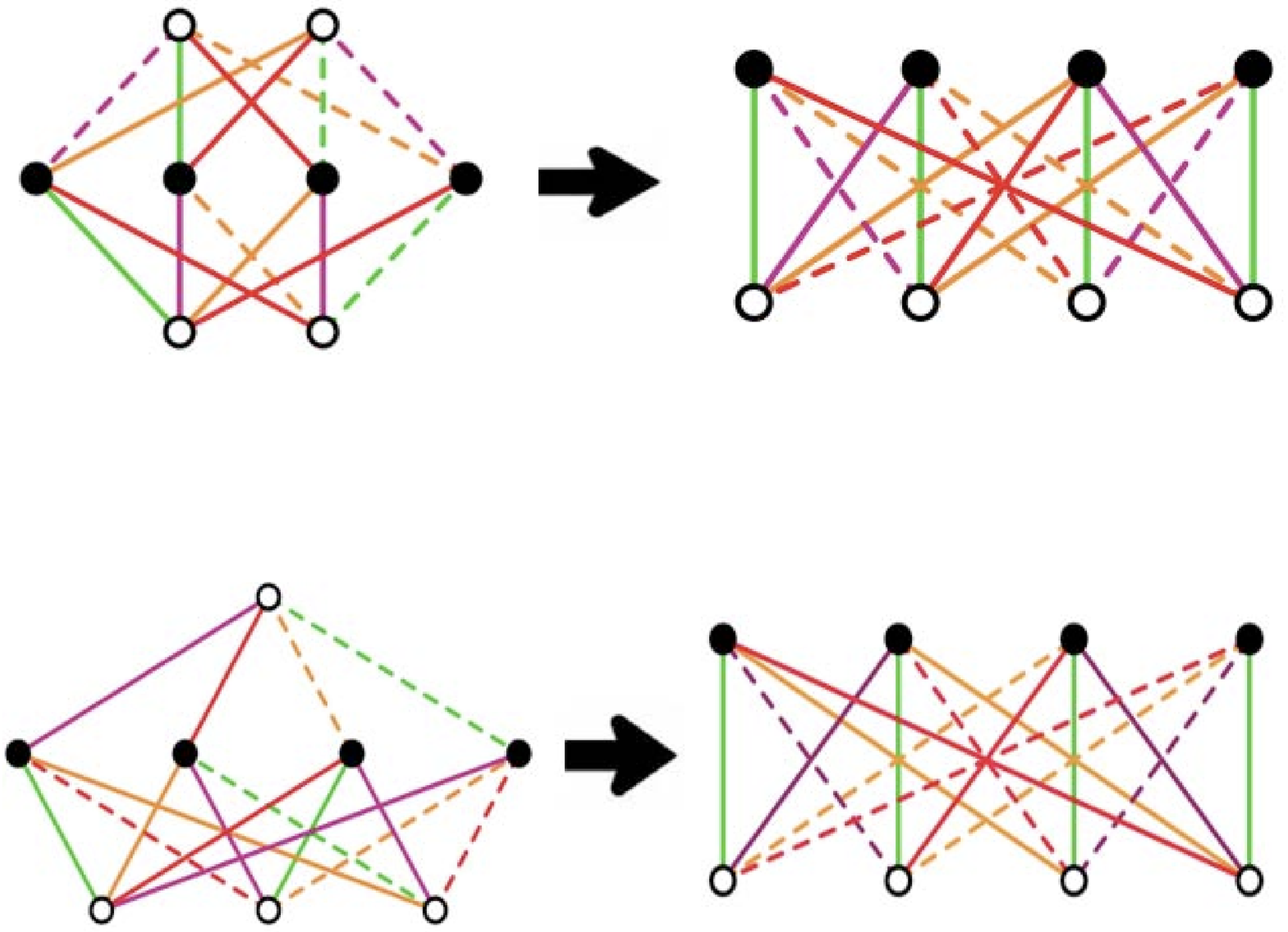}}
  \end{picture}}.
  \label{CisTrnsAdnks}
\ee \vskip 3.5in \noindent
The Adinkras on the right hand side of (\ref{CisTrnsAdnks}) can be used to `read off\footnote{See
the third appendix also.}' the L-matrices associated with each Valise.  The L-matrices associated 
with the uppermost Valise Adinkra are simply
\be  \eqalign{
 (\,{\rm L}_1\,) ~&=~ {\bf I}_4 ~,~  (\,{\rm L}_2\,) ~=~ i \, \s^3 \otimes \s^2
 ~,~  (\,{\rm L}_3\,) ~=~ i \, \s^2 \otimes {\bf I}_2  ~,~  (\,{\rm L}_4\,) 
 ~=~ -\, i \, \s^1 \otimes \s^2  ~,~~~
} \label{MxB8sis1}
\ee
and L-matrices associated with the lowermost Valise Adinkra are simply
\be  \eqalign{
 (\,{\rm L}_1\,) ~&=~ {\bf I}_4 ~,~  (\,{\rm L}_2\,) ~=~ i \, \s^3 \otimes \s^2
  ~,~  (\,{\rm L}_3\,) ~=~-\, i \, \s^2 \otimes {\bf I}_2  ~,~  (\,{\rm L}_4\,) ~=~ -\, i \, \s^1 \otimes \s^2
  ~~.
} \label{MxB8sis2}
\ee
It can be shown that the chromocharacters associated with these matrices agree with those in
(\ref{Chr0M3a}) and (\ref{Chr0M5}).  Also it can be shown there exists a sequence of Adinkra 
manipulations that take the Tensor Multiplet Adinkra into the lowermost Valise Adinkra.

At first it may seem puzzling that the two Valise Adinkras above can give different chromocharacters. 
In fact, there is a very small distinction between the two images.  It is seen that all the solid
orange lines in the first are replaced by dashed orange lines in the second (and vice-versa).
This is reflected in the differences in the signs of the L${}_3$ matrices in (\ref{MxB8sis1})
and (\ref{MxB8sis2}). All other colors and dashing match up perfectly.  It is apparent that 
$\chi_{{}_0}$ is keeping track of this property of the Adinkras!  This property of the Adinkra 
is correlated with 4D fields that carried vector indices versus those without such indices.

It may also seem puzzling that {\em {both}} sets of matrices in (\ref{MxB8sis1}) and (\ref{MxB8sis2}) 
are linearly related to {\em {only}} the $\b$-matrices in (\ref{Chr0M$a}).  This is due to a very
special element that exists among the $\cal X$ and $\cal Y$ matrices.  It can be shown
\be \eqalign{
  \D \,{\a}{}_{\bj I}  ~=~  {\b}{}_{\bj I} \,  \D   ~~~,~~~ (\D)^2 ~=~ {\bf I}_4  ~~~,
}
\label{eq:E18b}
\ee
where
\be \eqalign{    
\D ~&=~   \fracm 12 \,     \left[ \, {\bf I}_4 ~-~   {\vec {\a}} \, \cdot \, {\vec {\b}}  \, \right]   ~~.
}  \label{eq:E15b}
\ee
Thus, by choice of $\cal X$ and $\cal Y$, the $\a$'s can be `traded' for the $\b$'s and
vice-versa.

Finally, the when matrices in (\ref{MxB8sis1}) and (\ref{MxB8sis2}) are written explicitly, it
can be shown that they are far more symmetrical than the corresponding matrices in 
(\ref{chiD0F}) and (\ref{tenD0F}).  In general, were we to randomly lower the upper
nodes in the off-shell height-3 Adinkras shown in (\ref{Adnk1}), the resultant height-2
Adinkra would appear quite `cluttered' to the eye.  On the other hand, the Adinkras in
(\ref{CisTrnsAdnks}) appear quite orderly.  It is very satisfying to note that the more symmetrical
the matrices, the more symmetrical the Adinkras appear.  In fact, the basis used in the
Adinkras shown in (\ref{CisTrnsAdnks}) is a maximally symmetrical basis.  Calculations
are often simpler using such bases.

\newpage
\section{Comparisons To Known Results}

~~~~ The Adinkra which appears in the upper left of the diagram numbered as equation 
(\ref{CisTrnsAdnks}) has been named the (2, 4, 2) representation of 1D, $\cal N$ = 4 
supersymmetry (e.g. see the works of \cite{OthReps}).  In a similar manner, the Adinkra 
which appears in the lower left of the diagram numbered as equation (\ref{CisTrnsAdnks}) 
has been named the (3, 4, 1) representation of 1D, $\cal N$ = 4 supersymmetry.  Adinkras 
have the property that when `flipped' about a horizontal axis through the Adinkra there results 
in a new Adinkra that also describes a supermultiplet.  Applying this `flipping' operation 
to the (3, 4, 1) representation results in a (1, 4, 3) representations.  Although these
representations have been given these names in the works of \cite{OthReps}, these
representations have been known to one of the current authors (SJG) since the 
presentation of the formula (58) in the work of \cite{ENUF}.
 
The possibility to change the height of nodes (although not expressed using this
language) was first discovered in 1994 and shortly thereafter presented in the
literature \cite{Ultra}. Later taking advantage of this possibility, the concept of the 
``root superfield'' was introduced \cite{ENUF}.  The original definition of this 
concept\footnote{Other authors \cite{Root} while retaining the terminology of
`root multiplet', have 
changed \newline $~~~~~~$ the meaning of the term to only refer to valise 
Adinkras and associated superfields.} was an expression containing exponents 
whose values determine the height at which nodes appear in a corresponding 
Adinkra.  Our explicit reduction of the component fields of a 4D, $\cal N$ = 1 chiral 
multiplet yields a (2, 4, 2) (see also \cite{ENUF}). The reduction of the component 
fields of a 4D, $\cal N$ = 1 vector multiplet yields a (3, 4, 1).  

We do not see how the analyses in \cite{OthReps} capture a critical point.  If there 
were a unique 1D, $\cal N$ = 4 valise (a (4, 4, 0) or root in their conventions), then 
by raising one node, it could be turned into a (3, 4, 1).  Or if two nodes of a unique 
1D, $\cal N$ = 4 valise were raised, it would turn into a (2, 4, 2).  Thus, if one made 
the assumption of a unique 1D, $\cal N$ = 4 valise, then its two distinct raised-node 
relatives must be the dimensional reduction of the component fields of a 4D, $\cal 
N$ = 1 vector multiplet and the dimensional reduction of the component fields of a 
4D, $\cal N$ = 1 chiral scalar multiplet respectively.
 
Instead  what our calculations show is that the dimensional reduction of the 
component fields of a 4D, $\cal N$ = 1 chiral scalar multiplet leads to the
Adinkra on the upper left hand side of the image numbered as equation 
(\ref{CisTrnsAdnks}) in this current paper.  While the dimensional reduction 
of the component fields of a 4D, $\cal N$ = 1 vector multiplet leads to the 
Adinkra on the lower left hand side of the image numbered as equation 
(\ref{CisTrnsAdnks}) in the current paper.  

The Adinkras on the right hand side of (\ref{CisTrnsAdnks}) are distinct, there are 
no field redefinitions or rearrangements of the bosons among themselves (and
the same for the fermions) which will map one of these Adinkras into the other.   
The degeneracy of the (4,4,0) representation (and corresponding node lifts) is
difficult to see in the analyses of  \cite{OthReps}.  In fact, the distinction 
between the two valises is reflected in the distinct values found for $ \chi{}_{{}_0}$ 
and is exactly the distinction between chiral and twisted chiral multiplets known 
in 2D, $\cal N$ = 2 theories.  This result had been {\em {surmised}} in other work 
by the DFGHILM collaboration.  The calculations in this paper are the first to {\em 
{prove}} this is the case and shows the value of why explicit calculations need 
to be performed to support the many conjectures made solely by looking at the 
1D structure of these theories.
 
One other matter we will attempt to make clear for our readers what is the meaning
of root superfields, {\em {as originally defined in}} \cite{ENUF} and how are
these related to higher 4D, $\cal N$ = 1 representations.   The original meaning
of a root multiplet or root superfield is that this term refers to set of distinct ordinary
superfields that form part of a web obtained by raising and lowering nodes.  Thus
the complete root superfield associated with the upper part of the diagram in
equation (\ref{CisTrnsAdnks}) takes the form
\be
\vCent
 {\setlength{\unitlength}{1mm}
  \begin{picture}(-20,-140)
   \put(-68,-110){\includegraphics[width=4.8in]{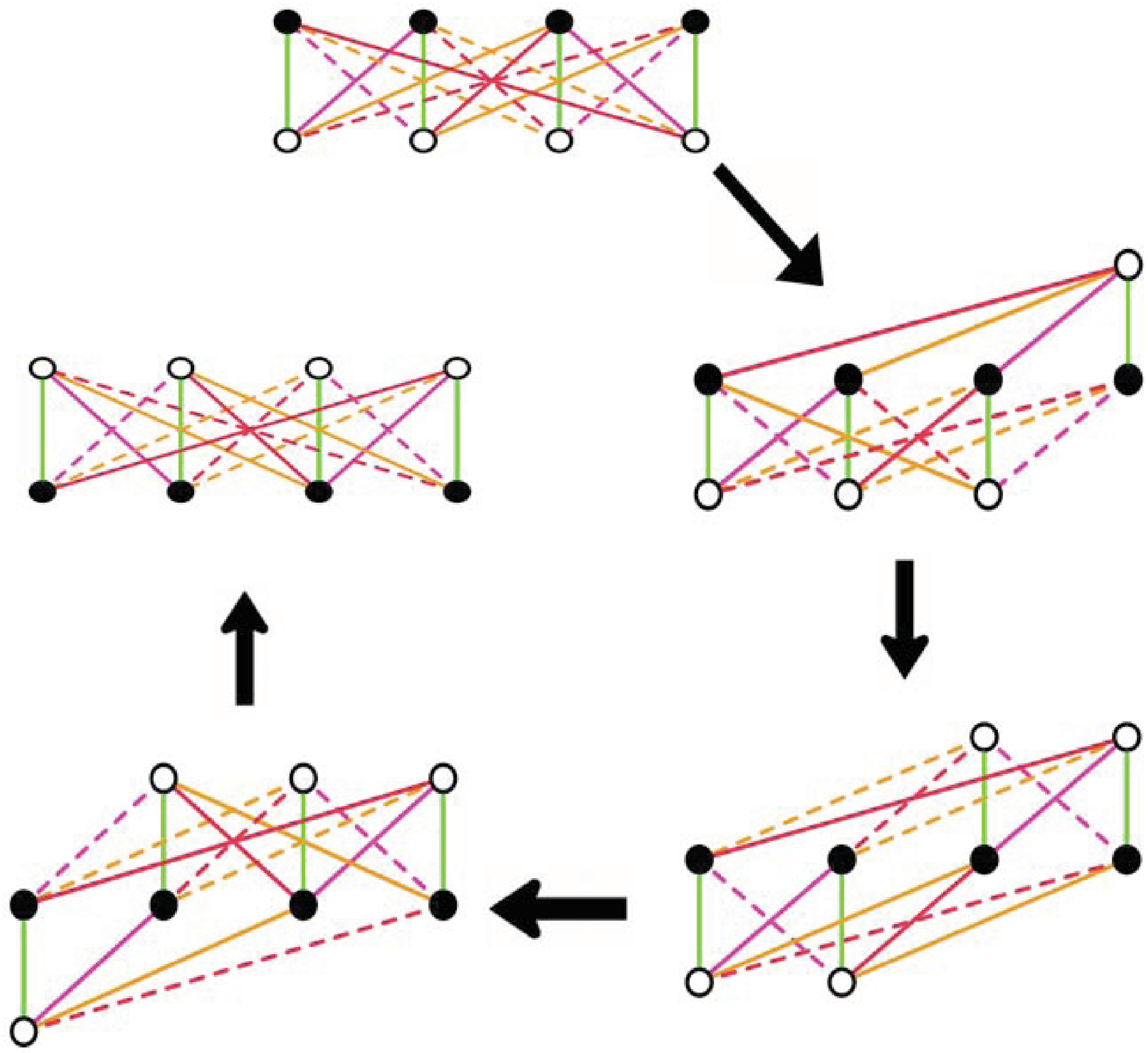}}
  \end{picture}}.
  \label{cisRootSF}
\ee \vskip 1.5in \newpage \noindent
and this diagram explicitly shows the transformations
\be
 (4, \, 4, \, 0){}_c ~~\to~~ (3, \,4, \,1){}_c ~~\to~~ (2, \,4, \,2){}_c ~~\to~~ (1,\, 4,\, 3)
 {}_c ~~\to~~ (0, \,4,  \, 4){}_c ~~~.
 \label{cisRootSFeq}
\ee
In a similar manner the complete root superfield associated with the lower part of the diagram in
equation (\ref{CisTrnsAdnks}) takes the form
\be
\vCent
 {\setlength{\unitlength}{1mm}
  \begin{picture}(-20,-140)
   \put(-68,-110){\includegraphics[width=4.8in]{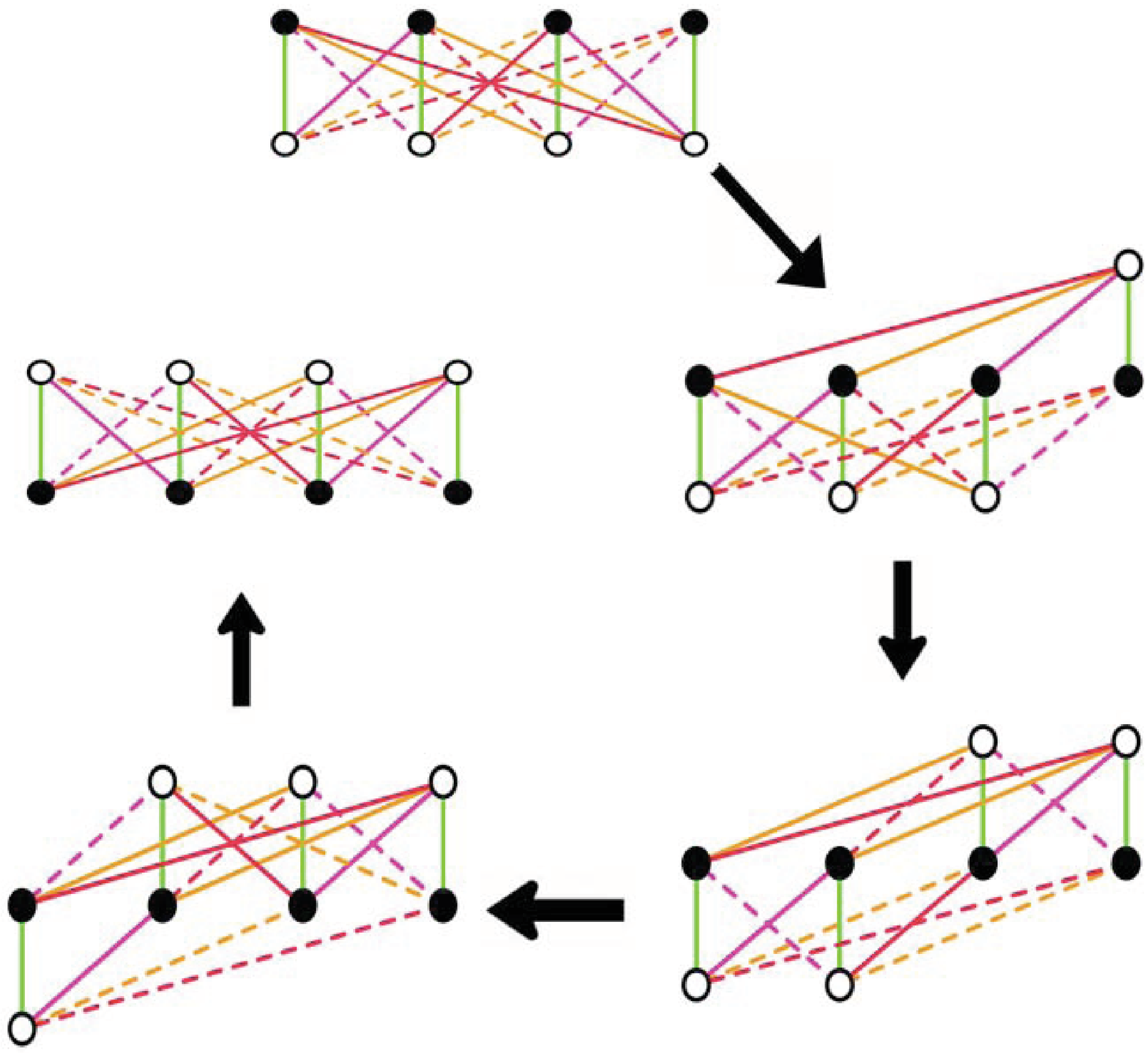}}
  \end{picture}}.
  \label{transRootSF}
\ee \vskip 4.0in \noindent
and this diagram explicitly shows the transformations
\be
 (4, \, 4, \, 0){}_t ~~\to~~ (3, \,4, \,1){}_t ~~\to~~ (2, \,4, \,2){}_t ~~\to~~ (1,\, 4,\, 3)
 {}_t ~~\to~~ (0, \,4,  \, 4){}_t ~~~.
 \label{transRootSFeq}
\ee

The works of \cite{Holog} show that there is an exclusion principle-like nature to
lifting these multiplets to 4D.  One can only `oxidize' the (2, 4, 2)${}_c$ to become a 4D,
$\cal N$ = 1 chiral scalar multiplet and one can only `oxidize' the (3, 4, 1)${}_t$ to 
become a 4D, $\cal N$ = 1 vector multiplet.  This sort of behavior is what was
anticipated in \cite{FauxG}.  Only a very limited number of representations in the 
lower dimension can be oxidized among members of a root superfield.  The only
ambiguity found is one that amounts to a re-definition of the relation of which of
two right hand Adinkras in (\ref{CisTrnsAdnks}) is chosen as a starting point.
 
\newpage
\section{Conclusion}

~~~~ In this present work, there has been presented a survey of features that occur
in the study of embedding 4D, $\cal N$ = 1 supersymmetrical systems into the context
of Adinkras and Garden Algebras.  We have explicitly demonstrated that off-shell 
supersymmetrical multiplets lead, upon reduction on 0-branes,  to a universal algebraic
structure described by (\ref{GarDNAlg1}) and (\ref{GarDNAlg2}) that we refer to as 
defining a mathematical structure denoted by $\cal {GR}$(d, $\cal N$).  On the
other hand, we have shown that on-shell theories typically lead to an algebraic
characterization in terms of  $\cal {GR}$(d${}_L$, d${}_R$, $\cal N$).

The structures we have discussed allow for a completely algebraic characterization of  
``The Fundamental Supersymmetry Challenge'' (see final work in \cite{GRana}).  The 0-brane 
reduction of all supersymmetrical theories (including all ten and eleven dimensional ones) 
is conceptually no different from the exercises undertaken in the third chapter for the on-shell 
chiral multiplet (equations (\ref{chi4}) - (\ref{chi6}) \& (\ref{chiD0M}) - (\ref{chiD0O})) and 
vector multiplets (equations (\ref{V4}) - (\ref{V6}) \& (\ref{V1D0G}) - (\ref{V1D0H})).  Thus, 
ten and eleven dimensional on-shell supersymmetrical multiplets possess derivable $\cal 
{GR}$(d${}_L$, d${}_R$, $\cal N$) representations in terms of L-matrices and R-matrices 
similar to those in (\ref{chiD0N}), (\ref{chiD0O}), and (\ref{V1D0H}).  In the case of the 
on-shell chiral and vector multiplets, their L-matrices and R-matrices ((\ref{chiD0N}), 
(\ref{chiD0O}), and (\ref{V1D0H})) can be embedded into the L-matrices and R-matrices 
((\ref{chiD0F}), (\ref{chiD0K}), and (\ref{V1D0D})) of the off-shell chiral and vector multiplets.

We can thus state the first part of the fundamental supersymmetry challenge solely 
as a algebraic problem: `When can a given representation of $\cal {GR}$(d${}_L$, 
d${}_R$, $\cal N$) be embedded into $\cal {GR}$(d, $\cal N$)?  The answer to this
question may hold a key to obtaining some interesting results.

With regard to $\cal {GR}$(d, $\cal N$) versus $\cal {GR}$(d${}_L$, d${}_R$, $\cal N$),
we have been able to advance the state-of-the-art understanding.  From the part of our survey comparing
off-shell versus on-shell multiplets, we have found that when viewed from the perspective
of one dimension, the main difference between them lies in regard to a chiral SU(2) $\times$
SU(2) theory.  Off-shell theories possess full invariance with regard to all of the 
chiral SU(2) $\times$ SU(2) group, while on-shell theories possess symmetry only
with respect to a broken sub-group.

The two Adinkras in (\ref{CisTrnsAdnks}) show a remarkable resemblance to the cis-trans 
isomerism well known in chemistry.  Specifically in fact, we can refer to the uppermost Adinkra 
as the 4D, $\cal N$ = 1 cis-Valise\footnote{In recognition that the sign of the $\e$-term is
the same as the first term in the expression \newline $~~~~~~$
 this is appropriate.}  and the second Adinkra as the 4D, $\cal N$ = 1 trans-Valise.  
We believe these are to 4D, $\cal N$ = 1 representation theory as quarks 
and anti-quarks are to SU(3).  

However, we know from the current understanding of the work of the DFGHILM collaboration,
that the analogs of higher $\cal N$ studies show an incredible proliferation
of representations that valise Adinkras produce.  This rich spectrum of representations
is more reminiscent of biology and genetics instead of the representation theory
normally seen in physics.  Because of this, we have been influenced in our studies by genomics
in particular.  From this vantage point, it would perhaps be appropriate to refer to the cis-Valise 
and trans-Valise as `genes.'
 
This leads us to a final  conjectures:

 $~~~$ {\it {The cis-Valise and trans-Valise  are the fundamental 4D, $\cal N$ = 1 genes}} 
\newline \indent
$~~~$ {\it {from which all off-shell 4D, $\cal N$ = 1 supersymmetry representations can}} 
\newline \indent
$~~~$ {\it {be derived.}}  
\vskip .005in
 \noindent
Should this conjecture be true, it implies that for the genetic classification of all 4D, $\cal N$ = 1 
supersymmetry representations, at least the two integers $n_t$ and $n_c$ (which give the 
number of trans-Valises and cis-Valises contained in a general representation) are required.

In a number of presentations by one of the authors (SJG), the expression, `the DNA of Reality,'
has been used.  Our current work provides the most detailed explanation to date for why this 
may be more than merely metaphorical.

 \vspace{.05in}
 \begin{center}
 \parbox{4in}{{\it ``My methods are really methods of working and thinking; $\,~~$
 this is why they have crept in everywhere anonymous-\\ $\,~~$
 ly.''}\,\,-\,\, Emmy Noether}
 \end{center}

 \noindent
{\bf Added Note In Proof}\\[.1in] \indent
 After the conclusion of this work, two papers have appeared on the arXiv which
 provide some specific examples of how Adinkras via the Garden Algebras provide
 a 1D holographic description of 4D, $\cal N$ = 1 supermultiplets.  These works
 can be found in papers cited as \cite{Holog}.

 \noindent
{\bf Acknowledgements}\\[.1in] \indent
This work was partially supported by the National Science Foundation grants 
PHY-0652983 and PHY-0354401. This research was also supported in part by the 
endowment of the John S.~Toll Professorship and the University of Maryland Center for 
String \& Particle Theory.  We thank
M. Kulikova for collaboration in the early stages of this work.  Additional acknowledgment 
is given by the students for the hospitality  of the University of Maryland and in particular 
of the Center  for String and Particle Theory, as well as recognition for their participation 
in 2008 SSTPRS (Student Summer Theoretical Physics Research Session).  Adinkras 
were drawn with the aid of the {\em  Adinkramat\/}~\copyright\,2008 by G.~Landweber.  
S.J.G. also wishes to acknowledge the many discussions within the DFGHILM collaboration 
that shaped his conception of this work,  and he appreciates their consideration in allowing it
to be undertaken.

\newpage
\noindent
{\Large\bf Appendix A: Conventions for Gamma Matrices}

Our conventions for the four dimensional discussion are such that we use real
four component spinors (when their indices are in an up position).  Our choice of
Minkowski metric is the `mostly plus metric.'

We use the outer product to write our 4 x 4 matrices in terms of 2 x 2 matrices.
If $M$ and $N$ are two such matrices where
$$  M ~=~ \left(\begin{array}{cc}
~m_{11} & ~m_{12}\\
~m_{21} & ~m_{22} \\
\end{array}\right) ~~,~~~
N ~=~ \left(\begin{array}{cc}
~n_{11} & ~n_{12}\\
~n_{21} & ~n_{22} \\
\end{array}\right)
\eqno(A.1) 
$$
then we choose our conventions so that
$$  \eqalign{  {~~~~~~}
M \otimes N ~=&~ \left(\begin{array}{cc}
~m_{11}  \left(\begin{array}{cc}
~n_{11} & ~n_{12}\\
~n_{21} & ~n_{22} \\
\end{array}\right) & ~m_{12}  \left(\begin{array}{cc}
~n_{11} & ~n_{12}\\
~n_{21} & ~n_{22} \\
\end{array}\right)\\
~m_{21}  \left(\begin{array}{cc}
~n_{11} & ~n_{12}\\
~n_{21} & ~n_{22} \\
\end{array}\right) & ~ 
m_{22}  \left(\begin{array}{cc}
~n_{11} & ~n_{12}\\
~n_{21} & ~n_{22} \\
\end{array}\right) \\
\end{array}\right)  \cr
~=&~ \left(\begin{array}{cc}
~ \begin{array}{cc}
~ m_{11} n_{11} & ~ m_{11}  n_{12}\\
~ m_{11} n_{21} & ~ m_{11} n_{22} \\
\end{array}\ & ~  \begin{array}{cc}
~ m_{12} n_{11} & ~ m_{12} n_{12}\\
~ m_{12} n_{21} & ~ m_{12} n_{22} \\
\end{array} \\
~ \begin{array}{cc}
~ m_{21}  n_{11} & ~ m_{21}  n_{12}\\
~ m_{21}  n_{21} & ~ m_{21} n_{22} \\
\end{array} & ~
\begin{array}{cc}
~ m_{22} n_{11} & ~ m_{22} n_{12}\\
~ m_{22} n_{21} & ~ m_{22} n_{22} \\
\end{array} \\
\end{array}\right)  ~~~.
} \eqno(A.2)   $$

In this notation, the four dimensional gamma matrices we use are defined by\\
$$ \eqalign{
&~~~~~~~~~~~~~~~~~~~ {(\gamma^0)}{}_a{}^b  = i ( \sigma^3
 \otimes \sigma^2  ){}_a{}^b 
~~~~,~~~~~~ {(\gamma^1)}{}_a{}^b  = ({\bf I}_2 
\otimes \sigma^1 ){}_a{}^b ~~~~~, \cr
&~~~~~~~~~~~~~~~~~~~ {(\gamma^2)}{}_a{}^b  = (\sigma^2 
\otimes \sigma^2 ){}_a{}^b ~~~~~,~~~~~~ 
  {(\gamma^3)}{}_a{}^b  = ({\bf I}_2 
\otimes \sigma^3  ){}_a{}^b  ~~~~~.
} 
\eqno(A.3) $$
which can all be seen to be purely imaginary.  The corresponding gamma-5 matrix
is given by\\
$$~~~~~~~
 {(\gamma^5)}{}_a{}^b  = -(\sigma^1 \otimes \sigma^2 ){}_a{}^b ~~~~~. 
 \eqno(A.4)$$
 
 Some useful Identities then follow
$$ \eqalign{
&\gamma^\mu \, \gamma^\nu ~+~  \gamma^\nu \, \gamma^\mu ~=~
2 \, \eta{}^{\mu \, \nu} \, {\bf I}_4  ~~,~~ \gamma^\mu \, \gamma_\mu ~=~
4 \, {\bf I}_4   ~~,~~   \gamma^\mu \, \gamma_\alpha \, \gamma_\mu ~=~- \, 2 \,
\gamma_\alpha ~~~,
\cr
 {~~~~~~~~}
&\gamma^5 \,[ \, \gamma^\a ~,~ \gamma^\b \, ] ~=~ - i \fracm 12 \, \e^{\a \, \b
\, \m \, \n } \, [\,  \gamma_\m ~,~ \gamma_\n \, ]  ~~~,~~~
 \gamma^\mu \,  [ \, \gamma^\a ~,~ \gamma^\b \, ]   \, \gamma_\mu ~=~ 0
~~~, \cr
{~~~~~~~~}
&\gamma^\m \,[ \, \gamma^\a ~,~ \gamma^\b \, ] ~=~  2 \, [~  \eta^{\m \, \a}
 \, \gamma^\b ~-~    \eta^{\m \, \b} \, \gamma^\a ~] ~+~ i \, 2\,  \e^{\a \, \b \m \, \n}
 \g^5 \g_{\n}   ~~~,
\cr
 {~~~~~~~~}
& [ \, \gamma^\a ~,~ \gamma^\b \, ] \, \gamma^\m~=~ - \, 2 \, [~  \eta^{\m \, \a}
 \, \gamma^\b ~-~    \eta^{\m \, \b} \, \gamma^\a ~] ~+~ i \, 2\,  \e^{\a \, \b \m \, \n}
 \g^5 \g_{\n}   ~~~. }
\eqno(A.5) $$

In order to raise and lower spinor indices, we define a spinor metric by \\ 
$$~~~~~~~~~~  C_{ab} \equiv -i (\sigma^3 \otimes \sigma^2)_{a b} 
~=~ \left(\begin{array}{cccc} 0 & -1 & 0 & 0\\ 1 & 0 & 0 & 0\\ 0 & 0 & 0 & 1\\ 
0 & 0 & -1 & 0\end{array}\right)~~~\to ~~ C_{ab} = -C_{ba} ~~~.
\eqno(A.6)$$\\
The inverse spinor metric is defined by the condition $C^{ab}C_{ac} = \delta{}_c{}^b$.

The second rank anti-symmetric matrix is defined by
$$ \eqalign{ 
{(\sigma^{\mu\nu})}{}_a{}^b  \,&\equiv\, \frac{i}{2} [ {(\gamma^\mu)}{}_a{}^c  {(\gamma^\nu)}
{}_c{}^b  ~-~   {(\gamma^\nu)}{}_a{}^c  {(\gamma^\mu)}{}_c{}^b]  ~~~.
 }
\eqno(A.7) 
 $$
Next a direct set of calculations show the following properties: \\
$$ \eqalign{ 
{(\gamma^\mu)}{}_a{}^c  C_{cb} \,&=\,  {(\gamma^\mu)}{}_b{}^c  C_{ca}  
~~~.}
\eqno(A.8) 
$$
$~$ \newline 
$$ \eqalign{ 
(\sigma^{\mu\nu})_{ab} \,&=\, (\sigma^{\mu\nu})_{ba} ~~~,
 }
\eqno(A.9) 
 $$
 $~$ \newline 
$$ \eqalign{ 
 {(\gamma^5\gamma^0)}{}_a{}^b  \,&=\,  -(\sigma^2 \otimes{\bf I}_2  ){}_a{}^b  ~~~, \cr
 {(\gamma^5\gamma^1)}{}_a{}^b  \,&=\,  i(\sigma^1 \otimes \sigma^3   ){}_a{}^b ~~~, \cr
 {(\gamma^5\gamma^2)}{}_a{}^b  \,&=\, -i(\sigma^3 \otimes{\bf I}_2   ){}_a{}^b ~~~, \cr
 {(\gamma^5\gamma^3)}{}_a{}^b  \,&=\, -i(\sigma^1 \otimes \sigma^1   ){}_a{}^b ~~~,
 }
\eqno(A.10) 
$$
$~$ \newline 
$$ \eqalign{ 
 {(\gamma^5\gamma^\mu)}{}_a{}^c  C_{cb} = - {(\gamma^5\gamma^\mu)}{}_b{}^c  C_{ca}
 ~~~.
 }
\eqno(A.11) 
$$

\newpage
\noindent
{\Large\bf Appendix B: ${ \bf{ {\cal GR} ({\rm d}_L, \, {\rm d}_R, \,  {\cal N})}}$ Closure Terms}

In the case of the on-shell Chiral Multiplet we find
$$ \eqalign{
{~~~~~}
(\,{\rm R}_\rJ\,)_\hi{}^j\>(\,{\rm L}_\rI\,)_j{}^\hk + (\,{\rm R}_\rI\,)_\hi{}^j\>(\,{\rm L}_\rJ\,)_j{}^\hk
&=~  \d{}_{\rI\rJ} \, ({\bf I})_\hi{}^\hk 
 ~+~  \ [\, {\vec {\a}} {\b}^1 \, ]_{\rI\rJ}\,  \cdot \,
(\, {\vec {\a}} {\b}^1 \, )_\hi{}^\hk   
~~.
}  \label{NotGDNAlg1}
\eqno(B.1)
$$
in place of the second equation of (\ref{GarDNAlg1}).
 
In the case of the Double Tensor Multiplet we will calculate the left hand sides of both 
the first and second equations in (\ref{GarDNAlg1}).  We find
$$ \eqalign{  {~~~~}
 (\,{\rm L}_\rI\,)_i{}^\hj\>(\,{\rm R}_\rJ\,)_\hj{}^k + (\,{\rm L}_\rJ\,)_i{}^\hj\>(\,{\rm R}_\rI\,)_\hj{}^k
&=  ~  2 \,  \d_{\rI\rJ}\,   (\,  {\bf I}_2 \otimes {\bf I}_3  
\, )_i{}^k ~-~ 2 \,   [\,  {\vec \a}\,  \b^1 \, ]_{\rI\rJ}\, \cdot \,  (\,  \s^2  \otimes   {\vec J} \, )_i{}^k
  ~~,    
}  \label{NotGDNAlg2}
 \eqno(B.2)
 $$
 where in writing this expression, we have introduced the dimensionless generators of 
 spin-1 angular momentum denoted by $J_1$, $J_2$ and $J_3$.  We simply note
 $$ \eqalign{
 J_1 ~&=~ \left(\begin{array}{ccc}
~0 & ~~0 &  ~~  i\\
~0 & ~~0 &  ~~0 \\
~ -\, i & ~~0 &  ~~0\\
\end{array}\right)  ~~,~~   J_2 ~=~ \left(\begin{array}{ccc}
~0 & ~~ i &  ~~0\\
~ -\, i & ~~0 &  ~~0 \\
~0 & ~~0 &  ~~0\\
\end{array}\right)  ~~,   
 \cr
 &~~~~~~~~~~~~~~~~~~
  J_3 ~~=~ \left(\begin{array}{ccc}
~0 & ~~0 &  ~~0\\
~0 & ~~0 &  ~~ i \\
~0 & ~~ -\, i &  ~~0\\
\end{array}\right) ~~.
  }  \label{Jmatrix}
 \eqno(B.3)
 $$
satisfy the commutation relationships
$$
\left[ \, J_i ~,~ J_j   \, \right] ~=~  i \, \e{}_{i \, j \, k} J_k ~~.
\eqno(B.4)
 $$
These relations are recognized as those for the usual generators of angular momentum.  We can 
continue and find the result
$$ \eqalign{
{~~~~~}
(\,{\rm R}_\rJ\,)_\hi{}^j\>(\,{\rm L}_\rI\,)_j{}^\hk + (\,{\rm R}_\rI\,)_\hi{}^j\>(\,{\rm L}_\rJ\,)_j{}^\hk
&=~ 3 \,  \d{}_{\rI\rJ} \, ({\bf I})_\hi{}^\hk 
 ~-~  \ [\, {\vec {\a}} {\b}^1 \, ]_{\rI\rJ}\,  \cdot \,
(\, {\vec {\a}} {\b}^1 \, )_\hi{}^\hk   
~~,
}  \label{NotGDNAlg3}
 \eqno(B.5)
 $$
 which is very similar to the case of the on-shell chiral multiplet given above
 (\ref{NotGDNAlg1}).  This similarity is so striking that one might hope for its
universality.  All such hopes vanish from the same calculation in the context
of the on-shell Vector Multiplet.  

For the second equation in (\ref{GarDNAlg1}) evaluated on the Vector Multiplets
we find the result
$$ \eqalign{
{~~~~~}
(\,{\rm R}_\rJ\,)_\hi{}^j\>(\,{\rm L}_\rI\,)_j{}^\hk + (\,{\rm R}_\rI\,)_\hi{}^j\>(\,{\rm L}_\rJ\,)_j{}^\hk
&=~ \fracm 32 \, \d_{\rI\rJ}\,  (\,  {\bf I}_4  \,)_\hi{}^\hk 
~-~ \fracm 12 \,  [\,  {\vec \a} \,  \b^2  \, ]_{\rI\rJ}\, \cdot \,  (\,  {\vec \a} \,  \b^2 
\, )_\hi{}^\hk  \cr
&~~~~+~ \fracm 12 \,  [\,   {\vec \a} \,  \b^1 \, ]_{\rI\rJ}\, \cdot \,  (\,  {\vec \a} \, 
\b^1   \, )_\hi{}^\hk  \cr
&~~~~+~ \fracm 12 \,  [\,   {\vec \a} \,  \b^3 \, ]_{\rI\rJ}\, \cdot \,  (\,  {\vec \a} \, 
\b^3   \, )_\hi{}^\hk  ~~.
}  \label{NotGDNAlg4}
 \eqno(B.6)
 $$
Interestingly, these calculations show some general regularities though still to 
a large degree, the exact nature of ${ { {\cal GR} ({\rm d}_L, \, {\rm d}_R, \,  {\cal 
N})}}$ for ${\rm d}_L$ $\ne $ ${\rm d}_R$ remains a mystery.

Aside from the identity matrix common to (B.1), (B.2), (B.5), and (B.6), there is an 
interesting similarity of the matrices that do appear on the right sides of the equations.  
These matrices that appear in (B.1), (B.2), (B.5), and (B.6), can be expressed in the
form utilizing matrix representations of SU(2) and are characteristic of theories 
realizing broken chiral SU(2) $\times$ SU(2) symmetries and none of the results 
in this appendix respect the full chiral SU(2) $\times$ SU(2) symmetry group seen 
in the off-shell theories.
\vskip.4in

\newpage \noindent
{\Large\bf Appendix C: A Primer On Adinkra Transforma- \\ $~~~~~~~~~~~~~~~~~~$
 tions}

In the discussion of chapter five, the two Adinkras shown in (\ref{CisTrnsAdnks}) were
used to generate the corresponding matrices in (\ref{MxB8sis1}) and (\ref{MxB8sis2})
without explanation of the intervening steps.  In an effort to be as transparent as
possible, in this short appendix we present an explanation on how to read an
Adinkra and generate the corresponding matrices.

A large class of the solutions to the conditions in (\ref{GarDNAlg1}) and (\ref{GarDNAlg2})
have the property that L-matrices and R-matrices contain rows and columns
with:
\vskip.05in
 \indent (a.) each row (when regarded as a vector) is a unit vector,
\vskip.05in
\indent (b.) each column (when regarded as a vector) is a unit vector, and
\vskip.05in
\indent (c.) the set of d row-vectors (or column-vectors) is an orthonormal set.
\vskip.05in
 \noindent
Taken together, these conditions imply the entries in these matrices are equal
to +1, 0, or  -1.  Our conventions are such that we use solid lines to indicate a
value of + 1, a dashed line to indicate a value of -1 and no line at all to indicate
a zero entry.  The conditions in (\ref{MxB8sis1}) and (\ref{MxB8sis1}) require
$\cal N$ linearly independent matrices in order for the representation to be
faithful.  For this purpose, $\cal N$ distinct colors are used in an Adinkra.

Rather than continue with a recitation of rules, it is easier to begin with a simple
example.  The basic $\cal N$ = 2 Adinkra appears as below.
$$
\vCent
{\setlength{\unitlength}{1mm}
\begin{picture}(-20,-140)
\put(-30,-50){\includegraphics[width=2.5in]{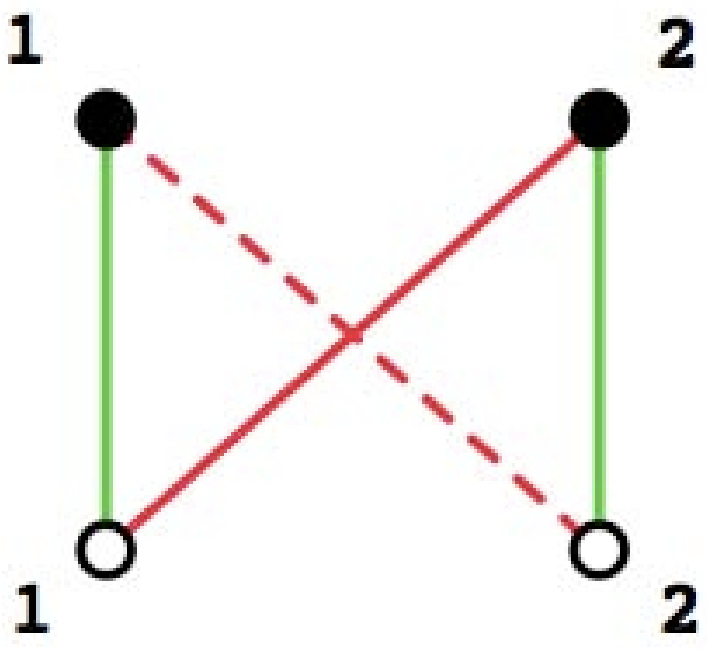}}
\end{picture}}.
\label{1M}
\eqno(C.1)
$$ \vskip 1.6in
\noindent
For the purpose of this appendix, we have numbered the white nodes and the
black nodes.  

Instead of regarding the white nodes as bosons and the black nodes as fermions, we can 
instead think of the white nodes as being associated with the rows in a matrix and the black 
nodes as being associated with the columns in a matrix.  This is a two-color Adinkra.  So it 
is necessarily associated with two matrices that we can denote by L${}_1$ and L${}_2$.  
We have a choice on how to associate which matrix with which color so we choose to 
associate the green edges with L${}_1$ and the red edges with L${}_2$.  

In order to concentrate on L${}_1$, the Adinkra (C.1) may be viewed through a ``green-pass'' 
filter that only allows the green edges to show.  Thus we arrive at the image below.
$$
\vCent
{\setlength{\unitlength}{1mm}
\begin{picture}(-20,-140)
\put(-40,-50){\includegraphics[width=2.5in]{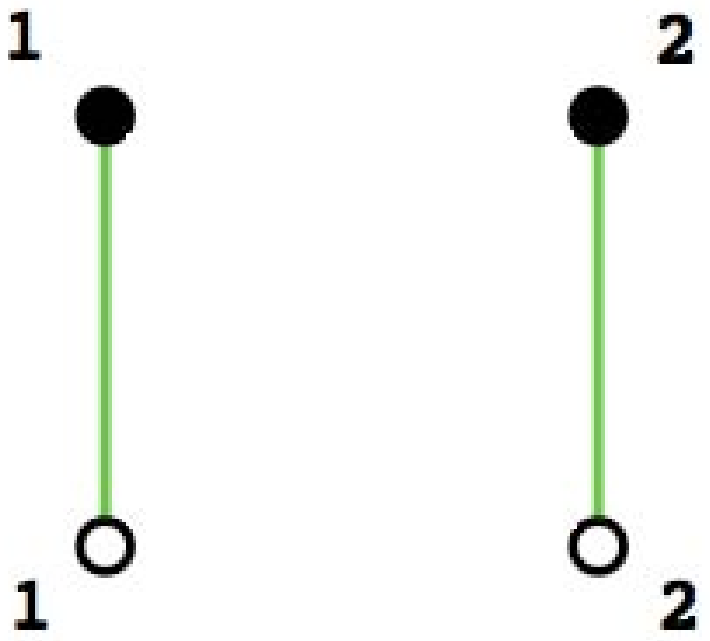}}
\end{picture}}.
\label{1Ma}
\eqno(C.2)
$$  \vskip 1.6in
\noindent
The information contained in this image is a factor of 1 appears in the first row and first
column of the matrix as well as a factor of 1 appears in the second row and second column 
of the matrix.  In other words this is the identity matrix ${\bf I}_2$.

In order to concentrate on L${}_2$, the Adinkra (C.1) may be viewed through a ``red-pass'' 
filter that only allows the red edges to show.  Thus we arrive at the image below.
$$
\vCent
 {\setlength{\unitlength}{1mm}
  \begin{picture}(-20,-140)
   \put(-40,-50){\includegraphics[width=2.5in]{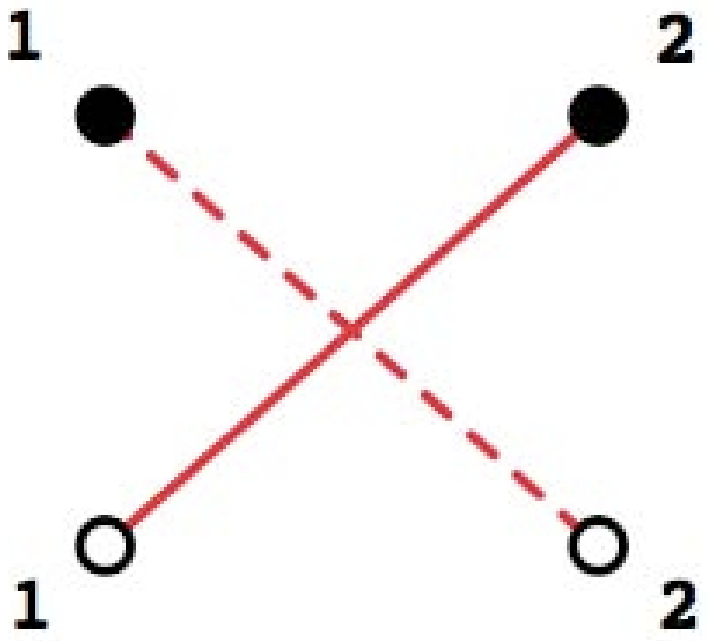}}
  \end{picture}}.
  \label{1Mb}
  \eqno(C.3)
$$   \vskip 1.6in
\noindent
The information contained in this image is that a factor of 1 appears in the first row and 
second column of the matrix as well as a factor of $-$ 1 appears in the second row and 
first column of the matrix.  In other words this is the matrix $i \, \s^2$.
So the Adinkra in (C.1) is associated with L${}_1$ and L${}_2$ via the equation
$$
\left( \, {\rm L}_1 , \, {\rm L}_2 \, \right) ~=~ \left( \, {\bf I}_2 , \, i{\s^2} \, \right) ~~.
\label{1Meq}
\eqno(C.4)
$$
It is notable that with complete fidelity, all the information of the matrices are contained in the
Adinkra.  In other words the Adinkra is a faithful representation of these matrices.

In a similar manner, the Adinkra whose image appears immediately below
$$
\vCent
{\setlength{\unitlength}{1mm}
\begin{picture}(-20,-140)
\put(-30,-50){\includegraphics[width=2.5in]{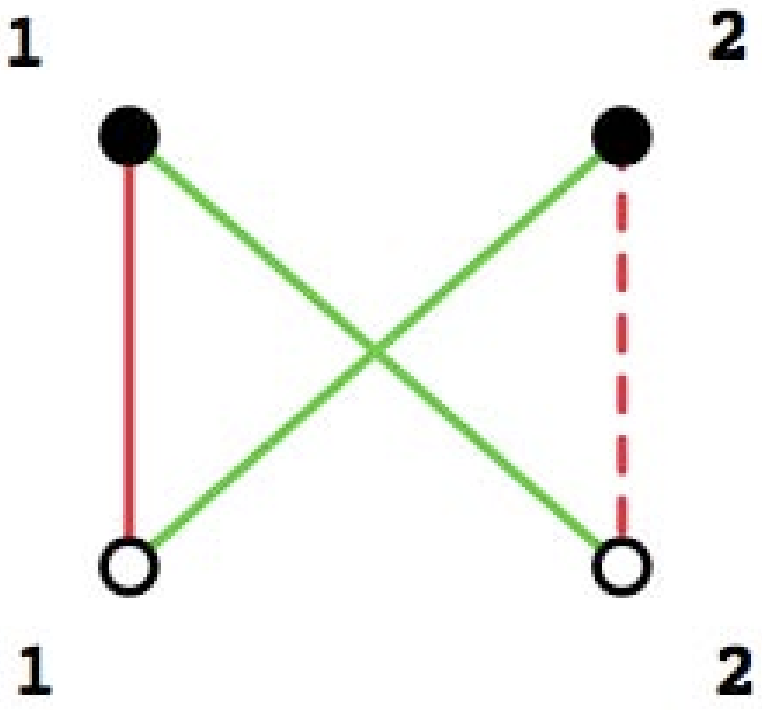}}
\end{picture}}.
\label{2M}
\eqno(C.5) 
$$    \vskip 1.6in
\noindent
possesses a `green-pass' filtered image of the form
$$
\vCent
 {\setlength{\unitlength}{1mm}
  \begin{picture}(-20,-140)
   \put(-30,-52){\includegraphics[width=2.5in]{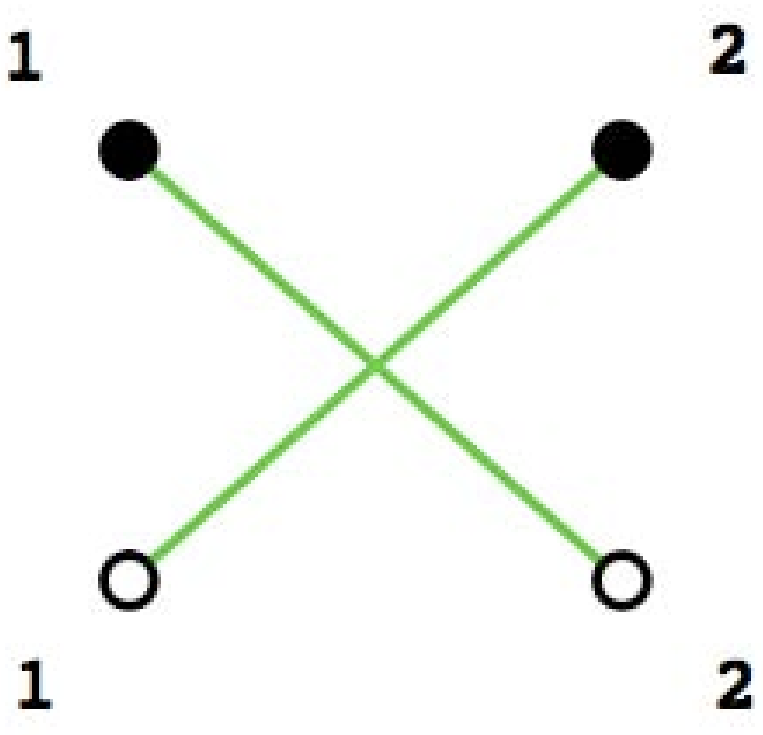}}
  \end{picture}}.
  \label{2Ma}
  \eqno(C.6) 
$$     \vskip 1.8in
\noindent
and possesses a `red-pass' filtered image of the form below.
$$
\vCent
 {\setlength{\unitlength}{1mm}
  \begin{picture}(-20,-140)
   \put(-30,-49.5){\includegraphics[width=2.5in]{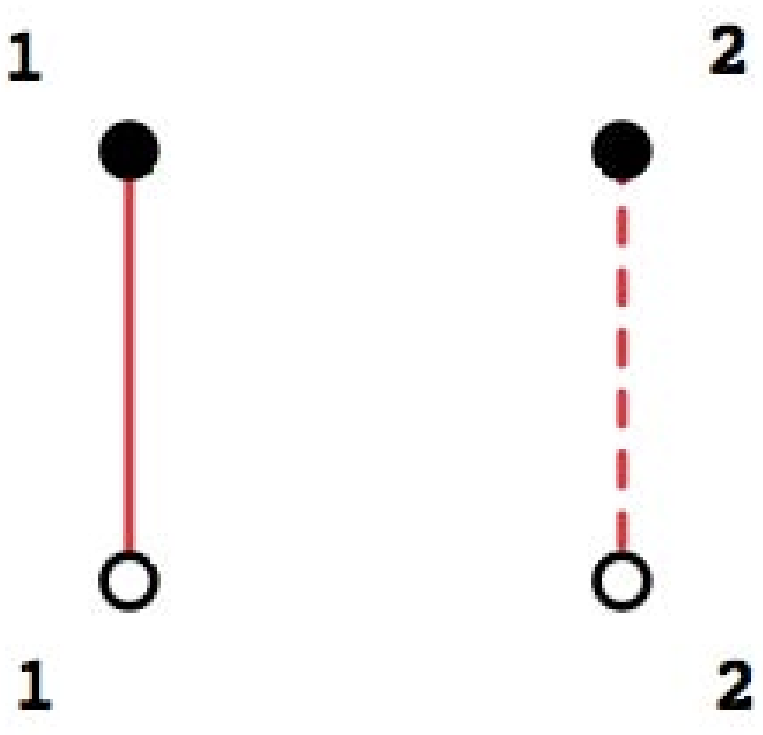}}
  \end{picture}}.
  \label{2Mb}
    \eqno(C.7) 
$$    \vskip 1.6in

Clearly, the Adinkra in (C.1) is different from the one in (C.5).  So the matrices associated with the
latter cannot be the same as those associated with the former.  We will denote the matrices associated
with the latter by ${\Hat {\rm L}}_1$ and ${\Hat {\rm L}}_2$.  By using the same logic that led to (C.4)
we find
$$
\left( \, {\Hat {\rm L}}_1 , \, {\Hat {\rm L}}_2 \, \right) ~=~ \left( \,  {\s^1},  \, \s^3 \, \right) ~~.
\label{2Meq}
\eqno(C.8)
$$
\noindent
However, there is a visual relation between (C.1) and (C.5).  If the two black nodes at the top of the
first Adinkra are exchanged, then Adinkra (C.1) changes into Adinkra (C.5).  Furthermore, it can be
verified that the sets of matrices given in (C.4) and (C.8) satisfy the conditions in (\ref{eq:E01}) and
(\ref{eq:E02}).

In equations (\ref{eq:E03}) and (\ref{eq:E04}) there were defined matrices $\cal X$ and $\cal Y$ that
transform L-matrices and R-matrices along orbits and define a class structure.  It might be possible
to work out the explicit forms of $\cal X$ and $\cal Y$ to relate the matrices in (C.4) to those in
(C.8).  It is straightforward calculation to show the required matrices take the forms
$$ \eqalign{
{\cal  X} ~=~ { {k_1 \, {\bf I} ~+~ i\,  k_2 \, \s^2 } \over {\sqrt {~ k_1^2 ~+~ k_2^2 ~}} } ~~~,~~~
{\cal  Y} ~=~  { { k_1 \, \s^1  ~-~    k_2 \, \s^3  }  \over {\sqrt {~k_1^2 ~+~ k_2^2~ }} }   ~~~,
}
\label{eq:E10}
\eqno(C.9)
$$
where $k_1$ and $k_2$ are arbitrary real parameters.  Notice for the choice $k_2$ = 0, this set of
transformation corresponds to the identity map acting on the white nodes and a pure exchange on
the black nodes as was the visual intuition gained by comparing (C.1) to (C.5).  The two matrices
in (C.9) effectuate the exchange of the two closed nodes that occur in the transformation from (C.1)
to (C.5).  

Two additional $\cal N$ = 2 Adinkras are shown in (C.10).
$$
\vCent
 {\setlength{\unitlength}{1mm}
  \begin{picture}(-20,-140)
   \put(-70,-51){\includegraphics[width=2.5in]{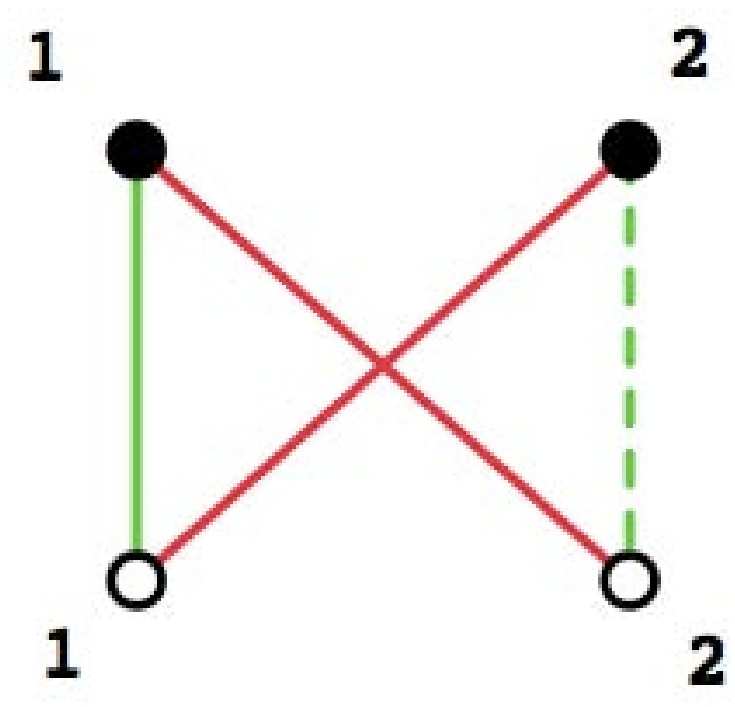}}
  \end{picture}}
  \label{3M}
$$

$$
\vCent
{\setlength{\unitlength}{1mm}
\begin{picture}(-20,-140)
\put(0,-32.5){\includegraphics[width=2.6in]{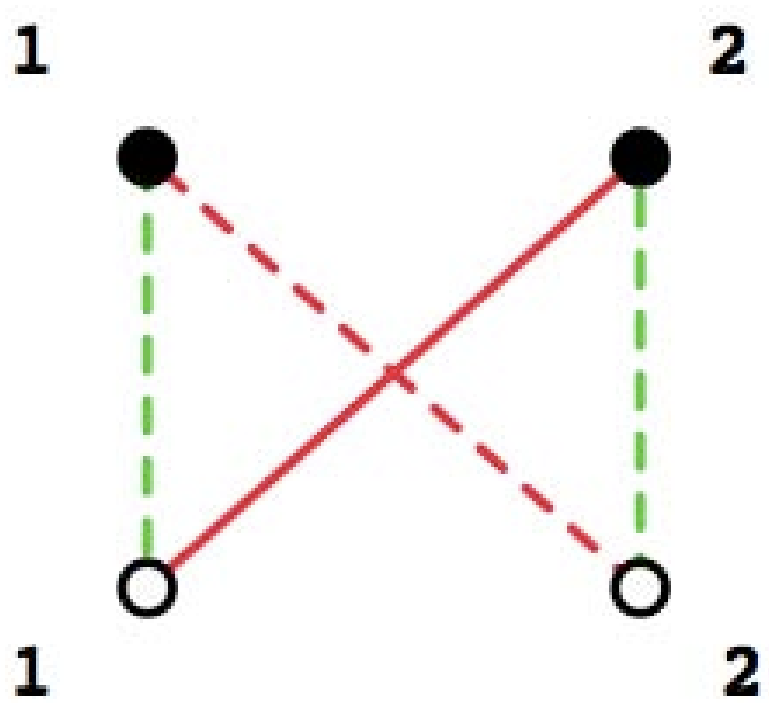}}
\end{picture}}
\label{4M}
\eqno(C.10)
$$ \vskip 1in \noindent
For the leftmost image, we have
$$
\left( \, {\Tilde {\rm L}}_1 , \, {\Tilde {\rm L}}_2 \, \right) ~=~ \left( \,  {\s^3},  \, \s^1 \, \right) ~~,
\label{3-4Meqa}
\eqno(C.11)
$$
and for the rightmost image, there is
$$
\left( \, {\Bar {\rm L}}_1 , \, {\Bar {\rm L}}_2 \, \right) ~=~ \left( \,  - \, {\bf I}_2,  \, i \, \s^2 \, \right) ~~.
\label{3-4Meqb}
\eqno(C.12)
$$
One choice of $\cal X$ and $\cal Y$ which relates the first of these to (C.1) is given by
$\cal X$ $=$ $\s^3$ and $\cal Y$ $=$ ${\bf I}_2$.  This effectuates a change of sign 
to the links attached to the open node at position 2 in the image of (C.1).  For the second
in (C.10) one set of matrices we see $\cal X$ $=$ $- \, \s^3$ and $\cal Y$ $=$ $\s^3$ will 
relate it to (C.1). This effectuates a change of sign to the links attached to the open
node at position 1 {\em {and}}  a change of sign to the links attached to the open
closed node at position 2.

With a bit of practice, it become very simple to use an Adinkra to generate a corresponding
set of matrices. However, the real advantage of Adinkras, used in the work of
the DFGHILM collaboration, is the ability to visually manipulate (using the Adinkramat) these
images to change basis and generally investigate the Garden Algebra matrices.

The Adinkra of (C.1) is also associated with a collection of superfields and spinorial 
differential equations that relate them.
$$
\eqalign{ 
{~~~~~~~~~~~~~~~~~~~~~~~~~~~~~} {~~}&{~~} \cr
{\rm D}{}_{1} \Phi_1 ~&=~i \, \Psi_1  ~~~, ~~  {\rm D}{}_{2} \Phi_1 ~=~ i \,  \Psi_2 ~~,  \cr
{\rm D}{}_{1} \Phi_2 ~&=~i \, \Psi_2  ~~~, ~~  {\rm D}{}_{2} \Phi_2 ~=~- \, i \,  \Psi_1 ~~,  \cr
 {\rm D}{}_{1} \Psi_1 ~&=~ \pa_{\t} \Phi_1 ~~,~~   {\rm D}{}_{2} \Psi_1 ~=~-\, \pa_{\t} \Phi_2 ~~, \cr
 {\rm D}{}_{1} \Psi_2 ~&=~ \pa_{\t} \Phi_2 ~~,~~   {\rm D}{}_{2} \Psi_2 ~=~ \pa_{\t} \Phi_1 ~~. \cr 
}
\vCent
{\setlength{\unitlength}{1mm}
\begin{picture}(-20,-140)
\put(-132,-30){\includegraphics[width=2.3in]{1M}}
\end{picture}}
\label{1MaX}
\eqno(C.13)
$$  \vskip .2in 
\noindent
Here the bosonic superfields $\Phi_1$ and $\Phi_2$ are associated with the open nodes
\# 1 and \# 2 at the lowest level of the Adinkra.  The fermionic superfields $\Psi_1$ and $\Psi_2$ 
are associated with the closed nodes \# 1 and \# 2 at the highest level of the Adinkra.  The spinorial 
derivative D${}_1$ is associated with green edges and the spinorial derivative D${}_2$ is associated 
with  red edges.

The process of ``lifting a node'' can be shown by first making one local redefinition
and one the non-local redefinition; $\Phi_1 ~\to ~ A$, $\Phi_2 ~\to ~\pa_{\t}^{-1} F$. 
Due to the second equation, the field $F$ has a higher engineering dimension than $\Phi_2$ and
accordingly the node associated with it is lifted in the Adinkra which can be redrawn as

$$
\eqalign{ 
{~~~~~~~~~~~~~~~~~~~~~~~~~~~~~~~~~~~~~~} 
&~~~~~~{\rm D}{}_{1} A ~=~i \, \Psi_1  ~~~~~~, ~~  {\rm D}{}_{2} A ~=~ i \,  \Psi_2 ~~,  \cr
&~~~~~~{\rm D}{}_{1} F ~=~i \, \pa_{\t} \Psi_2  ~~~, ~~  {\rm D}{}_{2} F ~=~- \, i \, \pa_{\t} \Psi_1 ~~,  \cr
&~~~~~~{\rm D}{}_{1} \Psi_1 ~=~ \pa_{\t} A ~\,~~~,~~   {\rm D}{}_{2} \Psi_1 ~=~- \, F ~~, \cr
&~~~~~~{\rm D}{}_{1} \Psi_2 ~=~ F ~~~~~~~,~~   {\rm D}{}_{2} \Psi_2 ~=~  \pa_{\t} A  ~~. \cr 
}
\vCent
{\setlength{\unitlength}{1mm}
\begin{picture}(-20,-140)
\put(-134,-30){\includegraphics[width=2.3in]{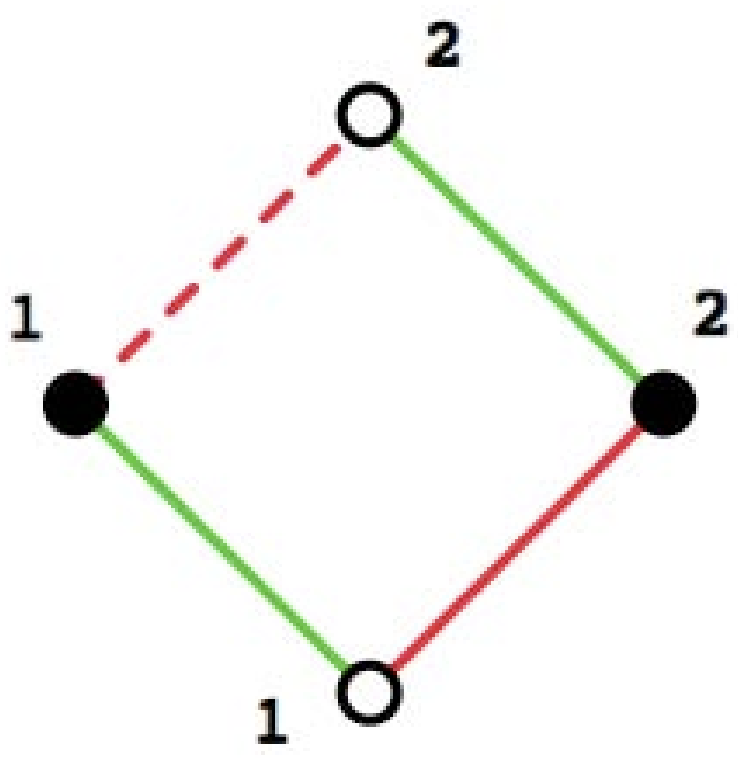}}
\end{picture}}
\label{AdnkLFT}
\eqno(C.14)
$$  \vskip .2in 
\noindent
The association of one superfield with each node in an Adinkra and the association of each
Adinkra color-edge with a distinct D-operator was implicitly introduced in the work of
the DFGHILM collaboration seen in \cite{AdnkDynam}.

\newpage \noindent
{\Large\bf Appendix D: SU(2) $\otimes$ SU(2) Symmetry \&  \\ $~~~~~~~~~~~~~~~~~~$
 Quartic Chromocharacters}

For convenience of this discussion, let us begin by gathering the quartic chromocharacters
here below
$$ \eqalign{
\varphi^{(2)} {}_{\bj I }  {}^{\bj J }  {}_{\bj K }  {}^{\bj L } (I) ~&=~  4 \, 
\left[~ \delta {}_{\bj I }  {}^{\bj J }  \delta{}_{\bj K }  {}^{\bj L }  ~+~  
[\vec {\b}]  {}_{\bj I }  {}^{\bj J }  \cdot \,   [\vec {\b}]{}_{\bj K }  {}^{\bj L }   ~ 
\right] ~~~, \cr
{~~~~~~} \varphi^{(2)}
{}_{\bj I }  {}^{\bj J }  {}_{\bj K }  {}^{\bj L } 
(III) ~&=~ 4 \, 
\left[~ \delta {}_{\bj I }  {}^{\bj J }  \delta{}_{\bj K }  {}^{\bj L }  ~+~  
[\vec {\a}]  {}_{\bj I }  {}^{\bj J }  \cdot \,   [\vec {\a}]{}_{\bj K }  {}^{\bj L }   ~ 
\right] ~~~, \cr
\varphi^{(2)} 
{}_{\bj I }  {}^{\bj J }  {}_{\bj K }  {}^{\bj L } (V) ~&=~  4 \, 
\left[~ \delta {}_{\bj I }  {}^{\bj J }  \delta{}_{\bj K }  {}^{\bj L }  ~+~  
[\vec {\a}]  {}_{\bj I }  {}^{\bj J }  \cdot \,   [\vec {\a}]{}_{\bj K }  {}^{\bj L }   ~ 
\right] ~~~, \cr
\varphi^{(2)}
{}_{\bj I }  {}^{\bj J }  {}_{\bj K }  {}^{\bj L } (II)
~&=~  2\, \delta {}_{\bj I }  {}^{\bj J }  \delta{}_{\bj K }  {}^{\bj L }    
+~ 2 \, [\, \b^1  \, ]  {}_{\bj I }  {}^{\bj J }  \,  [\, \b^1 \, ]{}_{\bj K }  {}^{\bj L }  ~~~, \cr
\varphi^{(2)} 
{}_{\bj I }  {}^{\bj J }  {}_{\bj K }  {}^{\bj L } (IV)
~&=~  6 \, \delta {}_{\bj I }  {}^{\bj J }  \delta {}_{\bj K }  {}^{\bj L }  
  ~+~ 6 \, [\,  \b^1 \, ] {}_{\bj I }  {}^{\bj J } \,  [\,  \b^1 
\, ]{}_{\bj K }  {}^{\bj L }     \cr
&  ~~~~+~ 4 \, [\, {\vec {\a}} \, \b^1 \, ] {}_{\bj I }  {}^{\bj J } \,  \cdot \,  [\, {\vec {\a}} \, \b^1 \, ]
{}_{\bj K }  {}^{\bj L }
~+~ 4 \, [\, {\vec {\a}}  \, ] {}_{\bj I }  {}^{\bj J }   
\,  \cdot \,  [\, {\vec {\a}} \, ]{}_{\bj K }  {}^{\bj L }   \cr
\varphi^{(2)} 
{}_{\bj I }  {}^{\bj J }  {}_{\bj K }  {}^{\bj L } (VI)
~&=~    3 \,  \delta {}_{\bj I }  {}^{\bj J }  \delta{}_{\bj K }  {}^{\bj L }  
~+~ 2 \, [\,  \a^2 \, ] {}_{\bj I }  {}^{\bj J }  \, 
[\, \a^2 \, ]{}_{\bj K }  {}^{\bj L }  ~+~ 2 \, [\, \b^2  \, ] {}_{\bj I }  {}^{\bj J }  \,  
[\, \b^2 \, ] {}_{\bj K }  {}^{\bj L }   \cr
&  ~~~
~+~ 2 \, [\,  \a^1 \, ] {}_{\bj I }  {}^{\bj J } \,  [\,  \a^1 \, ]{}_{\bj K }  {}^{\bj L } 
~~~~~,
} \eqno(D.1)
\label{D1}
$$
where we have used a Euclidean metric to raise a pair of indices.  We next observe
the relations
$$  \eqalign{
\Big[ \,   \a^{\rm A}  ~,~   \a^{\rm B}   \, \Big]   ~=~ i \, 2 \,  \e{}^{ \rm A \, \rm B \, \rm C} 
 &\a^{\rm C}   ~~~,~~~  \Big[ \,   \b^{\rm A}  ~,~   \b^{\rm B}   \, \Big]   ~=~ i \, 2 \,  \e{}^{ 
 \rm A \, \rm B \, \rm C}  \b^{\rm C}   ~~~, \cr 
 &\Big[ \,   \a^{\rm A}  ~,~   \a^{\rm B}   \, \Big]   ~=~ 0 ~~~. 
}
\label{D2}
\eqno(D.2)
$$
These imply that a group element of SU(2) $\otimes$ SU(2) denoted by
$\cal G$ can be written in the form
$$
\left[ {\cal G}(u, v) \right]{}_{\bj I }  {}^{\bj J }  ~=~ \left[ exp( i \fracm 12 \, 
u^{\rm A}  \a^{\rm A})   \,  exp( i \fracm 12 \, v^{\rm A}  \b^{\rm A})  \right]{}_{\bj I }  {}^{\bj J }   ~~~.
\eqno(D.3)
$$
We next calculate the following results
$$ \eqalign{
\left[ \varphi^{(2)} {}_{\bj I }  {}^{\bj J }  {}_{\bj K }  {}^{\bj L } (I) \right]^{\prime} ~&=~
\left[ {\cal G}(u, v) \right]{}_{\bj I }  {}^{\bj R } \left[ {\cal G}(u, v) \right]{}_{\bj J }  {}^{\bj T }  
\varphi^{(2)} {}_{\bj R }  {}^{\bj S }  {}_{\bj T }  {}^{\bj U } (I) \left[ {\cal G}{}^{-1}(u, v) 
 \right]{}_{\bj S }  {}^{\bj J }   \left[ {\cal G}{}^{-1}(u, v) \right]{}_{\bj U }  {}^{\bj L }  \cr
 &=~ \varphi^{(2)} {}_{\bj I }  {}^{\bj J }  {}_{\bj K }  {}^{\bj L } (I)  ~~~, \cr
 \left[ \varphi^{(2)} {}_{\bj I }  {}^{\bj J }  {}_{\bj K }  {}^{\bj L } (III) \right]^{\prime} ~&=~
\left[ {\cal G}(u, v) \right]{}_{\bj I }  {}^{\bj R } \left[ {\cal G}(u, v) \right]{}_{\bj J }  {}^{\bj T }  
\varphi^{(2)} {}_{\bj R }  {}^{\bj S }  {}_{\bj T }  {}^{\bj U } (III) \left[ {\cal G}{}^{-1}(u, v)
 \right]{}_{\bj S }  {}^{\bj J }  \left[ {\cal G}{}^{-1}(u, v) \right]{}_{\bj U }  {}^{\bj L }  \cr
 &=~ \varphi^{(2)} {}_{\bj I }  {}^{\bj J }  {}_{\bj K }  {}^{\bj L } (III)  ~~~, \cr
  \left[ \varphi^{(2)} {}_{\bj I }  {}^{\bj J }  {}_{\bj K }  {}^{\bj L } (V) \right]^{\prime} ~&=~
\left[ {\cal G}(u, v) \right]{}_{\bj I }  {}^{\bj R } \left[ {\cal G}(u, v) \right]{}_{\bj J }  {}^{\bj T }  
\varphi^{(2)} {}_{\bj R }  {}^{\bj S }  {}_{\bj T }  {}^{\bj U } (V)
 \left[ {\cal G}{}^{-1}(u, v) \right]{}_{\bj S }  {}^{\bj J }   \left[ {\cal G}{}^{-1}(u, v) \right]{
 }_{\bj U }  {}^{\bj L }  \cr
 &=~ \varphi^{(2)} {}_{\bj I }  {}^{\bj J }  {}_{\bj K }  {}^{\bj L } (V)  ~~~,
}
\eqno(D.4)
$$
which show that the quartic chromocharacters associated with the cases $I$,
$III$, and $V$, possess the full SU(2) $\otimes$ SU(2) symmetry under the group
element defined by (D.3).  On the otherhand, we also see
$$ \eqalign{
\left[ \varphi^{(2)}
{}_{\bj I }  {}^{\bj J }  {}_{\bj K }  {}^{\bj L } (II)  \right]^{\prime} ~&=~
 2\, \delta {}_{\bj I }  {}^{\bj J }  \delta{}_{\bj K }  {}^{\bj L }    
+~ 2 \, [\, {\Tilde \b}^1  \, ]  {}_{\bj I }  {}^{\bj J }  \,  [\,  {\Tilde \b}^1 
 \, ]{}_{\bj K }  {}^{\bj L }  ~~~, \cr
 \left[ \varphi^{(2)}
{}_{\bj I }  {}^{\bj J }  {}_{\bj K }  {}^{\bj L } (IV)  \right]^{\prime} ~&=~
  6 \, \delta {}_{\bj I }  {}^{\bj J }  \delta {}_{\bj K }  {}^{\bj L }  
  ~+~ 6 \, [\, {\Tilde \b}^1  \, ] {}_{\bj I }  {}^{\bj J } \,  [\,  {\Tilde \b}^1 
\, ]{}_{\bj K }  {}^{\bj L }     \cr
&  ~~~~+~ 4 \, [\, {\vec {\a}} \, {\Tilde \b}^1  \, ] {}_{\bj I }  {}^{\bj J } \,  \cdot \,  [\, {\vec {\a}} \, 
{\Tilde \b}^1  \, ]{}_{\bj K }  {}^{\bj L } ~+~ 4 \, [\, {\vec {\a}}  \, ] {}_{\bj I }  {}^{\bj J }   
\,  \cdot \,  [\, {\vec {\a}} \, ]{}_{\bj K }  {}^{\bj L } 
}
\eqno(D.5)
$$
where
$$
[\,   {\Tilde \b}^1   \, ]  {}_{\bj I }  {}^{\bj J }  ~=~
 [\, e^{ i \fracm 12 \, v^{\rm A}  \b^{\rm A} } \b^1 e^{- i \fracm 12 \, v^{\rm A}  \b^{\rm A} } 
 \, ]  {}_{\bj I }  {}^{\bj J }   ~~~.
 \eqno(D.6)
$$
The results in (D.5) show that the quartic chromocharacters in the cases of $II$ and
$IV$ only possess an SU(2) $\otimes$ U(1) symmetry.

We may write the transformed final chromocharacter for case-$VI$ in the form
$$
 \eqalign{  {~~~~~~}
\left[ \varphi^{(2)}
{}_{\bj I }  {}^{\bj J }  {}_{\bj K }  {}^{\bj L } (VI)  \right]^{\prime} ~&=~    
3 \,  \delta {}_{\bj I }  {}^{\bj J }  \delta{}_{\bj K }  {}^{\bj L }  
~+~ 2 \, [\,  {\vec \a} \, ] {}_{\bj I }  {}^{\bj J }  \, \cdot \,
[\, {\vec \a} \, ]{}_{\bj K }  {}^{\bj L } 
 ~+~ 2 \, [\, {\Tilde \b}^2  \, ] {}_{\bj I }  {}^{\bj J }  \,  
[\, {\Tilde \b}^2 \, ] {}_{\bj K }  {}^{\bj L }   \cr
&  ~~~~-~ 2 \, [\,  {\Tilde \a}^3 \, ] {}_{\bj I }  {}^{\bj J } \,  [\,  {\Tilde \a}^3 \, ]{}_{\bj K }  {}^{\bj L } 
~~~~~,
} 
\eqno(D.7)
$$ 
where
$$
[\,   {\Tilde \a}^3   \, ]  {}_{\bj I }  {}^{\bj J }  ~=~
 [\, e^{ i \fracm 12 \, u^{\rm A}  \a^{\rm A} } \a^3 e^{- i \fracm 12 \, u^{\rm A}  \a^{\rm A} } 
 \, ]  {}_{\bj I }  {}^{\bj J }   ~~~.
 \eqno(D.6)
$$
This proves that at most this chromocharacter possess a U(1) $\otimes$ U(1) symmetry.

\end{document}